\documentclass[onecolumn,notitlepage,letterpaper,superscriptaddress,nofootinbib,longbibliography]{revtex4-2}

\usepackage{amsmath}
\allowdisplaybreaks[4]

\usepackage{graphicx}
\usepackage{amsfonts}
\usepackage{latexsym}
\usepackage{bbold}
\usepackage{calligra}
\usepackage{float}
\usepackage{ulem}
\usepackage{inputenc}
\usepackage{xspace}
\usepackage{url}
\usepackage{epstopdf}
\usepackage{tikz}
\usepackage{amsthm}
\usepackage{cancel}
\usepackage{comment}
\usepackage{bigints}
\usepackage{ragged2e}
\usepackage[pdftoolbar = true, pdfstartview = FitH, pdfmenubar = true]{hyperref}
\hypersetup{
    colorlinks=true,
    linkcolor=blue,
    filecolor=magenta,      
    citecolor=blue
}

\usepackage{mathtools}

\newcommand{\be}{\begin{equation}}
\newcommand{\ee}{\end{equation}}
\newcommand{\beq} {\begin{equation}}
\newcommand{\eeq} {\end{equation}}
\newcommand{\ba}{\begin{eqnarray}}
\newcommand{\ea}{\end{eqnarray}}

\usepackage{subcaption}

\usepackage{mathtools}
\usepackage{tabularx}
\usepackage{array}
\allowdisplaybreaks

\usepackage{placeins}

\newcommand{\ABV}{\texorpdfstring{\ensuremath{\mathcal{A}\mathcal{B}\mathcal{V}}}{ABV}}

\newcommand{\ABVC}{\texorpdfstring{\ensuremath{\mathcal{A}\mathcal{B}\mathcal{V}\mathcal{C}}}{ABVC}}

\newcommand{\ABVa}{\texorpdfstring{\ensuremath{\mathcal{A}\mathcal{B}\mathcal{V}a}}{ABVa}}

\newcommand{\ABVb}{\texorpdfstring{\ensuremath{\mathcal{A}\mathcal{B}\mathcal{V}b}}{ABVb}}

\newcommand{\ABVab}{\texorpdfstring{\ensuremath{\mathcal{A}\mathcal{B}\mathcal{V}ab}}{ABVab}}

\newcommand{\ABVCabc}{\texorpdfstring{\ensuremath{\mathcal{A}\mathcal{B}\mathcal{V}\mathcal{C}abc}}{ABVCabc}}

\begin{document}
\title{Scalar field with nonminimal couplings to metric-affine geometry}
	
\author{Ilaria Andrei}
\email{ilaria.andrei@ut.ee}
\affiliation{Laboratory of Theoretical Physics, Institute of Physics, University of Tartu, W. Ostwaldi 1, 50411 Tartu, Estonia}

\author{Damianos Iosifidis}
\email{damianos.iosifidis@ut.ee}
\affiliation{Scuola Superiore Meridionale di Napoli (SSM),\\ Largo S. Marcellino, 10, 80138 Napoli, Naples, Italy}

\author{Laur Järv}
\email{laur.jarv@ut.ee}
\affiliation{Laboratory of Theoretical Physics, Institute of Physics, University of Tartu, W. Ostwaldi 1, 50411 Tartu, Estonia}

\author{Margus Saal}
\email{margus.saal@ut.ee}
\affiliation{Laboratory of Theoretical Physics, Institute of Physics, University of Tartu, W. Ostwaldi 1, 50411 Tartu, Estonia}

\date{\today}

\begin{abstract}
We study a scalar field nonminimally coupled to metric-affine gravity within actions linear in the affine curvature and containing all independent parity-even and parity-odd terms quadratic in torsion and nonmetricity, including mixed contractions. We also include Nieh-Yan-like derivative couplings between the scalar-field derivative and the four torsion and nonmetricity vectors. We derive the connection, metric, and scalar-field equations and, since the connection equation is algebraic, eliminate the independent connection on the generic nondegenerate branch to obtain an equivalent metric scalar-tensor theory in which the non-Riemannian interactions are encoded in an effective kinetic function. We classify several sectors and find that the pure quadratic nonmetricity and pure quadratic torsion sectors separately leave the Einstein-frame kinetic function unchanged relative to the simplest metric-affine scalar-tensor model, whereas their simultaneous presence, the derivative couplings, and generic mixed torsion-nonmetricity interactions can modify it. In the derivative-coupling sector, projective consistency imposes a relation among the couplings. We then consider polynomial coupling functions and study the resulting canonical field redefinition and Einstein-frame potentials. In particular, we illustrate how derivative and mixed torsion--nonmetricity couplings reshape quadratic and quartic potentials in canonical-field space, and show that a quadratic Jordan-frame potential can be mapped, for a suitable choice of derivative and nonminimal couplings, into a natural-inflation potential after canonical normalization. Finally, we separately impose the cosmological principle, determine the reduced combinations of quadratic couplings and the scalar field hypermomentum, and obtain the corresponding modified cosmological equations in the simplest sectors.
\end{abstract}
	
	\maketitle

\clearpage

\setcounter{tocdepth}{1}

\vspace*{0.8cm}

\begin{center}
{\large\bfseries Table of contents}
\end{center}

\vspace{2cm}

\begingroup
\setlength{\columnsep}{1.2cm}
\twocolumngrid

\makeatletter
\@starttoc{toc}
\makeatother

\onecolumngrid
\endgroup

\clearpage

\section{Introduction}
\label{intro}

General relativity (GR) provides an extraordinarily successful description of gravitation across a wide range of scales. Nevertheless, the physical origin of the accelerated phases in the history of the Universe, as well as the problem of consistently connecting gravity with quantum field theory, motivate the systematic study of extensions of GR \cite{CosmoVerseNetwork:2025alb, CANTATA:2021asi}. The primordial accelerated expansion associated with inflation was introduced in several works \cite{Starobinsky:1980te, Guth:1980zm, Linde:1981mu}, while its predictions are now tightly constrained by observations of the cosmic microwave background \cite{Planck:2018jri}. Numerous single scalar field inflationary models \cite{Martin:2013tda} can be classified according to the potential of the canonically normalized field in the minimally coupled (Einstein) frame \cite{Jarv:2016sow}.

In GR, spacetime is described by a Lorentzian metric $g_{\mu\nu}$ and its uniquely associated Levi-Civita connection $\tilde{\Gamma}^\lambda_{\phantom{\lambda}\mu\nu}$, 
which is both torsion-free and metric-compatible. The underlying geometry is therefore pseudo-Riemannian. Metric-affine gravity (MAG) generalizes this construction by treating the metric and the affine connection as independent fundamental variables \cite{Hehl:1976my, Hehl:1976kj, Hehl:1994ue, Sotiriou:2006qn, Olmo:2011uz, Vitagliano:2010sr, Iosifidis:2019dua, Iosifidis:2021pta}. On the geometric side, the independent connection introduces two tensors in addition to curvature: torsion, which is associated with the antisymmetric part of the connection and with the infinitesimal non-closure of parallelograms, and nonmetricity, which describes the change of the metric under parallel transport and may therefore modify the norms and relative angles of transported vectors. On the matter side, the independent connection is sourced by the hypermomentum tensor, defined as the response of the matter action to variations of the affine connection \cite{Hehl:1977fj,Hehl:1994ue}. Hypermomentum generalizes the spin current: its antisymmetric part describes intrinsic spin, its trace describes intrinsic dilation, and its symmetric traceless part describes intrinsic shear. It therefore encodes microscopic properties of matter that cannot be represented by the usual energy-momentum tensor alone. The role of these currents and their possible cosmological realizations has been investigated through hyperfluid models \cite{Iosifidis:2020gth, Iosifidis:2021nra, Obukhov:2023yti, Andrei:2024vvy}. The presence of independent curvature, torsion, and nonmetricity tensors permits the construction of a very large number of gravitational invariants \cite{Baldazzi:2021kaf}. Higher-order curvature terms in a general metric-affine theory may propagate additional degrees of freedom and can suffer from ghost or gradient instabilities unless suitable degeneracy conditions or additional symmetries are imposed \cite{BeltranJimenez:2019acz, BeltranJimenez:2020sqf, Percacci:2020ddy}. In the present work, we restrict ourselves to actions that are linear in the affine curvature and contain terms quadratic in torsion and nonmetricity \cite{Iosifidis:2020dck, Iosifidis:2021fnq, Iosifidis:2024bsq}, without direct couplings between curvature and torsion or nonmetricity. Within this class, the connection equations are algebraic in torsion and nonmetricity. On a nondegenerate branch, they can therefore be solved, and the independent connection can be integrated out, yielding an equivalent metric theory with a modified matter sector \cite{Iosifidis:2021tvx, Iosifidis:2021bad, Pradisi:2022nmh}.

Scalar fields nonminimally coupled to gravity provide an especially relevant application of this framework \cite{German:1985tuo,Kim:1986hh,Smalley:1986tr,Berthias:1993aa,Kozak:2018vlp,Aoki:2019rvi,BeltranJimenez:2020sih, Rigouzzo:2023sbb}. Already in the simpler Palatini formulation, treating the connection independently can lead to inflationary predictions that differ significantly from those of the corresponding metric theory \cite{Bauer:2008zj,Tenkanen:2020dge,Jarv:2020qqm,Racioppi:2019jsp,Gialamas:2020snr,Kubota:2020ehu,Jarv:2024krk,Gialamas:2024jeb}. More general metric-affine scalar theories permit couplings to torsion, nonmetricity, and parity-odd structures, thereby modifying the effective scalar dynamics and the relation between the original scalar field and its canonically normalized counterpart. Inflationary models in metric-affine, Einstein-Cartan, conformal metric-affine, and quadratic metric-affine settings have consequently received increasing attention \cite{Shimada:2018lnm,Mikura:2020qhc,Langvik:2020nrs,Rigouzzo:2022yan,Gialamas:2022xtt,Shaposhnikov:2020gts,Piani:2022gon,He:2024wqv}. When the connection is auxiliary and can be integrated out, these non-Riemannian interactions are transferred to the effective scalar kinetic function. They can therefore modify the canonical field redefinition and the shape of the potential when expressed in terms of the canonically normalized field.

In this work, we consider a scalar field coupled to a general metric-affine action that is linear in curvature and contains all independent parity-even and parity-odd terms quadratic in torsion and nonmetricity, including their mixed contractions. The corresponding coefficients are allowed to be arbitrary functions of the scalar field. We additionally include Nieh-Yan-like couplings between the scalar-field derivative and the four independent torsion and nonmetricity vectors. The general class of scalar-field metric-affine actions considered here is equivalent, up to integrations by parts and a change of basis, to that introduced in Ref.~\cite{Rigouzzo:2022yan}. Here, we develop the dynamical analysis further by deriving the complete connection, metric, and scalar-field equations and by systematically studying the different connection branches and quadratic sectors. Since the connection equations are algebraic, the affine connection can be integrated out on the nondegenerate branches, leading to an equivalent metric scalar-tensor theory. We determine which classes of non-Riemannian couplings generate genuine modifications of the effective kinetic function, study their consequences for canonical normalization and reshaping of the potential, and separately investigate the restrictions imposed by the cosmological principle and the resulting cosmological dynamics.

The paper is organized as follows. In Sec.~\ref{generalprocedure}, we introduce the geometric conventions and the general action, derive the connection, metric, and scalar-field equations, construct the dynamically equivalent metric theory, and introduce the polynomial couplings functions used in the subsequent analysis. In Sec.~\ref{ABVsection}, we study the basic metric-affine scalar-tensor model containing only the nonminimal coupling to curvature $\mathcal{A}$, kinetic function $\mathcal{B}$, and potential $\mathcal{V}$. In Sec.~\ref{ABVCsection}, we add the scalar field derivative couplings to the torsion and nonmetricity vectors and determine their effects on the field equations and the effective kinetic function. In Sec.~\ref{Alltheothercases}, we analyze the full quadratic theory with a generic nondegenerate set of solutions to the connection equation, and study its Riemannian, torsionless, and metric-compatible branches. In Sec.~\ref{subcase2}, we classify representative quadratic sectors, including the pure nonmetricity, pure torsion, combined pure torsion and nonmetricity terms, and mixed torsion--nonmetricity sectors, together with a universal coupling example. In Sec.~\ref{canonicalredefinition}, we study how the resulting Einstein-frame kinetic functions affect the canonical field redefinition and reshape the scalar potential, first through illustrative derivative-coupling and mixed torsion--nonmetricity examples and then through an explicit reconstruction of a natural-inflation potential. In Sec.~\ref{CPgeneral}, we separately impose the cosmological principle, determine its restrictions on torsion, nonmetricity, and the quadratic couplings, and study the resulting cosmological dynamics in representative sectors. Finally, in Sec.~\ref{conclusions}, we summarize our results and discuss possible directions for further work.

\section{General action and field equations}
\label{generalprocedure}

\subsection{Setup}\label{setup}
As fundamental geometric objects in our four-dimensional non-Riemannian space, we have the metric $g_{\mu\nu}$ and an independent affine connection which can be expanded as
\begin{equation}
\label{decgamma}
    \Gamma^{\lambda}{}_{\mu \nu} = \tilde{\Gamma}^{\lambda}{}_{\mu \nu} + N^{\lambda}{}_{\mu \nu}\,,
\end{equation}
where $\tilde{\Gamma}^\lambda_{\phantom{\lambda}\mu\nu}$ is the Levi-Civita connection
\beq
\label{lcconn}
    \tilde{\Gamma}^\lambda_{\phantom{\lambda}\mu\nu} = \frac{1}{2} g^{\rho\lambda}
    \left(\partial_\mu g_{\nu\rho} + \partial_\nu g_{\rho\mu} - \partial_\rho g_{\mu\nu}\right)
\eeq
with its associated covariant derivative denoted as $\tilde{\nabla}$.
The tensor $N^\lambda{}_{\mu\nu}$, which encodes the non-Riemannian part of the connection, is called distortion. It can be decomposed into
\beq
\label{distortion}
    N^\lambda{}_{\mu\nu} = {\frac12 g^{\rho\lambda}\left(Q_{\mu\nu\rho} + Q_{\nu\rho\mu}
    - Q_{\rho\mu\nu}\right)} - {g^{\rho\lambda}\left(S_{\rho\mu\nu} +
    S_{\rho\nu\mu} - S_{\mu\nu\rho}\right)} \,
\eeq
by introducing two fundamental tensors: nonmetricity and torsion.
The nonmetricity is defined as
\beq
    Q_{\alpha\mu\nu}:= -\nabla_{\alpha}g_{\mu\nu}  = -\partial_{\alpha}g_{\mu\nu} + \Gamma^{\lambda}_{\phantom{\lambda}\mu\alpha}g_{\lambda\nu}
    +\Gamma^{\lambda}_{\phantom{\lambda} \nu\alpha}g_{\lambda\mu}\,.
\eeq
This equation also fixes our index convention for the covariant derivative, i.e.\ in contrast to some conventions in the literature, the derivative index is the second lower index of the connection coefficients, see Schouten \cite{schouten1954ricci}. We define the raised covariant derivative as $\nabla^\alpha\equiv g^{\alpha\beta}\nabla_\beta$. The torsion tensor is defined as
\beq  S_{\mu\nu}^{\phantom{\mu\nu}\lambda}:=\Gamma^{\lambda}_{\phantom{\lambda}[\mu\nu]} \,,
\eeq
where the square brackets denote antisymmetrization ($A_{[ab]}\equiv\frac{1}{2} (A_{ab} - A_{ba})$). By contraction, their respective vectors are defined as
\begin{equation}
\label{QSvectors}
    t^{\alpha } := \varepsilon^{\alpha \beta \gamma \rho} \,S_{ \beta \gamma \rho}, \hspace{1 cm}    
    S_{\alpha } := S_{\alpha \beta }{}^{\beta }, \hspace{1 cm}
    Q_{\mu }:= g^{\alpha \lambda } Q_{\mu \alpha \lambda },  \hspace{1 cm} 
    q_{\mu } := g^{\alpha \lambda } Q_{\alpha \lambda \mu } \,,
\end{equation}
where $\varepsilon_{\alpha \beta \gamma \rho}$ is the Levi-Civita tensor.
The non-Riemannian curvature tensor of the affine connection is defined by
\begin{equation}\label{Riemanntensor}
R^\mu{}_{\nu\alpha\beta} := \partial_\alpha \Gamma^\mu{}_{\nu\beta} 
- \partial_\beta \Gamma^\mu{}_{\nu\alpha} 
+ \Gamma^\mu{}_{\lambda\alpha}\Gamma^\lambda{}_{\nu\beta} 
- \Gamma^\mu{}_{\lambda\beta}\Gamma^\lambda{}_{\nu\alpha}\,.
\end{equation}
Its contractions define three rank-2 tensors
\beq
    R_{\nu \beta}:=R^{\mu}_{\phantom{\mu} \nu \mu \beta}\,, \qquad 
    \widehat{R}_{\alpha \beta}:=R^{\mu}_{\phantom{\mu} \mu \alpha \beta} \,,\qquad \breve{R}^{\lambda}_{\phantom{\lambda} \kappa}
    :=R^{\lambda}_{\phantom{\lambda} \mu\nu\kappa}g^{\mu\nu} \,.
\eeq
Here, the first one is the generalized Ricci tensor, the second is the homothetic curvature, and the third is called the co-Ricci tensor.
The only non-trivial scalar that can be formed from the above Ricci tensors is the non-Riemannian  scalar curvature (or generalized Ricci scalar):
\beq \label{Ricci scalar}
    R:=R_{\mu\nu}g^{\mu\nu}=-\breve{R}_{\mu\nu}g^{\mu\nu}\,, \qquad  \widehat{R}_{\mu\nu}g^{\mu\nu}=0 \,.
\eeq
With the above conventions, the generalized Ricci scalar $R$ decomposes into its Riemannian part $\tilde{R}$ and distortion contributions as
\begin{equation}
\label{Rnonriem}
\begin{aligned}
   R &= \tilde{R} +\frac{1}{4} Q_{\mu \nu \sigma } Q^{\mu \nu \sigma } -  \frac{1}{2} Q_{\mu \nu \sigma } Q^{\nu \mu \sigma } 
   + \frac{1}{2} q_{\mu } Q^{\mu } 
   - \frac{1}{4} Q_{\mu } Q^{\mu } + 2 \,S_{\mu \sigma }{}^{\nu } S^{\mu }{}_{\nu }{}^{\sigma }\\
   &\qquad + S_{\mu \nu }{}^{\sigma } S^{\mu \nu }{}_{\sigma } + 2 Q_{\mu \nu \sigma } S^{\mu \nu \sigma } + 2  \,q_{\mu } S^{\mu } 
   - 2  \,Q_{\mu } S^{\mu } - 4  \,S_{\mu } S^{\mu } + \tilde{\nabla}_{\lambda }q^{\lambda } 
   -  \tilde{\nabla}_{\lambda }Q^{\lambda } - 4 \, \tilde{\nabla}_{\lambda }S^{\lambda } \,,
   \end{aligned}
\end{equation} 
where $\tilde{\nabla}$ is the Levi-Civita covariant derivative.

It is an interesting geometric fact that the Ricci scalar $R$ is invariant under projective transformation, which is defined as a vector shift of the affine connection \cite{Weyl1921,eisenhart1929non,Iosifidis:2019fsh}:
\beq
\label{projective}
    \Gamma^{\lambda}{}_{\mu\nu} \mapsto \Gamma^{\lambda}{}_{\mu\nu} + \delta^{\lambda}_{\mu}\xi_{\nu} \,.
\eeq
Whenever it is a symmetry of the action, the projective transformation \eqref{projective} represents a gauge freedom of the affine connection. This symmetry will play an important role below, since the connection field equations must be compatible with the projective invariance of the Ricci scalar unless the additional terms in the action explicitly break it.

We use the mostly plus convention for the metric signature $(-1,+1,+1,+1)$ and consider natural units: $\hbar=c=1$. Along the paper, derivatives with respect to the scalar field $\phi$ are indicated by the prime symbol.

\subsection{Action}
\label{subs:action}

We consider the most general action within the class that is linear in the affine curvature and quadratic in torsion and nonmetricity, each term endowed with scalar field dependent coefficients, while excluding direct couplings between curvature and the quadratic torsion or nonmetricity invariants,\footnote{The corresponding theory without the scalar field was studied in Actions 3.1--3.3 of Ref.~\cite{Iosifidis:2024bsq} and with a scalar field but in a different parametrization in Ref.\ \cite{Rigouzzo:2022yan}. The precise map to the parametrization used in Ref.\ \cite{Rigouzzo:2022yan} is given in Appendix \ref{AppendixC}. As a matter of fact, one may consider also another term linear in the connection that is parity odd. It is given by $\varepsilon^{\alpha\beta\gamma\delta} \,R_{\alpha\beta\gamma\delta}$ and is frequently called the Holst or Hojman term. However, performing a post-Riemannian expansion on the latter, its contributions can be reabsorbed in certain quadratic torsion and nonmetricity terms already included in our action. Furthermore, allowing in addition direct curvature--torsion or curvature--nonmetricity couplings would introduce many additional invariants and may lead to additional propagating degrees of freedom, whose stability requires a separate analysis beyond the scope of the present work.}
\begin{equation}
    \begin{aligned}
     S[g, \Gamma, \phi]
	=\frac{1}{2 \kappa}\int \mathrm{d}^{4}x \sqrt{-g} \Big[ & \mathcal{A}(\phi) R-\kappa \mathcal{B}(\phi) \partial_\mu \phi \partial^\mu \phi -2 \kappa\mathcal{V}(\phi) + \kappa\left(\mathcal{C}_1(\phi) Q^{\mu}+\mathcal{C}_2(\phi) q^{\mu} + \mathcal{C}_3(\phi)  S^{\mu}+\mathcal{C}_{4}(\phi) t^{\mu}\right)\partial_{\mu}\phi 
     \\&\\&
    +a_{1}(\phi)Q_{\alpha\mu\nu}Q^{\alpha\mu\nu} +
	a_{2}(\phi) Q_{\alpha\mu\nu}Q^{\mu\nu\alpha} +
	a_{3}(\phi) Q_{\mu}Q^{\mu}+
	a_{4}(\phi) q_{\mu}q^{\mu}\\
    &+
	a_{5}(\phi) Q_{\mu}q^{\mu}+a_{6}(\phi) \varepsilon^{\alpha\beta\gamma\delta}Q_{\alpha\beta\mu}Q_{\gamma\delta}{}^{\mu} \\&\\&
	+b_{1}(\phi)S_{\alpha\mu\nu}S^{\alpha\mu\nu} +
	b_{2}(\phi)S_{\alpha\mu\nu}S^{\mu\nu\alpha} +
	b_{3}(\phi)S_{\mu}S^{\mu} +b_4(\phi) S_{\mu}t^{\mu}+b_5(\phi) \varepsilon^{\alpha\beta\gamma\delta}S_{\alpha\beta\mu}S_{\gamma\delta}{}^{\mu} \\&\\&
	+c_{1}(\phi) Q_{\alpha\mu\nu}S^{\alpha\mu\nu}+
	c_{2}(\phi) Q_{\mu}S^{\mu} +
	c_{3}(\phi) q_{\mu}S^{\mu}\\
    &+c_{4}(\phi) Q_{\mu}t^{\mu}+c_{5}(\phi)q^{\mu}t_{\mu}+c_{6}(\phi) \varepsilon^{\alpha\beta\gamma\delta}Q_{\alpha\beta\mu}S_{\gamma\delta}{}^{\mu}
	\Big] \,.\label{action}   
    \end{aligned}
\end{equation}

The action generalises the Jordan-frame scalar-tensor theory \cite{Flanagan:2004bz, Jarv:2014hma, Jarv:2014laa, Jarv:2016sow} by allowing nonminimal couplings to the non-Riemannian Ricci scalar $R$, which is related to the metric Ricci scalar $\tilde{R}$ through \eqref{Rnonriem}, as well as to the nonmetricity and torsion quadratic invariants pure and mixed (respectively the terms proportional to $a_i$, to $b_i$ and to $c_i$), and in addition also derivative couplings to the nonmetricity and torsion vectors (the terms proportional to $\mathcal{C}_i$). Within the class of actions considered here, all independent parity-even and parity-odd terms quadratic in torsion and nonmetricity are included \cite{Iosifidis:2024bsq, Iosifidis:2020dck}. 
The parity-odd terms contain the Levi-Civita tensor $\varepsilon^{\alpha\beta\gamma\delta}$ explicitly or implicitly, as in the definition of $t^\mu$ in \eqref{QSvectors}. As in the usual scalar-tensor case, the scalar field acts as a source for the geometric gravitational fields, while simultaneously the coupling function $\mathcal{A}(\phi)$ determines the effective strength of the gravitational interaction. Therefore, the conceptual distinction of the terms in \eqref{action} into the gravitational and matter sectors remains ambiguous.
For the purpose of identifying the hypermomentum source in the connection equation, we consider the definition \eqref{hypedef}, that is we define what we consider matter after having performed the variation with respect to the connection and having set to zero torsion and nonmetricity. The remaining terms are regarded as belonging to the gravitational part. This split is a convention and does not affect the form of the full action.

The dimensions of the geometric quantities and the scalar coefficients in \eqref{action} are
\begin{equation}
\begin{aligned}
     & [g_{\mu\nu}] = M^0, \hspace{1.75 cm}  [\Gamma^{\alpha}{}_{\mu\nu}] = M^1, \hspace{2 cm} [\kappa] = M^{-2}, \hspace{2.1 cm} [\phi] = M^{1}, \\
     &[R] =M^{2}, \hspace{1.96 cm}[S^{\alpha}{}_{\mu\nu}] = M^{1},  \hspace{1.5 cm}    
     [Q^{\alpha}{}_{\mu\nu}] = M^{1},\\
     & [\mathcal{A}(\phi)] = M^{0}, \hspace{1.5 cm}[\mathcal{B}(\phi)] = M^{0}, \hspace{1.5 cm}  [\mathcal{V}(\phi)] = M^{4},  \\& [\mathcal{C}_i(\phi)] = M^{1}, \hspace{1.5 cm} [a_{i}(\phi)] =[b_{i}(\phi)]=[c_{i}(\phi)]= M^{0}\,\,\,\,.
    \end{aligned}
\label{eq: ABCVabc dimensions}
\end{equation}
The coupling functions $\mathcal{A}$, $\mathcal{B}$, $a_{i}$, $b_{i}$ and $c_{i}$ are dimensionless, while the potential $\mathcal{V}$ has a mass dimension four, and the derivative couplings $\mathcal{C}_i$ have a mass dimension one. The gravitational constant is defined, as usual, as $\kappa = 8 \pi G_N$. 

Let us finally comment on the projective symmetry \eqref{projective}. The Ricci scalar $R$ is projectively invariant, but the additional torsion and nonmetricity invariants in \eqref{action} generally break this symmetry (aside from the terms proportional to $a_{6}(\phi)$ and $\mathcal{C}_{4}(\phi)$ that are projectively invariant individually). 
Thus, the full gravitational sector is generally not projectively invariant. In the \ABV\, and \ABVC \, subcases (where the functions $a_i$, $b_i$, $c_i$ vanish), however, the restricted gravitational sector is projectively invariant, and projective consistency then imposes conditions on the matter couplings. 
After this side-note, we now derive the equations of motion obtained by varying the action with respect to the metric, the affine connection, and the scalar field.

\subsection{Connection field equations}
\label{subs:cfe}

Firstly, we compute the connection field equations, i.e.\ the equations obtained by varying the action \eqref{action} with respect to the affine connection $\Gamma^\lambda{}_{\mu\nu}$,
\begin{equation}
\begin{aligned}
\label{Gfieldeqs}
    \,P^{(1)}{}_\lambda{}^{\mu\nu} + P^{(2)}{}_\lambda{}^{\mu\nu} = \kappa \,\hat{\Delta}_{\lambda}{}^{ \mu\nu} \,,
\end{aligned}
\end{equation}
where we have also divided by $\mathcal{A}$, assuming $\mathcal{A} \neq 0$. On the left-hand side of \eqref{Gfieldeqs} we have the quantities $P^{(1)}{}_\lambda{}^{\mu\nu}$ and $ P^{(2)}{}_\lambda{}^{\mu\nu}$. The Palatini tensor $ P^{(1)}{}_\lambda{}^{\mu\nu}$ is defined as the variation of the non-Riemannian $R$ with respect to the affine connection $\Gamma^\lambda{}_{\mu\nu}$, and can be written as
\begin{equation}\label{palatinite}
    P^{(1)}{}_\lambda{}^{\mu\nu}:= \left( \frac{Q_{\lambda}}{2}+2 S_{\lambda}\right)g^{\mu\nu}
    - (Q_{\lambda}{}^{\mu\nu}+2 S_{\lambda}{}^{\mu\nu})
    +\left( q^{\mu} -\frac{Q^{\mu}}{2}-2 S^{\mu}\right)\delta_{\lambda}^{\nu}\,.
\end{equation}
The tensor $P^{(2)}{}_\lambda{}^{\mu\nu}$ is defined from the variations of the rest of the quadratic torsion and nonmetricity terms in the action \eqref{action} and normalized by the factor $\mathcal{A}^{-1}$. Its full form is reported in Appendix \ref{Aizi0}.

The right-hand side of \eqref{Gfieldeqs} contains the rescaled hypermomentum tensor $\hat{\Delta}_{\lambda}{}^{ \mu\nu} $. As mentioned in Sec.\ \ref{subs:action}, what we consider matter are the terms remaining nonzero after having performed the connection variation and having set to zero torsion and nonmetricity:
\begin{equation}
\label{hypedef}
     \hat{\Delta}_{\lambda}{}^{ \mu\nu}  
     := -\frac{2}{\sqrt{-g}\mathcal{A}}\frac{\delta ( \sqrt{-g} \,\mathcal{L} )}{\delta \Gamma^{\lambda}_{\phantom{\lambda}\mu\nu}}\Big|_{S=0=Q} \,,
\end{equation}
while the standard prototype hypermomentum (unscaled) version reads without the $\mathcal{A}$. For the action \eqref{action}, this gives
\begin{equation}\label{hyper}
    \begin{aligned}
    \hat{\Delta}_{\lambda}{}^{ \mu\nu}  =  & \frac{1}{ \kappa\,\mathcal{A}} \Big( \bigl(g^{\mu \nu } \partial_{\lambda }\phi - \delta_{\lambda }{}^{\nu } \partial^{\mu }\phi\bigr)\, \mathcal{A}^{\prime} 
    - \frac{1}{2} \kappa\bigl( \mathcal{C}_3(\phi)+2\, \mathcal{C}_2(\phi)\bigr) \delta_{\lambda }{}^{\nu } \partial^\mu \phi \\
    &- \frac{1}{2} \kappa\bigl(-  \mathcal{C}_3(\phi) + 4 \,\mathcal{C}_1(\phi)\bigr) \delta_{\lambda }{}^{\mu } \partial^\nu \phi - \kappa \mathcal{C}_2(\phi) \,g^{\mu \nu }  \partial_\lambda 
    \phi+\kappa \mathcal{C}_4(\phi) \,\varepsilon_{\lambda }{}^{\mu \nu \alpha} \partial_\alpha \phi\Big)\,,
    \end{aligned}
\end{equation}
where we used the definition \eqref{hypedef}. Note that the first term in the expression \eqref{hyper} also exists in scalar-tensor theory when the Ricci scalar is given by a general affine connection, see Sec.\ \ref{ABVsection}, while the remaining terms arise from the scalar field derivative couplings (Nieh-Yan-like contributions).

The connection equations \eqref{Gfieldeqs} can, in principle, be solved directly for the torsion and nonmetricity tensors following the method of Ref.~\cite{Iosifidis:2021ili}. In practice, the direct solution of the full system is algebraically cumbersome. For the \ABV\, and \ABVC \, subcases considered in Secs.~\ref{ABVsection} and \ref{ABVCsection}, we solve the connection equations directly. For the generic nondegenerate solution of the full theory, covariance and the available tensorial building blocks $\partial_\mu\phi$, $g_{\mu\nu}$, and $\varepsilon_{\mu\nu\alpha\beta}$ imply that the solutions can be written in the form
\beq
    S_{\mu\nu\alpha} = 2 A_{1}(\phi)(\partial_{[\mu}\phi)g_{\nu]\alpha} + A_{2}(\phi)\varepsilon_{\mu\nu\alpha\rho}\partial^{\rho} \phi 
\label{torphi}
\eeq
and
\beq 
\label{nonmetphi}
    Q_{\alpha \mu \nu}  =  A_{3}(\phi) g_{\mu \nu}\partial_{\alpha}\phi +  A_{4}(\phi) g_{\alpha(\mu} \partial_{\nu)}\phi  \,.
\eeq
Here $A_i(\phi)$ are elaborate expressions with the dimension $M^{-1}$. Round and square brackets denote symmetrization and antisymmetrization, respectively. From \eqref{torphi} and \eqref{nonmetphi}, $A_1=A_2=0$ corresponds to vanishing torsion, whereas $A_3=A_4=0$ corresponds to vanishing nonmetricity.

Here and in what follows, we call a connection branch nondegenerate when the determinant of the coefficient matrix of the algebraic connection system is nonzero, so that the connection can be uniquely solved for. When this determinant vanishes, we refer to the branch as degenerate; in that case some connection components may remain undetermined or additional consistency conditions may arise.

For the general nondegenerate branch solution with at least one of $a_i$, $b_i$, or $c_i$ nonzero, the explicit functions $A_i(\phi)$ are reported in Appendix \ref{AiziPart1}. The corresponding expressions for $a_i=b_i=c_i=0$, relevant to the \ABV\, and \ABVC \, sectors, are given in Appendix \ref{AiziPart2}.

\subsection{Metric field equations}\label{MetricFE}

The metric equations of motion, i.e., what we obtain after variation of the action \eqref{action} with respect to the metric $g^{\mu\nu}$ (the affine connection $\Gamma^\lambda{}_{\mu\nu}$ and the scalar field $\phi$ are kept fixed during the variation), are
\begin{equation}\label{metriceqs}
  \mathcal{A}\,  \mathcal{G}_{\mu\nu} +  \mathcal{G}^{(\mathcal{C}_i)}{}_{\mu\nu} +\mathcal{G}^{(a_i)}{}_{\mu\nu}+\mathcal{G}^{(b_i)}{}_{\mu\nu}
  +\mathcal{G}^{(c_i)}{}_{\mu\nu}+ \kappa  \,\mathcal{G}^{(\phi)}{}_{\mu\nu}=0\,.
\end{equation}
Here $\mathcal{G}_{\mu\nu} \equiv \frac{1}{2} (R_{\mu \nu} + R_{\nu \mu} - g_{\mu \nu} R)$ is the analog of the Einstein tensor but built with the affine Ricci scalar and the Ricci tensor (which is not symmetric in general) instead of the Levi-Civita ones. The remaining tensors consist of metric variations of the different sectors of the action: $\mathcal{G}^{(\mathcal{C}_i)}{}_{\mu\nu}$ comes from the derivative couplings, $\mathcal{G}^{(a_i)}{}_{\mu\nu}$ from the pure nonmetricity sector, $\mathcal{G}^{(b_i)}{}_{\mu\nu}$ from the pure torsion sector, $\mathcal{G}^{(c_i)}{}_{\mu\nu}$ from the mixed torsion--nonmetricity sector, and
$\mathcal{G}^{(\phi)}{}_{\mu\nu}$ from the scalar kinetic and potential terms. These terms are reported in Appendix \ref{Aizi0}.

Throughout the paper, we will rewrite this equation and its relevant subcases in terms of Levi-Civita quantities. To this purpose, we here report the relation between the Einstein tensor in affine quantities and its form in Levi-Civita ones when torsion and nonmetricity are given by \eqref{torphi} and \eqref{nonmetphi}:
\begin{equation}
\begin{aligned}\label{e2}
\frac12 \mathcal A \bigl(R_{\mu\nu}+R_{\nu\mu}-g_{\mu\nu}R\bigr)
&=
\frac12 \mathcal A \bigl(\tilde R_{\mu\nu}+\tilde R_{\nu\mu}-g_{\mu\nu}\tilde R\bigr)
+\frac{\mathcal A}{4}\,(16A_1+4A_3-A_4)\,g_{\mu\nu}\widetilde{\square}\phi -\frac{\mathcal A}{2}\,(8A_1+2A_3+A_4)\,\tilde{\nabla}_\mu \tilde{\nabla}_\nu\phi \\
&+\frac{\mathcal A}{2}\,\partial_\mu\phi\,\partial_\nu\phi\,
\Bigl(
16A_1^2-4A_2^2+8A_1A_3+A_3^2+4A_1A_4+A_3A_4-\frac12 A_4^2
-8A_1^{\prime}-2A_3^{\prime}-A_4^{\prime}
\Bigr) \\
&+\frac{\mathcal A}{2}\,g_{\mu\nu}\,\partial_\rho \phi\,\partial^\rho \phi\,
\Bigl(
8A_1^2-2A_2^2+4A_1A_3+\frac12 A_3^2-4A_1A_4-A_3A_4-\frac14 A_4^2
+8A_1^{\prime}+2A_3^{\prime}-\frac12 A_4^{\prime}
\Bigr)\,,
\end{aligned}
\end{equation}
where $\widetilde{\square} \phi \equiv \tilde{\nabla}_\rho \tilde{\nabla}^\rho \phi$.

\subsection{Scalar field equations}

Varying the action \eqref{action} with respect to the scalar field $\phi$, while keeping the metric and affine connection fixed, we obtain
\begin{equation}\label{scalareq}
    \frac{1}{2 \kappa} R \,\mathcal{A}^{\prime} + \mathcal{B} (\nabla_\mu \nabla^\mu \phi -  N^{\mu }{}_{\nu \mu } \nabla^\nu\phi) 
    + \frac{1}{2} \mathcal{B}^{\prime}\,\nabla_\mu\phi \nabla^\mu\phi  - \mathcal{V}^{\prime}
    + \frac{1}{2 \kappa} (\Phi^{(a_i)} +\Phi^{(b_i)} + \Phi^{(c_i)}) + \frac{1}{2}\Phi^{(\mathcal{C}_i)} = 0\,.
\end{equation}
The first four terms are the direct scalar-field contributions from the nonminimal coupling, the kinetic sector, and the potential. The quantities $\Phi^{(a_i)}$, $\Phi^{(b_i)}$, and $\Phi^{(c_i)}$ collect the contributions obtained by differentiating the scalar-dependent coefficients of the quadratic nonmetricity, torsion, and mixed sectors, respectively, while the $\Phi^{(\mathcal{C}_i)}$ are the contributions to the scalar equation coming from the derivative couplings between the scalar field and the torsion/ nonmetricity vectors,
\begin{equation}
\begin{aligned}
  & \Phi^{(a_i)} = Q_{\mu \nu \alpha } Q^{\mu \nu \alpha } a_1^{\prime} + Q^{\mu \nu \alpha } Q_{\nu \mu \alpha } a_2^{\prime} 
     + Q_{\mu } Q^{\mu } a_3^{\prime} + q_{\mu } q^{\mu } a_4^{\prime} + q^{\mu } Q_{\mu } a_5^{\prime} 
     + \varepsilon^{\mu \alpha \beta \gamma } Q_{\beta \nu \gamma } Q_{\mu }{}^{\nu }{}_{\alpha } a_6^{\prime},  \\
     & \Phi^{(b_i)} = - S_{\mu \beta }{}^{\alpha } S^{\mu }{}_{\alpha }{}^{\beta } b_2^{\prime} + S_{\mu } \bigl(S^{\mu } b_3^{\prime} 
     + t^{\mu } b_4^{\prime}\bigr) + S_{\mu \alpha }{}^{\beta } \bigl(S^{\mu \alpha }{}_{\beta } b_1^{\prime} 
     + \varepsilon^{\mu \alpha \gamma \delta } S_{\gamma \delta \beta } b_5^{\prime}\bigr),\\
     & \Phi^{(c_i)} =q_{\mu } S^{\mu } c_3^{\prime} + Q_{\mu } \bigl(S^{\mu } c_2^{\prime} + t^{\mu } c_4^{\prime}\bigr) + q^{\mu } t_{\mu } c_5^{\prime} 
     + Q_{\mu \nu \alpha } \bigl(S^{\mu \nu \alpha } c_1^{\prime} + \varepsilon^{\mu \alpha \beta \gamma } S_{\beta \gamma }{}^{\nu } c_6^{\prime}\bigr),\\
     &  \Phi^{(\mathcal{C}_i)} = + \mathcal{C}_1 (N^{\mu }{}_{\nu \mu } Q^{\nu } - \nabla_\mu Q^{\mu })+ \mathcal{C}_2 (q^{\mu } N^{\nu }{}_{\mu \nu } - \nabla_\mu q^{\mu }{}) 
     + \mathcal{C}_3 ( N^{\mu }{}_{\nu \mu } S^{\nu } - \nabla_\mu S^{\mu }{}) +  \mathcal{C}_4 ( N^{\nu }{}_{\mu \nu } t^{\mu } - \nabla_\mu t^{\mu }{})\,.
\end{aligned}
\end{equation}
Like the metric equations \eqref{metriceqs} with quantities defined in Appendix \ref{Aizi0}, the scalar field equation \eqref{scalareq} is written in terms of the affine connection ($\nabla_\mu$). Once the connection solution is substituted, it can be rewritten entirely in terms of $g_{\mu\nu}$, $\phi$, and the Levi-Civita covariant derivative.

\subsection{Metric theory equivalent action}\label{metrictheoryequiv}

We can write our metric-affine gravity action \eqref{action} as an equivalent metric action with an effective modified kinetic term for the scalar field. In order to do so, we substitute the solutions of the connection equations, \eqref{nonmetphi} and \eqref{torphi}, back into the action \eqref{action}. This can be done because the connection field equations \eqref{Gfieldeqs} are algebraic. We obtain the equivalent metric action
\beq\label{action22}
S[g,\phi]=\frac{1}{2 \kappa}\int \mathrm{d}^{4}x \sqrt{-g}\Big[\mathcal{A}\tilde{R}- \kappa\,\mathcal{K}(\phi) \partial_\mu \phi \partial^\mu \phi -2\kappa \mathcal{V}(\phi) + \tilde{\nabla}_\lambda (\mathcal{A}(- 4\, S^{\lambda} + q^{\lambda } - Q^{\lambda }))\Big]\,.
\eeq
For completeness, we show the boundary term $\tilde{\nabla}_\lambda (\mathcal{A}(- 4\, S^{\lambda} + q^{\lambda } - Q^{\lambda }))$. Given that we are interested in the bulk dynamics, this term can be omitted. The Ricci scalar $\tilde{R}$ is the usual Riemannian one defined by the Levi-Civita connection. When rewriting the action \eqref{action} using \eqref{nonmetphi} and \eqref{torphi}, new terms proportional to $\partial_\mu \phi \partial^\mu \phi$ appear and enter into the kinetic term. The function $\mathcal{K}(\phi)$ is the effective kinetic coefficient in the Jordan-frame
\begin{equation}\label{kinetic}
\begin{aligned}
\mathcal{K}(\phi) =&\mathcal{B} -\frac{1}{\kappa} \Big[3 \kappa\,\,\mathcal{C}_3 \,A_1 +A_1^2(- 24 \,\mathcal{A} + 6 b_1- 3 \,b_2+ 9 \,b_3 )+ 6\, \kappa\,\mathcal{C}_4\, A_2\\
&\qquad \qquad+ \,A_1\, A_2(18\, b_4 + 24\, b_5) +A_2^2( 6\, \mathcal{A}  - 6\, b_1 - 6\, b_2 ) + \kappa\,A_3( 4\,\mathcal{C}_1 + \mathcal{C}_2)\\
&\qquad \qquad +A_1 \,\,A_3(- 12\, \mathcal{A} + 3\, c_1 + 12 \,\,\,c_2 + 3\, c_3)+ A_2 A_3 (24 \,c_4+ 6\, c_5 + 6\, c_6) \\
&\qquad \qquad + A_3^2(-  \frac{3}{2} \mathcal{A} + 4\, a_1 + a_2 + 16 \,a_3 + a_4 + 4 \,a_5 )+ \kappa A_4( \mathcal{C}_1 + \frac{5}{2}\, \mathcal{C}_2)\\
&\qquad \qquad + A_1\, A_4( 6 \,\mathcal{A} - \frac{3}{2}\, c_1 + 3\, c_2 + \frac{15}{2} \,c_3) +A_2\, A_4( 6\, c_4 + 15\, c_5 - 3 \,c_6 ) \\
&\qquad \qquad +A_3 \,A_4( \frac{3}{2} \,\mathcal{A} + 2 \,a_1 + 5\, a_2 + 8 \,a_3 + 5\, a_4 + 11 \,a_5) \\
&\qquad \qquad + A_4^2( \frac{3}{4} \,\mathcal{A} + \frac{5}{2}\, a_1 + \frac{7}{4}\, a_2+ a_3+ \frac{25}{4} \,a_4 + \frac{5}{2} \,a_5)  + \mathcal{A}^{\prime}(12 \,A_1 + 3 \,A_3 - \frac{3}{2}\, A_4 ) \Big]\,.
\end{aligned}
\end{equation}
The functions $A_i(\phi)$ determine the torsion and nonmetricity through \eqref{torphi} and \eqref{nonmetphi}, while all the remaining quantities are the coefficient functions of the original action \eqref{action}.

Performing a conformal transformation of the metric, we can rewrite our action \eqref{action22} in the Einstein-frame, following the standard scalar-tensor construction \cite{Jarv:2014hma}
\begin{equation}
\label{metrictransf}
    g_{\mu\nu_{JF}} \rightarrow \mathcal{A}^{-1}\, g_{\mu\nu_{_{EF}}},  \hspace{1 cm}  
    g^{\mu\nu}_{_{JF}}\rightarrow \mathcal{A}\, g^{\mu\nu}_{_{EF}}, \hspace{1 cm} 
    \sqrt{-g_{_{JF}}} \rightarrow \frac{1}{\mathcal{A}^2} \sqrt{-g_{_{EF}}}\,.
\end{equation}
The Riemannian Ricci scalar $\tilde{R}$ (defined with the Levi-Civita symbols $\tilde{\Gamma}^\lambda{}_{\mu\nu}$) transforms with the conformal metric transformations \eqref{metrictransf}, up to a total derivative, as
\begin{equation}
\label{Rjftoef}
    \sqrt{-g}_{_{JF}}\,\mathcal{A}\,\tilde{R}_{_{JF}} \rightarrow \sqrt{-g_{_{EF}}} \Big(\tilde{R}_{_{EF}} 
    -  \frac{3}{2} \frac{(\mathcal{A}^{\prime})^2}{\mathcal{A}^2} \partial_\mu \phi\, \partial^\mu \phi\Big)\,,
\end{equation}
where $\mathcal{A}$ should be strictly positive.
Under the transformation \eqref{metrictransf}, the action \eqref{action22} becomes
\begin{equation}\label{actionEF}
S[g_{_{EF}},\phi]=\frac{1}{2 \kappa}\int \mathrm{d}^{4}x \sqrt{-g_{_{EF}}}\Big[\tilde{R}_{_{EF}} - \kappa\, \mathcal{K}_{_{EF}}(\phi)\, \partial_\mu \phi \partial^\mu \phi -2\kappa \mathcal{V}_{_{EF}}(\phi)\Big]\,,
\end{equation}
where the Einstein-frame Ricci scalar is related to the Jordan-frame one by the relation \eqref{Rjftoef}.
The new kinetic term and potential in the Einstein-frame are
\begin{equation}\label{EFkinetic}
 \mathcal{K}_{_{EF}}(\phi) = \frac{3}{2} \frac{(\mathcal{A}^{\prime})^2}{\kappa \mathcal{A}^2}+ \frac{\mathcal{K}(\phi)}{\mathcal{A}}
\,, \qquad \mathcal{V}_{_{EF}}(\phi)=\frac{\mathcal{V}(\phi)}{\mathcal{A}^2}\,.
\end{equation}
Thus, all effects of the non-Riemannian couplings in \eqref{action} are encoded in the effective Einstein-frame kinetic function \eqref{EFkinetic}. Similar reductions of auxiliary metric-affine connections to effective metric theories, in which the non-Riemannian interactions modify the scalar kinetic sector, have been discussed in Refs.~\cite{Rigouzzo:2022yan,Pradisi:2022nmh,Iosifidis:2021bad}.


\subsection{Polynomial coupling functions}\label{dim}

In the rest of the paper, we will sometimes consider specific expressions for the coefficient functions in our action \eqref{action} and study the kinetic term \eqref{EFkinetic} of its Einstein-frame equivalent \eqref{actionEF}. Motivated by dimensional analysis and by a low-order polynomial parametrization of the scalar-field dependence, we take the coefficient functions to be polynomials in $\phi$, inserting appropriate powers of $\kappa$ so that all coefficients $\xi$ are dimensionless. Keeping the dimensions explicit with $[\kappa]\sim M^{-2}$ and recalling the dimensions presented in \eqref{eq: ABCVabc dimensions}, we choose the following expansions

\begin{subequations}\label{expansion}
\begin{alignat}{3}
\mathcal{A}(\phi) &= \xi_{\mathcal{A}_0}+\sqrt{\kappa}\,\xi_{\mathcal{A}_1}\phi+\kappa\,\xi_{\mathcal{A}_2}\phi^2\,, \qquad&
\mathcal{B}(\phi) &= \xi_{\mathcal{B}_0}+\kappa\,\xi_{\mathcal{B}_2}\phi^2\,, \qquad&
\mathcal{V}(\phi) &= \frac{m^2\phi^2}{2}+\frac{\lambda\phi^4}{4}\,, \label{expansionABV}\\
\mathcal{C}_i(\phi) &= \xi_{\mathcal{C}_{i_1}}\phi+\sqrt{\kappa}\,\xi_{\mathcal{C}_{i_2}}\phi^2\,, \label{expansionC}\\
a_i(\phi) &= \xi_{a_{i_0}}+\sqrt{\kappa}\,\xi_{a_{i_1}}\phi+\kappa\,\xi_{a_{i_2}}\phi^2\,, \qquad&
b_i(\phi) &= \xi_{b_{i_0}}+\sqrt{\kappa}\,\xi_{b_{i_1}}\phi+\kappa\,\xi_{b_{i_2}}\phi^2\,, \label{expansionab}\\
c_i(\phi) &= \xi_{c_{i_0}}+\sqrt{\kappa}\,\xi_{c_{i_1}}\phi+\kappa\,\xi_{c_{i_2}}\phi^2\,. \label{expansionc}
\end{alignat}
\end{subequations}

The index range of $\mathcal{C}_i$'s is  $i=1,\ldots,4$,  while $a_i$, $c_i$ have $i=1,\ldots,6$, and $b_i$ have $i=1,\ldots,5$. All coupling coefficients $\xi$ are dimensionless, while in the potential, the parameters $m$ and $\lambda$ are real, positive, and of dimension $M^1$ and $M^0$, respectively. In the low energy regime, when the field comes to rest at the minimum of its (effective) potential at $\phi=0$, the reduced Planck mass is given by $M_{\mathrm{Pl}}^2=\frac{\xi_{\mathcal{A}_0}}{\kappa}$. Without loss of generality, we can set $\xi_{\mathcal{A}_0}=1$ by absorbing it into the definition of $\kappa$, and also take $\xi_{\mathcal{B}_0}=1$ by redefining the scalar field $\phi$ suitably. Often the linear term with $\, \xi_{\mathcal{A}_1}$ is omitted to keep the coupling function $\mathcal{A}$ positive over the relevant field range. We retain it, while still assuming $\mathcal{A} > 0$ over the field range of interest, because at high field values in combination with the $m$ term, it can generate a plateau in the effective potential and produce good observational predictions for inflation  \cite{Kallosh:2025rni}. We do not include a term linear in $\phi$ in $\mathcal B$, since such a term may change the sign of the kinetic coefficient in part of field space. Excluding it is a simple way to avoid this possible ghost-like behavior in the scalar sector. To keep the potential nonnegative, we omit odd powers of $\phi$ in the potential $\mathcal V(\phi)$. This leaves the standard mass-plus-quartic form. In contrast to the $Z_2$-symmetric polynomial parametrization adopted in Ref.~\cite{Rigouzzo:2022yan}, we keep both linear and quadratic terms in the couplings $a_i$, $b_i$, $c_i$, and $\mathcal{C}_i$. In the absence of an additional symmetry such as $\phi\to-\phi$, there is no general reason to exclude the linear terms in these non-Riemannian couplings. However, for simplicity we exclude the constant terms in the derivative couplings $\mathcal{C}_i$.

For generic nondegenerate connection solutions, nonzero quadratic coefficients and in the absence of cancellations, the small and large-field behaviors of the Jordan-frame kinetic function \eqref{kinetic} and its corresponding function in the Einstein-frame \eqref{EFkinetic} with the expansions \eqref{expansion} are
\begin{equation}\label{csmallEF}
\begin{aligned}
&\mathcal{K}(\phi)=\mathcal{O}(1)\,,\qquad \mathcal{K}_{_{EF}}(\phi)=\mathcal{O}(1)\,,\qquad &&\phi\to0\,,\\
&\mathcal{K}(\phi)=\mathcal{O}(\phi^2)\,,\qquad \mathcal{K}_{_{EF}}(\phi)=\mathcal{O}(1)\,,\qquad &&\phi\to\infty\,.
\end{aligned}
\end{equation}

\section{Metric-affine scalar-tensor theory (\ABV\,)}\label{ABVsection}
In this section, we consider the simplest scalar-tensor sector of the general action \eqref{action}, obtained by keeping only the functions $\mathcal A(\phi)$, $\mathcal B(\phi)$, and $\mathcal V(\phi)$. This gives a nonminimally coupled scalar field to the affine Ricci scalar. All the results in this section follow from the general procedure in Sec.\ \ref{generalprocedure} by taking $a_i = b_i = c_i = \mathcal{C}_i= 0$. The action \eqref{action} becomes
\begin{equation}
    \begin{aligned}
     S[g, \Gamma, \phi]
	=\frac{1}{2 \kappa}\int \mathrm{d}^{4}x \sqrt{-g} \Big[ & \mathcal{A}(\phi)R-\kappa\mathcal{B}(\phi) \partial_\mu \phi \partial^\mu \phi -2 \kappa \mathcal{V}(\phi) \Big] \label{action0}\,.
    \end{aligned}
\end{equation}

The general connection field equations \eqref{Gfieldeqs} here reduce to
\begin{equation}\label{connABV}
   \,P^{(1)}{}_\lambda{}^{\mu\nu} = \kappa\, \Delta_{\lambda }{}^{\mu \nu }\,,
\end{equation}
where the hypermomentum, recalling the discussion in Sec.\ \ref{subs:cfe}, arises from the coupling of the scalar field to the connection in $R$ as
\begin{equation}\label{hyper1}
    \Delta_{\lambda }{}^{\mu \nu } = \frac{\mathcal{A}^{\prime}}{\kappa\,\mathcal{A}}( g^{\mu \nu } \partial_{\lambda }\phi -  \delta^{\nu }{}_{\lambda } \,\partial^{\mu}\phi)\,.
\end{equation}
The connection equations \eqref{connABV} are solved by \eqref{torphi} and \eqref{nonmetphi} with the specific functions $A_i$
\begin{equation}\label{Aiex1g1}
    \begin{aligned}
    &A_3 + 4 A_1 = \frac{\mathcal{A}^{\prime}}{\mathcal{A}} \,,\hspace{1.5 cm}  A_2 =0 \,, \hspace{1.5cm}  A_4=0\,.
    \end{aligned}
\end{equation}

The \ABV\, connection equations \eqref{connABV} are projectively invariant. Consequently, one vectorial component of the connection remains undetermined and can be fixed by a projective gauge choice. Choosing to set the torsion vector to zero corresponds to setting  $A_{1}=0$. Moreover, as we see from \eqref{Aiex1g1}, setting $A_{1}=0$ also implies that the full torsion tensor is zero. Alternatively, choosing the nonmetricity vector to zero corresponds to setting $A_{3}=0$. This, from \eqref{Aiex1g1}, also implies that the full nonmetricity tensor vanishes. This reflects the torsion/nonmetricity duality \cite{Berthias:1993aa} that appears generally in metric-affine $f(R)$ theories \cite{Iosifidis:2019dua, Afonso:2017bxr}. The metric and scalar equations, as well as the effective kinetic term, are independent of this gauge choice.
The metric equations in this case are found from the general ones given in Sec.\ \ref{MetricFE} by setting $a_i = b_i = c_i = \mathcal{C}_i = 0$, considering the specific functions \eqref{Aiex1g1} and going to Levi-Civita quantities (for the Einstein tensor the conversion is facilitated by the formula \eqref{e2}):
\begin{equation}\label{metricABVgen}
\mathcal{A} \,\tilde{G}_{\mu\nu} + \kappa g_{\mu \nu } \mathcal{V} +\mathcal{A}^{\prime}( g_{\mu \nu } \,\widetilde{\square} \phi  -  \tilde{\nabla}_{\mu }\tilde{\nabla}_{\nu }\phi) 
+ \partial_{\mu }\phi \partial_{\nu }\phi \Bigl(- \kappa\mathcal{B} + \frac{3 (\mathcal{A}^{\prime})^2}{2\mathcal{A}} -  \mathcal{A}^{\prime\prime}\Bigr) + g_{\mu \nu } \partial_{\rho}\phi \partial^{\rho}\phi \Bigl(\frac{1}{2} \kappa\mathcal{B} -  \frac{3 (\mathcal{A}^{\prime})^2}{4\mathcal{A}} + \mathcal{A}^{\prime\prime}\Bigr) = 0\,,
\end{equation}
where $\tilde{G}_{\mu\nu} \equiv \tilde{R}_{\mu \nu } -  \frac{1}{2} g_{\mu \nu } \tilde{R}$ is the metric Einstein tensor.
The scalar equation \eqref{scalareq} after substituting the connection solution, eliminating $\tilde R$ with the metric trace, and taking $a_i = b_i = c_i = \mathcal{C}_i = 0$ becomes
\begin{equation}\label{scalarABVgen}
   \mathcal{B} \,\widetilde{\square}\phi +  \partial_{\rho}\phi \partial^{\rho}\phi \Big(\frac{\mathcal{B} \mathcal{A}^{\prime}}{2 \mathcal{A}} + \frac{1}{2}\mathcal{B}^{\prime}\Big)+ \frac{2 \mathcal{V} \mathcal{A}^{\prime}}{\mathcal{A}} - \mathcal{V}^{\prime} = 0\,.
\end{equation}
The choice of projective gauge on the coefficients \eqref{Aiex1g1}, for instance $A_1=0$ or $A_3=0$, does not affect the effective metric action. Following the general procedure of Sec.\ \ref{metrictheoryequiv}, we can rewrite the action \eqref{action0} as a metric action with a modified kinetic term. And the Jordan-frame and Einstein-frame kinetic functions are
\begin{equation}\label{K1}
  \mathcal{K}(\phi) = \mathcal{B} - \frac{3}{2} \frac{\bigl( \mathcal{A}^{\prime}\bigr)^2}{\kappa\,\mathcal{A}} \,, \hspace{2 cm} \mathcal{K}_{_{EF}}(\phi)=\frac{\mathcal{B}}{\mathcal{A}}\,.
\end{equation} 
We note how all dependence on the independent connection cancels from the Einstein-frame kinetic function except through the usual ratio $\mathcal B/\mathcal A$. We will use \eqref{K1} as a reference result in the following sections. Some extensions of \eqref{action0} leave this kinetic term unchanged, while others generate new contributions.

\paragraph{Polynomial coupling functions.}
In general, without the cosmological principle assumption, we can consider specific expressions for the functions in our action following the discussion in Sec.\ \ref{dim}. This will imply specific expressions for the kinetic term \eqref{K1}. 
If $\mathcal{A} = \mathcal{B} = 1$ we have
\begin{equation}
 \mathcal{K}(\phi) = 1 \,, \hspace{1 cm}\mathcal{K}_{_{EF}}(\phi) = 1\,.
\end{equation}
given that $\mathcal{A}^{\prime} = 0$. If instead $\mathcal{A} = 1+ \sqrt{\kappa}\,\, \xi_{\mathcal{A}_1}\, \phi  + \kappa\,\, \xi_{\mathcal{A}_2}\, \phi^2$, $\mathcal{B} = 1 + \kappa\, \xi_{\mathcal{B}_2} \,\phi^2 $ we have
\begin{equation}
  \begin{aligned}
    \mathcal{K}(\phi) =& 1 + \kappa \, \xi_{\mathcal{B}_{2}} \phi^2 -  \frac{3 (\sqrt{\kappa} \,\xi_{\mathcal{A}_{1}} 
    + 2 \,\kappa \,\xi_{\mathcal{A}_{2}} \phi)^2}{2 \kappa (1 + \sqrt{\kappa} \,\xi_{\mathcal{A}_{1}} \phi + \kappa \,\xi_{\mathcal{A}_{2}} \phi^2)}, \hspace{1 cm} 
   \mathcal{K}_{_{EF}}(\phi) = \frac{1 + \kappa \,\xi_{\mathcal{B}_{2}} \phi^2}{1 + \sqrt{\kappa} \,\xi_{\mathcal{A}_{1}} \phi + \kappa \,\xi_{\mathcal{A}_{2}} \phi^2}\,.
   \end{aligned}
   \label{sec:VI:K_EF}
\end{equation}
We can see how, for large scalar field values, the modified kinetic term is controlled by the coefficients $\xi_{\mathcal B_2}$ and $\xi_{\mathcal{A}_i}$ appearing in the expansion of $\mathcal{A}$ and $\mathcal{B}$.

We next examine three special realizations of the \ABV\, connection solution. Requiring both torsion and nonmetricity to vanish selects a special Riemannian branch, whereas the torsionless and metric-compatible cases correspond to two different projective gauges of the generic \ABV\, solution.

\subsection{Riemannian branch}\label{Riemmbranch}

In the \ABV\, model, the affine connection is an independent variable and can in general carry torsion and nonmetricity. In this subsection, we ask under which conditions the Levi-Civita connection is an admissible solution of the \ABV\, equations of motion. For a generic scalar configuration with $\partial_\mu\phi\neq0$, vanishing torsion and nonmetricity corresponds to $A_1=A_2=A_3=A_4=0$. From the solutions of the connection equation \eqref{Aiex1g1} we obtain the Riemannian branch by imposing that $A_1 = A_3 = 0$ (given that $A_2$ and $A_4$ are already found in \eqref{Aiex1g1} to be zero). This gives the requirement
\begin{equation}\label{Aiex1g4}
 \mathcal{A}^{\prime} = 0
\end{equation}
The same result can be seen directly starting from the general connection equations \eqref{Gfieldeqs}, setting there $A_1 = A_2 = A_3=A_4 =0 $ (zero torsion and zero nonmetricity) and $a_i = b_i = c_i = \mathcal{C}_i = 0$ in \eqref{Gfieldeqs}:
\begin{equation}\label{we}
    (- g^{\mu \nu } \partial_{\lambda } \phi + \delta_{\lambda }{}^{\nu } \partial^{\mu } \phi) \mathcal{A}^{\prime} = 0\,.
\end{equation}
We can take its traces in $\nu$ and $\lambda$, in $\mu$ and $\lambda$ (where this second vanishes identically being the \ABV\, model projective invariant). In this way, we obtain the following relations among parameters:
\begin{equation}\label{el}
    \begin{aligned}
        3\, \partial^\mu \phi{} \,\mathcal{A}^{\prime} = 0\,.
    \end{aligned}
\end{equation}
If $\partial^\mu \phi{}$ is nonzero then this implies $\mathcal{A}^{\prime} = 0$. Therefore, $\mathcal{A}$ is constant and the coupling $\mathcal{A} \tilde{R}$ is just the Einstein-Hilbert term with a rescaled gravitational constant. Hence, in the Riemannian branch, we obtain GR with a minimally coupled scalar field, possibly with a noncanonical kinetic coefficient $\mathcal{B}(\phi)$. Note that this study should be distinguished from imposing the Levi-Civita connection directly at the level of the action. In that case the independent connection is removed before variation and one obtains metric scalar-tensor gravity, where $\mathcal{A}(\phi)$ can remain nonconstant. Here instead the independent connection is varied first, and requiring its solution to be Levi-Civita imposes the additional condition $\partial^\mu\phi \mathcal{A}^{\prime}=0$.

We can also find the Riemannian metric equations from the general Sec.\ \ref{MetricFE} by again setting $A_1 = A_2 = A_3=A_4 =0 $ and $a_i = b_i = c_i = \mathcal{C}_i = 0$, going to Levi-Civita quantities (for the Einstein tensor the conversion is formula \eqref{e2}) and using the constraint found from the connection equations that $\mathcal{A}^{\prime} = 0$. We obtain
\begin{equation}
  \mathcal{A}\, \tilde{G}_{\mu\nu} +\kappa\, \mathcal{B}( -\partial_\mu\phi{} \partial_\nu\phi{} 
    + \frac{1}{2} g_{\mu \nu } \partial_\rho\phi{} \partial^\rho\phi{}) +\kappa\,g_{\mu \nu } \mathcal{V}=0\,.
\end{equation}
This is consistent with the literature, see formula (8) \cite{Jarv:2024krk}.
We can apply the same conditions to the scalar equation, i.e., setting $A_1 = A_2= A_3 =A_4 =0 $, $a_i = b_i = c_i = \mathcal{C}_i = 0$ and considering the connection solution $\mathcal{A}^{\prime} = 0$ in the scalar equation \eqref{scalareq}. We also eliminate $\tilde R$ with the trace of metric equation. In this way, we obtain
\begin{equation}
    \mathcal{B}\, \widetilde{\square} \phi + \frac{1}{2} \partial_\rho \phi \partial^\rho \phi \,\mathcal{B}^{\prime} - \mathcal{V}^{\prime} = 0\,.
\end{equation}
In summary, vanishing torsion and nonmetricity are possible only if hypermomentum \eqref{hyper1} is also zero, i.e.\ when the scalar field is either constant or minimally coupled to the Ricci scalar.

\subsection{Palatini projective gauge}\label{Palatini}

We now consider the torsionless representative of the \ABV\, connection solution. Unlike the Riemannian case discussed in Sec.\ \ref{Riemmbranch}, requiring only vanishing torsion does not impose an additional condition on $\mathcal{A}(\phi)$. This is a consequence of the projective invariance of the \ABV\, connection equations: the undetermined projective component of the connection can be fixed so that the torsion vanishes. 

For a generic scalar configuration with $\partial_\mu\phi\neq0$, the vanishing torsion Palatini equations are found from the general solutions of the connection equations \eqref{Aiex1g1} by choosing the gauge $A_1 = 0$. Since $A_2=0$ already, the full torsion tensor vanishes, while the nonmetricity remains present. The coefficients \eqref{Aiex1g1} are
\begin{equation}\label{Aiex1g2}
 A_1=A_2=A_4=0 \,, \hspace{1.7cm} A_3 = \frac{\mathcal{A}^{\prime}}{\mathcal{A}}\,.
\end{equation}
The same result can be obtained directly from the connection equation \eqref{Gfieldeqs} by imposing $A_1=A_2=0$ and $a_i=b_i=c_i=\mathcal{C}_i=0$. The resulting equation is
\begin{equation}
\begin{aligned}
  & - \nabla_\lambda \left( \mathcal{A} \sqrt{-g} \,g^{\mu \nu} \right) + \nabla_\rho \left( \mathcal{A} \sqrt{-g} \,g^{\mu \rho} \right) \delta^{\nu}_{\lambda}= 0
     \end{aligned}
\end{equation}
which can be rearranged as
  \begin{equation}
\begin{aligned}\label{Palatiniconnection}
  &\mathcal{A}\, q^{\mu } \delta_{\lambda }{}^{\nu } + \mathcal{A} (- Q_{\lambda }{}^{\mu \nu } + \frac{1}{2} g^{\mu \nu } Q_{\lambda }) -  \frac{1}{2} \mathcal{A}\, \delta_{\lambda }{}^{\nu } Q^{\mu } - \mathcal{A}^{\prime} g^{\mu \nu } \partial_\lambda \phi + \mathcal{A}^{\prime}\delta_{\lambda }{}^{\nu } \partial^\mu \phi= 0\,.
   \end{aligned}
\end{equation}
The last two terms in \eqref{Palatiniconnection} correspond to the \ABV\, hypermomentum contribution \eqref{hyper1}, which has here been moved to the left-hand side of the connection equation.

In order to connect with the usual torsionless Palatini scalar-tensor formulation, as considered, for example, in \cite{Jarv:2024krk}, we need to symmetrize equation \eqref{Palatiniconnection} in the indices $\mu$ and $\nu$. In that formulation, the independent connection is assumed to be symmetric from the outset, so that variation with respect to the connection directly yields the corresponding symmetrized connection equation. Substituting the nonmetricity ansatz \eqref{nonmetphi} into \eqref{Palatiniconnection} and taking the independent contractions reproduce the solution \eqref{Aiex1g2}, namely $A_3=\mathcal{A}^{\prime}/\mathcal{A}$ and $A_4=0$.

Substituting the torsionless conditions and the solution \eqref{Aiex1g2} into the general metric equation \eqref{metriceqs}, and rewriting it in Levi-Civita quantities using \eqref{e2}, gives precisely the \ABV\, metric equation \eqref{metricABVgen}. This is expected because \eqref{metricABVgen} is independent of the projective-gauge function $A_1$. The resulting metric equation is \eqref{metricABVgen} and we see that the Palatini case agrees with the literature, see for example equation (12) in \cite{Jarv:2024krk}.

We can also find the scalar equation \eqref{scalareq} in the Palatini projective gauge after substituting the connection solution, eliminating $\tilde R$ with the metric trace and taking again $A_1 = A_2 =0$ and $a_i = b_i = c_i = \mathcal{C}_i = 0$. We find that it is identical to the \ABV\, model one \eqref{scalarABVgen}. Again, this is because \eqref{scalarABVgen} is independent of the specific choice for the function $A_1$. This is expected because, after the connection equation has been solved, the scalar equation is independent of the projective gauge chosen for the connection. Therefore, the torsionless projective gauge gives the same scalar dynamics as the general \ABV\, solution.

The torsionless projective gauge therefore identifies the \ABV\, model with the usual Palatini scalar-tensor formulation. Eliminating the independent connection then maps the Palatini theory characterized by the functions $(\mathcal{A},\mathcal{B},\mathcal{V})$ into a dynamically equivalent metric scalar-tensor theory with the same nonminimal coupling and potential but with the shifted kinetic coefficient \eqref{K1}. This reproduces the relation between the Palatini and metric scalar-tensor formulations discussed in Ref.~\cite{Jarv:2024krk}. Notice that this correspondence does not identify Palatini and metric theories having the same function $\mathcal{B}$. Rather, the Palatini theory with the kinetic coefficient $\mathcal{B}$ is dynamically equivalent, after elimination of the independent connection, to the metric theory with the kinetic coefficient $\mathcal{K}$.

\subsection{Metric-compatible projective gauge}\label{metriccompatible}

We can alternatively use the projective freedom of the \ABV\, connection equations to choose a metric-compatible projective gauge. For a generic scalar configuration with $\partial_\mu\phi\neq0$, metric compatibility requires $A_3=A_4=0$. Since $A_4=0$ already follows from the general \ABV\, solution \eqref{Aiex1g1}, we use the projective freedom to set $A_3=0$. Equation \eqref{Aiex1g1} then gives
\begin{equation}\label{Aiex1g3}
A_2=A_3=A_4=0 \,, \hspace{1.7cm}    A_1 = \frac{1}{4} \frac{\mathcal{A}^{\prime}}{\mathcal{A}}\,,
\end{equation}
so that the nonmetricity vanishes. As in the torsionless case, this does not impose an additional restriction on $\mathcal{A}(\phi)$. Rather, it selects another representative of the same projective family of \ABV\, connection solutions. From the \ABV\, connection field equations \eqref{Gfieldeqs} we can arrive to the same solutions \eqref{Aiex1g3} by assuming $A_3 = A_4 =0$ ( zero nonmetricity) and $a_i = b_i = c_i = \mathcal{C}_i = 0$ and solving the so found connection equations to find $A_1$ and $A_2$. The connection equations are
\begin{equation}
    \begin{aligned}
        -2 \mathcal{A} \varepsilon_{\lambda }{}^{\mu \nu }{}_{\rho} A_2 \partial^\rho \phi + g^{\mu \nu } \partial_{\lambda} \phi \bigl(4 \mathcal{A} A_1 -  \mathcal{A}^{\prime}\bigr) + \delta_{\lambda }{}^{\nu } \partial^\mu \phi \bigl(-4 \mathcal{A} A_1 + \mathcal{A}^{\prime}\bigr) = 0\,.
    \end{aligned}
\end{equation}
Taking its traces and contracting with the Levi-Civita tensor reproduces the solution \eqref{Aiex1g3}.
The general metric equations \eqref{metriceqs} in the metric compatible projective gauge follows from taking again $A_3 = A_4 =0$ (zero nonmetricity) and $a_i = b_i = c_i = \mathcal{C}_i = 0$ and going to Levi-Civita quantities (for the Einstein tensor the conversion is formula \eqref{e2}) and with the solutions $A_1$, $A_2$ from \eqref{Aiex1g3}. We therefore recover the same metric equation as in both the Palatini limit and the general \ABV\, model, Eq.~\eqref{metricABVgen}, because this equation is independent of the specific choice of $A_1$.
For the scalar-field equation in the metric-compatible limit, we start from \eqref{scalareq}, substitute the connection solution, eliminate $\tilde{R}$ using the metric trace, and again set $A_3=A_4=0$ and $a_i=b_i=c_i=\mathcal{C}_i=0$. We then recover the Palatini/general-\ABV\, equation \eqref{scalarABVgen}, which is likewise independent of $A_1$.

The metric-compatible projective gauge is closely related to scalar-tensor theories formulated in Riemann--Cartan or Einstein--Cartan geometry. The equivalence between the metric-compatible torsionful and torsionless nonmetric formulations of this class of scalar-tensor theories was already traced to projective invariance early on \cite{Berthias:1993aa}. In particular, after accounting for differences in conventions, their metric-compatible solution exhibits the same vector-torsion structure as \eqref{Aiex1g3}, with the torsion proportional to the derivative of the logarithm of the nonminimal coupling. A direct comparison can also be made with the Einstein--Cartan--Brans--Dicke model of Ref.\ \cite{Shabani:2019cre}: in the absence of the spin contribution of matter and after accounting for their torsion convention, their solution gives precisely $A_1=\mathcal{A}^{\prime}/(4\mathcal{A})$. More generally, reference \cite{Shaposhnikov:2020frq} studies nonminimally coupled scalar fields in the metric-compatible Einstein--Cartan formulation and shows how the auxiliary torsion can be eliminated to obtain an equivalent metric theory. In the absence of fermions and of the additional Einstein--Cartan terms considered there, their theory reduces to the usual Palatini scalar-tensor formulation, consistently with the projective equivalence between the torsionless and metric-compatible gauges found here.

\section{Metric-affine scalar-tensor theory with Nieh-Yan-like couplings (\ABVC \,)}\label{ABVCsection}
In this section, we extend the \ABV\, model of Sec.\ \ref{ABVsection} by including the derivative couplings between $\partial_\mu\phi$ and the torsion and nonmetricity vectors, corresponding to the Nieh-Yan-like sector \cite{ Andrei:2026gmu}. The results follow from the general procedure outlined in Sec.\ \ref{generalprocedure}, taking $a_i = b_i = c_i = 0$. The action is as follows
\begin{equation}
    \begin{aligned}
     S[g, \Gamma, \phi]
	=\frac{1}{2 \kappa}\int \mathrm{d}^{4}x \sqrt{-g} \Big[ & \mathcal{A}(\phi) R-\kappa \mathcal{B}(\phi) \partial_\mu \phi \partial^\mu \phi -2 \kappa\mathcal{V}(\phi) 
    + \kappa\left(\mathcal{C}_1(\phi) Q^{\mu}+\mathcal{C}_2(\phi) q^{\mu} + \mathcal{C}_3(\phi)  S^{\mu}+\mathcal{C}_{4}(\phi) t^{\mu}\right)\partial_{\mu}\phi \Big]\,. \label{action1}
    \end{aligned}
\end{equation}
Some inflationary predictions for this \ABVC \, class were studied in detail in \cite{Andrei:2026gmu}.

\subsection{Connection equations}

The general connection field equations \eqref{Gfieldeqs} setting $a_i = b_i = c_i = 0$ reduce to 
\begin{equation}\label{ABVCconneq}
\,P^{(1)}{}_\lambda{}^{\mu\nu}= \kappa\,\hat{\Delta}_{\lambda}{}^{ \mu\nu} \,,
\end{equation}
where on the left-hand side we have the usual Palatini tensor \eqref{palatinite} and on the right-hand side we have the hypermomentum tensor \eqref{hypedef}.
The connection equation \eqref{ABVCconneq} is not projectively consistent for generic values of the derivative couplings. In fact, the Palatini tensor \eqref{palatinite} is projectively invariant while the generic hypermomentum \eqref{hypedef} is not. Following from the general comments given in Sec.\ \ref{subs:action}, before solving it we then have to find the condition to have projective consistency. This can be done by taking the contractions with $g^{\mu\nu}$, $g_{\lambda}{}^{\nu}$ and $g_{\lambda}{}^{\mu}$ of the connection equation \eqref{ABVCconneq} and solving the system obtained.
In this way, we find that the relation to impose among the coupling functions $\mathcal{C}_i$ in \eqref{ABVCconneq} to make it projectively coherent is
\begin{equation}\label{constr}
  \mathcal{C}_1 = \frac{1}{16}(-4\, \mathcal{C}_2 +3\, \mathcal{C}_3)\,.
\end{equation}

After imposing \eqref{constr}, the connection equation \eqref{ABVCconneq} is projectively invariant and therefore leaves one vectorial component of the connection undetermined. This projective freedom can be used to fix one vectorial component of the connection, i.e., to choose the projective gauge vector $\xi_\mu$ from \eqref{projective} in such a way that $q_\mu$ satisfies $q_\mu = 0$. This corresponds to choosing $\xi_\mu = - \frac{1}{2}q_\mu$. With this choice, the solutions of the connection equations \eqref{ABVCconneq} with the constraint \eqref{constr} are the \eqref{torphi}, \eqref{nonmetphi} with the coefficients $A_i$ here given by
\begin{equation}\label{eq: A_i when ABCV}
    \begin{aligned}
    A_1 &= - \frac{ \bigl(\kappa(6 \mathcal{C}_2 + \mathcal{C}_3) - 2 \mathcal{A}^{\prime}\bigr)}{8 \mathcal{A}} \,, \qquad & A_2 &=- \frac{\kappa \,\mathcal{C}_4 }{2 \mathcal{A}} \,, \\
   A_3 &= \frac{5 \kappa\,\bigl(4 \mathcal{C}_2 + \mathcal{C}_3\bigr) }{8 \mathcal{A}} \,, \qquad & A_4 &= -  \frac{\kappa\,\bigl(4 \mathcal{C}_2 + \mathcal{C}_3\bigr)}{4 \mathcal{A}}\,.
    \end{aligned}
\end{equation}
The couplings $\mathcal{C}_2$, $\mathcal{C}_3$, and $\mathcal{C}_4$ modify the torsion tensor through $A_1$ and $A_2$, whereas only $\mathcal{C}_2$ and $\mathcal{C}_3$ enter the nonmetricity coefficients $A_3$ and $A_4$. 

It is worth noting that the constraint \eqref{constr} arises because the purely geometric part of the action is projectively invariant, whereas the derivative-coupling sector proportional to $\mathcal{C}_i$ is not projectively invariant in general. Therefore, in the absence of additional matter fields coupled to the affine connection, projective invariance of the total action requires the condition \eqref{constr} \cite{Iosifidis:2019fsh}. If further matter sectors are present, their projective variation may compensate that of the scalar derivative-coupling sector, so that only the trace of the total hypermomentum is required to vanish. In that case, \eqref{constr} does not need to hold for the $\mathcal{C}_i$ sector separately.

It is interesting that the solution \eqref{eq: A_i when ABCV} exhibits a cross dependence between the derivative couplings and the non-Riemannian variables: the nonmetricity coupling $\mathcal{C}_2$ contributes to the torsion coefficient $A_1$, while the torsion-vector coupling $\mathcal{C}_3$ contributes to the nonmetricity coefficients $A_3$ and $A_4$. This is partly related to the projective consistency condition \eqref{constr}, which mixes the original vector couplings into projectively invariant combinations. More generally, however, the connection equations form a coupled algebraic system for torsion and nonmetricity, so that a coupling entering one sector of the action can also contribute to the solution of the other.

\subsection{Metric equations}

The general metric equations \eqref{metriceqs}, after setting $a_i=b_i=c_i=0$, imposing the projective consistency condition \eqref{constr}, going to Levi-Civita quantities (for the Einstein tensor the conversion formula is \eqref{e2}), and substituting the connection solution \eqref{eq: A_i when ABCV}, reduce to
\begin{equation}\label{metricBAVC}
\begin{aligned}
 & \mathcal{A} \,\tilde{G}_{\mu \nu } + \kappa g_{\mu \nu } \mathcal{V} +(-  \tilde{\nabla}_{\mu }\tilde{\nabla}_{\nu }\phi + g_{\mu \nu } \,\widetilde{\square}\phi) \mathcal{A}^{\prime} \\ 
 &+ \partial_{\mu }\phi \partial_{\nu }\phi \Bigl(-  \kappa \mathcal{B} + \frac{3 (\mathcal{A}^{\prime})^2}{2 \mathcal{A}} -  \mathcal{A}^{\prime\prime} + \mathcal{F}_{1} \Bigr)
  + g_{\mu \nu } \partial_{\rho }\phi \partial^{\rho }\phi \Bigl(\frac{1}{2} \kappa \mathcal{B} -  \frac{3 (\mathcal{A}^{\prime})^2}{4 \mathcal{A}} + \mathcal{A}^{\prime\prime} + \mathcal{F}_{2}\Bigr) &= 0\,,
    \end{aligned}
\end{equation}
where $\mathcal{F}_{1}= \mathcal{F}_{1}(\mathcal{C}_i, \mathcal{C}_i^{\prime},\mathcal{A}, \mathcal{A}^{\prime})$ and $\mathcal{F}_{2}= \mathcal{F}_{2}(\mathcal{C}_i, \mathcal{A},\mathcal{A}^{\prime})$ and their explicit expressions are reported in Appendix \ref{AiziPart3}. We see that the derivative couplings $\mathcal{C}_i$ do not introduce new tensorial structures compared with the \ABV\, scalar-tensor metric equation \eqref{metricABVgen}. Their effect is to modify the coefficients of the two tensorial terms $\partial_{\mu }\phi \partial_{\nu }\phi$ and $ g_{\mu \nu } \partial_{\rho }\phi \partial^{\rho }\phi$. We note that equation \eqref{metricBAVC} can also be found by varying the action in the Jordan-frame \eqref{action22} with kinetic term \eqref{kinetic} after removing all the coefficients other than \ABVC \, ones.

\subsection{Scalar equations}
The general scalar equations \eqref{scalareqlevi}, after setting $a_i=b_i=c_i=0$, imposing the projective consistency condition \eqref{constr}, and substituting the connection solution \eqref{eq: A_i when ABCV}, reduce to
\begin{equation}\label{scalarABVC}
    \begin{aligned}
        \widetilde{\square}\phi \,\bigl( \mathcal{B} +\mathcal{F}_{3}\bigr) + \partial_{\rho }\phi \partial^{\rho }\phi \bigl( \frac{\mathcal{B} \mathcal{A}^{\prime}}{2 \mathcal{A}} + \frac{1}{2} \mathcal{B}^{\prime} + \mathcal{F}_{4}\bigr) + \frac{2 \mathcal{A}^{\prime}}{\mathcal{A}} \mathcal{V}-  \mathcal{V}^{\prime} = 0\,,
    \end{aligned}
\end{equation}
where $\mathcal{F}_{3}=\mathcal{F}_{3}(\mathcal{C}_i,\mathcal{A}, \mathcal{A}^{\prime})$ and $\mathcal{F}_{4}=\mathcal{F}_{4}(\mathcal{C}_i, \mathcal{C}_i^{\prime},\mathcal{A}, \mathcal{A}^{\prime}, \mathcal{A}^{\prime \prime})$ are reported in Appendix \ref{AiziPart3}. Again we see that the derivative couplings $\mathcal{C}_i$ modify the coefficients of $\widetilde{\square}\phi$ and $\partial_{\rho }\phi \partial^{\rho }\phi$ but not the derivative structure itself. We note that equation \eqref{scalarABVC} can also be found by varying the action in the Jordan-frame \eqref{action22} with kinetic term \eqref{kinetic} after removing all the coefficients other than \ABVC \, ones.

\subsection{Kinetic term}

Following the general procedure described in Sec.\ \ref{metrictheoryequiv}, we can rewrite the action \eqref{action1} as a metric action with a modified kinetic term. The resulting Jordan- and Einstein-frame kinetic functions, previously derived in Ref.~\cite{Andrei:2026gmu}, are
\begin{equation}\label{EkC}
 \begin{aligned}
&\mathcal{K}(\phi) =  \mathcal{B} +\frac{9 \kappa \mathcal{C}_2^2}{8 \mathcal{A}} + \frac{9 \kappa \mathcal{C}_2 \mathcal{C}_3}{16 \mathcal{A}} -  \frac{3 \kappa \mathcal{C}_3^2}{128 \mathcal{A}} + \frac{3 \kappa \mathcal{C}_4^2}{2 \mathcal{A}} -  \frac{3 \mathcal{C}_3 \mathcal{A}^{\prime}}{4 \mathcal{A}} -  \frac{3 \bigl(\mathcal{A}^{\prime}\bigr)^2}{2 \kappa \mathcal{A}} \,,\\
&\mathcal{K}_{_{EF}}(\phi) = \frac{\mathcal{B}}{\mathcal{A}} + \frac{9 \kappa \mathcal{C}_2^2}{8 \mathcal{A}^2} + \frac{9 \kappa \mathcal{C}_2 \mathcal{C}_3}{16 \mathcal{A}^2} -  \frac{3 \kappa \mathcal{C}_3^2}{128 \mathcal{A}^2} + \frac{3 \kappa \mathcal{C}_4^2}{2 \mathcal{A}^2} -  \frac{3 \mathcal{C}_3 \mathcal{A}^{\prime}}{4 \mathcal{A}^2}\,,
\end{aligned}
\end{equation}
where the conditions for projective invariance \eqref{constr} have been used. Compared with the scalar-tensor metric \ABV\, result \eqref{K1}, the derivative couplings generate additional contributions to the Einstein-frame kinetic function. These can affect the inflationary predictions.

\subsection{Polynomial coupling functions}\label{sw}

We now fix $\mathcal{A}$, $\mathcal{B}$, and the derivative couplings $\mathcal{C}_i$ to the polynomial forms introduced in Sec.\ \ref{dim} and determine the corresponding effective kinetic functions when our theory is written as a metric theory as in Sec.\ \ref{metrictheoryequiv}. The expansion of the coupling $\mathcal{C}_1$ is always related to the other couplings, given the constraint \eqref{constr}.

First we keep in the expressions in Sec.\ \ref{dim} the lowest nonvanishing order in the expansions \eqref{expansionABV} and \eqref{expansionC}, with this choice the kinetic term $  \mathcal{K}(\phi)$ in Sec.\ \ref{metrictheoryequiv} becomes
\begin{equation}
  \begin{aligned}
     \mathcal{K}(\phi) =\mathcal{K}_{_{EF}}(\phi)= 1 + \frac{3}{128}\kappa (48 \,\xi_{\mathcal{C}_{2_{1}}}^2 + 24 \xi_{\mathcal{C}_{2_{1}}} \xi_{\mathcal{C}_{3_{1}}} - \xi_{\mathcal{C}_{3_{1}}}^2 + 64 \xi_{\mathcal{C}_{4_{1}}}^2) \phi^2\,.
   \end{aligned}
\end{equation}
We can also find the modified kinetic term for the full expansions introduced in Sec.\ \ref{dim}, \eqref{expansionABV} and \eqref{expansionC}. With these expressions, the kinetic term  $\mathcal{K}(\phi)$ becomes
\begin{equation}\label{KABVCsection}
 \begin{aligned}
&\mathcal{K}(\phi) = 1 + \kappa\, \xi_{\mathcal{B}_2} \,\phi^2 -  \frac{3 (64 \,\xi_{\mathcal{A}_{1}}^2 + \sqrt{\kappa} \,\xi_w \phi + \kappa \, \xi_y \phi^2 + \kappa^{3/2} \,\xi_z \phi^3 + \kappa^2 \,\xi_s \phi^4)}{128 (1 + \sqrt{\kappa}\, \xi_{\mathcal{A}_{1}} \phi + \kappa \,\xi_{\mathcal{A}_{2}} \phi^2)}\,,
\end{aligned}
\end{equation}
with 
\begin{equation}
\begin{aligned}
  \xi_w &= 32 \xi_{\mathcal{A}_{1}} (8 \xi_{\mathcal{A}_{2}} + \xi_{\mathcal{C}_{3_{1}}}),\\
  \xi_y &= 256 \xi_{\mathcal{A}_{2}}^2 - 48 \xi_{\mathcal{C}_{2_{1}}}^2 + \xi_{\mathcal{C}_{3_{1}}} (64 \xi_{\mathcal{A}_{2}} - 24 \xi_{\mathcal{C}_{2_{1}}} + \xi_{\mathcal{C}_{3_{1}}}) + 32 \xi_{\mathcal{A}_{1}} \xi_{\mathcal{C}_{3_{2}}} - 64 \xi_{\mathcal{C}_{4_{1}}}^2,\\
  \xi_z &= 2 \bigl(-48 \xi_{\mathcal{C}_{2_{1}}} \xi_{\mathcal{C}_{2_{2}}} - 12 \xi_{\mathcal{C}_{2_{2}}} \xi_{\mathcal{C}_{3_{1}}} + (32 \xi_{\mathcal{A}_{2}} - 12 \xi_{\mathcal{C}_{2_{1}}} + \xi_{\mathcal{C}_{3_{1}}}) \xi_{\mathcal{C}_{3_{2}}} - 64 \xi_{\mathcal{C}_{4_{1}}} \xi_{\mathcal{C}_{4_{2}}}\bigr),\\
  \xi_s &= -48 \xi_{\mathcal{C}_{2_{2}}}^2 + \xi_{\mathcal{C}_{3_{2}}} (-24 \xi_{\mathcal{C}_{2_{2}}} + \xi_{\mathcal{C}_{3_{2}}}) - 64 \xi_{\mathcal{C}_{4_{2}}}^2\,.
  \end{aligned}
\end{equation}
The small and large field behaviors of \eqref{KABVCsection} are
\begin{equation}
\mathcal{K}(\phi)=\mathcal{O}(1)\quad (\phi\to0)\,,\qquad \mathcal{K}(\phi)=\mathcal{O}(\phi^2)\quad (\phi\to\infty)\,.
\end{equation}
We also report the kinetic term in the Einstein-frame as from \eqref{actionEF} to be
\begin{equation}
 \begin{aligned}\label{kefC}
&\mathcal{K}_{_{EF}}(\phi) = \frac{1}{(1 + \sqrt{\kappa}\, \xi_{\mathcal{A}_{1}} \phi + \kappa \,\xi_{\mathcal{A}_{2}} \phi^2)}+  \frac{ ( \sqrt{\kappa} \,\xi_{w^{\prime}} \phi + \kappa \, \xi_{y^{\prime}} \phi^2 + \kappa^{3/2} \,\xi_{z^{\prime}} \phi^3 + \kappa^2 \,\xi_{s^{\prime}} \phi^4)}{ (1 + \sqrt{\kappa}\, \xi_{\mathcal{A}_{1}} \phi + \kappa \,\xi_{\mathcal{A}_{2}} \phi^2)^2}\,,
\end{aligned}
\end{equation}
with 
\begin{equation}
\begin{aligned}
 \xi_{w^{\prime}} &=- \frac{3}{128} \xi_w + 6 \xi_{\mathcal{A}_{1}} \xi_{\mathcal{A}_{2}}, \qquad &  \xi_{y^{\prime}} &=- \frac{3}{128} \xi_y + 6 \xi_{\mathcal{A}_{2}}^2 +\xi_{\mathcal{B}_2},\\
\xi_{z^{\prime}} &= - \frac{3}{128} \xi_z + \xi_{\mathcal{A}_{1}}\xi_{\mathcal{B}_2},  \qquad & \xi_{s^{\prime}} &=- \frac{3}{128} \xi_s + \xi_{\mathcal{A}_{2}} \xi_{\mathcal{B}_2}\,.
 \end{aligned}
\end{equation}

For $\xi_{\mathcal A_2}\neq0$ and in the absence of cancellations of the leading coefficients, the large-field behavior of \eqref{kefC} is
\begin{equation}
\mathcal{K}_{_{EF}}(\phi)=\frac{\xi_{s^{\prime}}}{\xi_{\mathcal A_2}^2}+\mathcal{O}(\phi^{-1}), \quad \phi\to\infty\,,
\end{equation}
and hence the Einstein-frame kinetic function approaches a constant.

\begin{figure}[H]
\centering
    \includegraphics[width=0.5\textwidth]{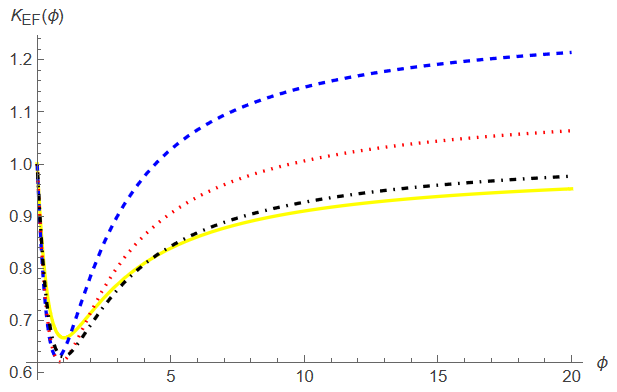}
\caption{Evolution of the Einstein-frame kinetic function $\mathcal{K}_{_{EF}}(\phi)$ for the quadratic ansatz, with $\kappa=1$ and $\xi_{\mathcal A_1}=\xi_{\mathcal A_2}=\xi_{\mathcal B_2}=1$. The coefficients $\xi_{\mathcal{C}_{i_j}}$, with $i=2,3,4$ and $j=1,2$, are taken to be equal to some $\xi_{\mathcal{C}}$ (and, as a consequence, $\xi_{\mathcal{C}_{1_1}}=\xi_{\mathcal{C}_{1_2}}=-\xi_{\mathcal{C}}/16$). The solid yellow curve corresponds to vanishing derivative couplings, whereas the black dot-dashed, red dotted, and blue dashed curves correspond to $\xi_{\mathcal{C}_{i_j}}=10^{-1}$, $2\times10^{-1}$, and $3\times10^{-1}$, respectively. The kinetic function initially decreases, reaches a positive minimum, and subsequently increases toward a finite large-field asymptotic value, $\mathcal{K}_{_{EF}}(\infty)=1$, $1.032$, $1.127$, and $1.285$, respectively. For the equal positive derivative-coupling coefficients considered here, increasing their common value raises this large-field asymptotic value.}\label{figureEKCAll}
\end{figure}

To illustrate how the derivative couplings modify the field dependence of the Einstein-frame kinetic function, let us consider a simple one-parameter slice of the coupling space. In particular, we can examine whether increasing the derivative-coupling strength changes the position and depth of the minimum of $\mathcal{K}_{_{EF}}$, whether the kinetic function remains positive, and how its large-field asymptotic value is affected. These features are relevant for the subsequent canonical field redefinition, which requires $\mathcal{K}_{_{EF}}>0$. The kinetic term \eqref{kefC} in Figure \ref{figureEKCAll} represents the case of the derivative coupling coefficients $\xi_{\mathcal{C}_{i_j}}$ with $i=2,3,4$ and $j=1,2$ being all positive and equal to some value $\xi_{\mathcal{C}}$. As a consequence, from \eqref{constr}, $\xi_{\mathcal{C}_{1_1}}=\xi_{\mathcal{C}_{1_2}}=-\xi_{\mathcal{C}}/16$. For the parameter choices used in Fig.~\ref{figureEKCAll}, $\kappa=1$ and $\xi_{\mathcal A_1}=\xi_{\mathcal A_2}=\xi_{\mathcal B_2}=1$, the kinetic function and its derivative w.r.t.\ $\phi$ satisfy
\begin{equation}
\mathcal{K}_{_{EF}}(0)=1 \,, \qquad \mathcal{K}_{_{EF}}^{\prime}(0)=-1-\frac{3}{4}\xi_{\mathcal C}\,.
\end{equation}
Therefore, for positive $\xi_{\mathcal{C}}$ the kinetic function initially decreases faster in comparison to the reference case without derivative couplings. However, its large-field asymptotic value is larger than the reference value without the couplings,
\begin{equation}
\lim_{\phi\to\infty}\mathcal{K}_{{EF}}(\phi)=1+\frac{405}{128}\xi_{\mathcal{C}}^2 \,.
\end{equation}
Thus, for nonzero positive $\xi_{\mathcal C}$, the asymptotic value is larger than the value at the origin. Together with the negative initial slope, this explains the decrease, the appearance of a minimum, and the subsequent increase observed in Fig.~\ref{figureEKCAll}. Increasing $\xi_{\mathcal C}$ also raises the large-field asymptotic value.

Although $\mathcal{K}_{_{EF}}$ remains positive for the parameter values shown in Fig.~\ref{figureEKCAll}, positivity is not guaranteed for arbitrary derivative couplings. However, regions in which $\mathcal{K}_{_{EF}}<0$ correspond to a wrong-sign scalar kinetic term in the Einstein-frame and must therefore be excluded.

\FloatBarrier

\section{General quadratic theory (\ABVCabc) }\label{Alltheothercases}
Having analyzed separately the subcases \ABV\,, in Sec.\ \ref{ABVsection}, and \ABVC \,, in Sec.\ \ref{ABVCsection}, we now consider the full action \eqref{action} including quadratic nonmetricity ($a_i$ couplings), torsion ($b_i$ couplings), and mixed sectors ($c_i$ couplings).

\subsection{Generic nondegenerate solution}

The connection field equation of the full action \eqref{action} is given in \eqref{Gfieldeqs}. Since it is algebraic and linear in torsion and nonmetricity, its generic solution can be written in the covariant forms \eqref{torphi} and \eqref{nonmetphi}. The corresponding functions $A_{i}(\phi)$ are reported in Appendix \ref{AiziPart1} for the case where at least one of $a_i$, $b_i$, or $c_i$ is nonzero. 
We now determine the conditions under which the connection equation \eqref{Gfieldeqs} admits Riemannian, torsionless, or metric-compatible solutions.

The metric equations of the full action \eqref{action} are reported in \eqref{metriceqs}. We proceed by substituting the covariant torsion and nonmetricity forms \eqref{torphi} and \eqref{nonmetphi} into the general metric equation \eqref{metriceqs}. On the generic nondegenerate connection solution, the functions $A_i$ are given in Appendix \ref{Aizi}. For the special branches considered below, the corresponding branch-specific connection solutions must be used. The resulting metric equation for the full ABVabcC theory is
\begin{equation}\label{metriceqsQS}
\begin{aligned}
  \mathcal{A}\Big(R_{(\mu \nu)} - \frac{1}{2}  g_{\mu \nu } R\Big) + (\mathcal{F}_{5} - \kappa\, \mathcal{B})\, \partial_\mu \phi\,  \partial_\nu \phi 
    + \Big(\mathcal{F}_{6}+ \frac{1}{2}\kappa\, \mathcal{B}\Big)\, g_{\mu \nu } \,\partial_\delta \phi\, \partial^\delta \phi + 2\mathcal{F}_{7} \,\nabla_{(\mu} \nabla_{\nu)} \phi\\
    +\mathcal{F}_{8}\, g_{\mu \nu } \,\nabla^\rho \nabla_\rho \phi -\kappa\, \mathcal{C}_1 g_{\mu \nu } \,\nabla_\rho \nabla^\rho \phi +\kappa g_{\mu \nu }\mathcal{V} &= 0\,,
    \end{aligned}
\end{equation}
where $\mathcal{F}_{5}(\mathcal{Z}_i)$, $\mathcal{F}_{6}(\mathcal{Z}_i)$, $\mathcal{F}_{7}(\mathcal{Z}_i,\mathcal{C}_2), \mathcal{F}_{8}(\mathcal{Z}_i)$ and the functions $\mathcal{Z}_i$ are defined in Appendix \ref{AiziPart3}.
In Eq.\ \eqref{metriceqsQS} we can see how the presence of torsion and nonmetricity has modified the metric equations with respect to the \ABV\, case studied in Sec.\ \ref{ABVsection}.
We can also rewrite \eqref{metriceqsQS} in terms of the Levi-Civita derivative $\tilde{\nabla}$. To this purpose, we need to use the relation among curvature tensors in MAG and in a metric theory \eqref{e2}, and using again \eqref{torphi}, \eqref{nonmetphi} we can rewrite the \eqref{metriceqsQS} in terms of the Levi-Civita derivative $\tilde{\nabla}$ as
\begin{equation}
\begin{aligned}\label{generalmetriceq}
&\mathcal{A} \,\tilde{G}_{\mu \nu } + \kappa g_{\mu \nu } \mathcal{V} + \mathcal{A}^{\prime} g_{\mu \nu } \widetilde{\square} \phi - \mathcal{A}^{\prime} \tilde{\nabla}_\mu \tilde{\nabla}_\nu \phi + \bigl(\mathcal{F}_{9} + \frac{1}{2} \kappa \mathcal{B}\bigr) g_{\mu \nu } \partial_\rho \phi \partial^\rho \phi + \bigl(\mathcal{F}_{10} -  \kappa \mathcal{B}\bigr) \partial_\mu \phi \partial_\nu \phi = 0\,,
    \end{aligned}
\end{equation}
where $\mathcal{F}_{9}(\mathcal{F}_{6}, \mathcal{F}_{7}, \mathcal{F}_{8}, \mathcal{A}, \mathcal{C}_1, A_i, A_i^{\prime}), \mathcal{F}_{10}(\mathcal{F}_{5}, \mathcal{F}_{7}, \mathcal{A}, A_i, A_i^{\prime})$ and the $\mathcal{Z}_{i}$ functions are defined in Appendix \ref{AiziPart3}. We note that equation \eqref{generalmetriceq} can also be found by varying the action in the Jordan-frame \eqref{action22} with kinetic term \eqref{kinetic} obtained after the algebraic connection solution had been substituted. In fact, the only terms that include a non-Riemannian contribution are the tensorial parts $g_{\mu \nu } \partial_\rho \phi \partial^\rho \phi$ and $\partial_\mu \phi \partial_\nu \phi$. Upon using the connection equations, the coefficients $\mathcal{F}_{9}$ and $\mathcal{F}_{10}$ satisfy $2 \mathcal{F}_{9}+\mathcal{F}_{10}=\mathcal{A}^{\prime\prime}$. These relations follow from the fact that, after eliminating the auxiliary connection, all non-Riemannian contributions to the scalar kinetic sector are encoded in the single effective function $\mathcal{K}(\phi)$.
The scalar equations related to the full action \eqref{action} are reported in \eqref{scalareq}. In terms of the Levi-Civita derivative $\tilde{\nabla}$ and of the expressions of torsion and nonmetricity \eqref{torphi}, \eqref{nonmetphi} the equation \eqref{scalareq} becomes
\begin{equation}\label{scalareqww}
\begin{aligned}
 & \frac{\tilde{R} \mathcal{A}^{\prime}}{2 \kappa}+ \bigl(\mathcal{F}_{11} + \mathcal{B}\bigr)\, \widetilde{\square} \phi + \partial_\rho \phi \partial^\rho \phi \bigl(\mathcal{F}_{12} + \frac{1}{2} \mathcal{B}^{\prime}\bigr) -  \mathcal{V}^{\prime} = 0\,,
   \end{aligned}
\end{equation}
where $\mathcal{F}_{11}(\mathcal{A}^{\prime},\mathcal{C}_i,A_i), \mathcal{F}_{12}(\mathcal{A}^{\prime},\mathcal{C}_i,a_i^{\prime},b_i^{\prime},c_i^{\prime},A_i,A_i^{\prime})$ are defined in Appendix \ref{AiziPart3}.
We may eliminate $\tilde{R}$ from the scalar equation \eqref{scalareqww} by using the trace of the Levi-Civita metric equation \eqref{generalmetriceq}:
\begin{equation}\label{scalareqlevi}
\begin{aligned}
&\Big(\frac{3 (\mathcal{A}^{\prime})^2}{2 \kappa \mathcal{A}} +\mathcal{F}_{11} + \mathcal{B}\Big) \widetilde{\square} \phi + \partial_\rho \phi \partial^\rho \phi \Big(\mathcal{F}_{13} + \frac{ \mathcal{B} \mathcal{A}^{\prime}}{2 \mathcal{A}}+\frac{1}{2}\mathcal{B}^{\prime}\Big) + \frac{2 \mathcal{V} \mathcal{A}^{\prime}}{\mathcal{A}} -  \mathcal{V}^{\prime} = 0\,,
  \end{aligned}
\end{equation}
where $\mathcal{F}_{11}(\mathcal{A}^{\prime},\mathcal{C}_i,A_i), \mathcal{F}_{13} (\mathcal{F}_{9},\mathcal{F}_{10},\mathcal{F}_{12},
\mathcal{A},\mathcal{A}^{\prime})$ are defined in Appendix \ref{AiziPart3}.

\subsection{Riemannian branch}

Setting both torsion and nonmetricity to zero, $S^\alpha{}_{\mu\nu}= Q^\alpha{}_{\mu\nu} = 0$ at the level of the equations of motion \eqref{Gfieldeqs} is equivalent to setting $A_1=A_2=A_3=A_4=0$ (these are, in fact, the coefficient functions for torsion and nonmetricity in \eqref{torphi}, \eqref{nonmetphi}), and we obtain
\begin{equation}\label{re0}
    \begin{aligned}
     - \mathcal{C}_4 \varepsilon_{\lambda }{}^{\mu \nu \rho} \partial_\rho \phi + \bigl(2 \mathcal{C}_1 -  \frac{1}{2} \mathcal{C}_3\bigr) \delta_{\lambda }{}^{\mu } \partial^\nu \phi + g^{\mu \nu } \partial_\lambda \phi \bigl(\mathcal{C}_2 -  \frac{\mathcal{A}^{\prime}}{\kappa}\bigr) + \delta_{\lambda }{}^{\nu } \partial^\mu \phi \bigl(\mathcal{C}_2 + \frac{1}{2} \mathcal{C}_3 + \frac{\mathcal{A}^{\prime}}{\kappa}\bigr) = 0\,.
    \end{aligned}
\end{equation}
Taking the traces in $\nu$ and $\lambda$, in $\mu$ and $\lambda$ and the contractions with the metric $g_{\mu\nu}$ and $\varepsilon_{\rho}{}^{\lambda}{}_{\mu \nu}$ give the following relations among the coupling functions:
\begin{equation}\label{Cconstra}
    \begin{aligned}
        & \mathcal{C}_1 = - \frac{\mathcal{A}^{\prime}}{\kappa} \,,\hspace{1 cm}  \mathcal{C}_2 = \frac{\mathcal{A}^{\prime}}{\kappa} \,, \hspace{1 cm} \mathcal{C}_3 = -4 \frac{\mathcal{A}^{\prime}}{\kappa} \,, \hspace{1 cm}\mathcal{C}_4 = 0\,.
    \end{aligned}
\end{equation}
It is expected, since the connection variation of every quadratic torsion/nonmetricity term is linear in S or Q and the respective contributions vanish when $S^{\alpha}_{\mu\nu}=Q^{\alpha}_{\mu\nu}=0$. Therefore only the derivative couplings and $\mathcal{A}^{\prime}$ remain in \eqref{re0}. Substituting these relations into \eqref{re0} makes the connection equation identically satisfied. The relations \eqref{Cconstra} should therefore be understood as the conditions required for the existence of a Riemannian branch with a generic scalar configuration, $\partial_\mu\phi\neq0$. They have been obtained after imposing $S^\alpha{}_{\mu\nu}=Q^\alpha{}_{\mu\nu}=0$ on the connection equations and do not imply, conversely, that, in general, imposing \eqref{Cconstra} on the unrestricted theory forces torsion and nonmetricity to vanish. If \eqref{Cconstra} holds over the field range explored by the scalar, the Levi-Civita connection is an admissible solution; possible additional non-Riemannian solutions are not excluded by this argument. The case $\partial_\mu\phi=0$ constitutes a separate branch and is not considered here. In the cosmological context, the former situation might allow synchronous oscillations of the scalar field, torsion, and nonmetricity around the Riemannian configuration.
Substituting the Riemannian conditions $A_1=A_2=A_3=A_4=0$ together with \eqref{Cconstra} into the metric equation \eqref{generalmetriceq} gives
\begin{equation}
    \mathcal{A} \tilde{G}_{\mu \nu } + \kappa g_{\mu \nu } \mathcal{V} + g_{\mu \nu } \,\widetilde{\square}\phi \mathcal{A}^{\prime} -  \tilde{\nabla}_{\mu }\tilde{\nabla}_{\nu }\phi \mathcal{A}^{\prime} + \partial_{\mu }\phi \partial_{\nu }\phi \bigl(- \kappa \mathcal{B} -  \mathcal{A}^{\prime\prime}\bigr) + g_{\mu \nu } \partial_{\rho}\phi \partial^{\rho}\phi \bigl(\tfrac{1}{2} \kappa \mathcal{B} + \mathcal{A}^{\prime\prime}\bigr) = 0\,.
\end{equation}
The scalar equation \eqref{scalareqlevi} similarly reduces to
\begin{equation}
 \Big(\frac{3 (\mathcal{A}^{\prime})^2}{2 \kappa \mathcal{A}} + \mathcal{B}\Big) \widetilde{\square} \phi + \partial_\rho \phi \partial^\rho \phi \Big(\frac{3 \mathcal{A}^{\prime\prime} \mathcal{A}^{\prime}}{2 \kappa \mathcal{A}} + \frac{\mathcal{B} \mathcal{A}^{\prime}}{2 \mathcal{A}}+\frac{1}{2}\mathcal{B}^{\prime}\Big) + \frac{2 \mathcal{V} \mathcal{A}^{\prime}}{\mathcal{A}} -  \mathcal{V}^{\prime} = 0\,.
\end{equation}
The Riemannian branch of the full ABVabcC theory differs from the corresponding Riemannian branch of the \ABV\, model discussed in Sec.\ \ref{Riemmbranch}. In the latter, the additional quadratic and derivative couplings are absent, $a_i=b_i=c_i=\mathcal{C}_i=0$, and the Levi-Civita connection requires $\mathcal{A}^{\prime}=0$. In the full theory, instead, the additional couplings remain present and the connection equations admit a Riemannian solution with nonconstant $\mathcal{A}$, provided the relations \eqref{Cconstra} are satisfied.

\subsection{Torsionless branch}
We can also consider the case of vanishing torsion at the level of the equations of motion $S^\alpha{}_{\mu\nu} = 0$ (i.e.\ setting $A_1 = A_2=0 $). Here the connection equation \eqref{Gfieldeqs} becomes
\begin{equation}\label{ConEqSzero}
    \begin{aligned}
      &  \bigl(2 a_5 -  \frac{1}{2} c_3\bigr) q^{\nu } \delta_{\lambda }{}^{\mu } + \bigl(\mathcal{A} + 2 a_4 + \frac{1}{2} c_3\bigr) q^{\mu } \delta_{\lambda }{}^{\nu } -  c_5 q^{\rho } \varepsilon_{\lambda }{}^{\mu \nu }{}_{\rho } + 2 a_4 q_{\lambda } g^{\mu \nu } \\
      &+ \bigl(- \mathcal{A} + 2 a_2\bigr) Q_{\lambda }{}^{\mu \nu } + \bigl(2 a_2 + \frac{1}{2} c_1\bigr) Q^{\mu }{}_{\lambda }{}^{\nu } + \bigl(4 a_1 -  \frac{1}{2} c_1\bigr) Q^{\nu }{}_{\lambda }{}^{\mu } \\
      &+ \bigl(-2 a_6 + c_6\bigr) \varepsilon^{\mu \nu \rho \sigma} Q_{\rho \lambda \sigma} - 2 a_6 \varepsilon_{\lambda }{}^{\nu \rho \sigma} Q_{\rho }{}^{\mu }{}_{\sigma} + \bigl(\frac{1}{2} \mathcal{A} + a_5\bigr) g^{\mu \nu } Q_{\lambda }\\
      &+ \bigl(- \frac{1}{2} \mathcal{A} + a_5 + \frac{1}{2} c_2\bigr) \delta_{\lambda }{}^{\nu } Q^{\mu } + \bigl(4 a_3 -  \frac{1}{2} c_2\bigr) \delta_{\lambda }{}^{\mu } Q^{\nu } -  c_4 \varepsilon_{\lambda }{}^{\mu \nu \rho } Q_{\rho } \\
      &+ \bigl(2 \mathcal{C}_1 -  \frac{1}{2} \mathcal{C}_3\bigr)\kappa \delta_{\lambda }{}^{\mu } \partial^\nu \phi -  \mathcal{C}_4 \kappa\varepsilon_{\lambda }{}^{\mu \nu \rho } \partial_\rho \phi + g^{\mu \nu } \partial_\lambda \phi \bigl(\mathcal{C}_2 \kappa-  \mathcal{A}^{\prime}\bigr) + \delta_{\lambda }{}^{\nu } \partial^\mu \phi \bigl(\mathcal{C}_2 \kappa+ \frac{1}{2} \mathcal{C}_3 \kappa + \mathcal{A}^{\prime}\bigr) = 0\,.
    \end{aligned}
\end{equation}
As expected, the pure torsion coefficients $b_i$ drop out of \eqref{ConEqSzero}, since the connection variation of quadratic terms in torsion is linear in torsion and therefore vanishes when $S^\alpha{}_{\mu\nu}=0$. The pure nonmetricity coefficients $a_i$ and, in general, the mixed coefficients $c_i$ remain and determine the allowed nonmetricity solution.

Substituting the nonmetricity ansatz \eqref{nonmetphi}, with $A_3$ and $A_4$ left undetermined, into \eqref{ConEqSzero}, we obtain
\begin{equation}\label{rep}
    \begin{aligned}
      \mathcal{F}_{14} \varepsilon_{\lambda }{}^{\mu \nu \rho} \partial_\rho \phi + \mathcal{F}_{15} g^{\mu \nu } \partial_\lambda \phi + \mathcal{F}_{16} \delta_{\lambda }{}^{\nu } \partial^\mu \phi + \mathcal{F}_{17} \delta_{\lambda }{}^{\mu } \partial^\nu \phi = 0 
    \end{aligned}\,,
\end{equation}
where $\mathcal{F}_{14}(\mathcal{C}_4,c_i,A_i), \mathcal{F}_{15}(\mathcal{A},\mathcal{A}^{\prime},\mathcal{C}_2,a_i,A_i), \mathcal{F}_{16}(\mathcal{A},\mathcal{A}^{\prime},\mathcal{C}_i,a_i,c_i,A_i), \mathcal{F}_{17}(\mathcal{A},\mathcal{C}_i,a_i,c_i,A_i)$ are defined in Appendix \ref{AiziPart3}. The functions $A_3$ and $A_4$ must be determined directly from the torsionless connection equations rather than simply taken from the generic expressions in Appendix~\ref{AiziPart1}. In fact, the latter were obtained by inverting the full algebraic system for $A_1,A_2,A_3,A_4$ under the assumption that its determinant is nonzero. Imposing $A_1=A_2=0$ reduces the connection system before this inversion and introduces a different nondegeneracy condition, together with branch-specific consistency relations. Therefore, the reduced system has to be solved independently. On the overlap where the generic full solution is nondegenerate and satisfies $A_1=A_2=0$, the two procedures are equivalent.

For a generic scalar configuration with $\partial_\mu\phi\neq0$, taking the traces of \eqref{rep} in $\nu$ and $\lambda$, in $\mu$ and $\lambda$, contractions with the metric $g_{\mu\nu}$ and with the tensor $\varepsilon^\lambda{}_{\mu\nu \sigma}$ we obtain $\mathcal{F}_{14}=\mathcal{F}_{15}=\mathcal{F}_{16}= \mathcal{F}_{17}=0$.
For the torsionless nondegenerate branch in which the determinant of the linear system defined by $\mathcal{F}_{14}=0$ and $\mathcal{F}_{15}=0$ is nonzero, these equations can be solved for $A_3$ and $A_4$. The expressions obtained are rational functions of the Lagrangian's coefficients:
\begin{equation}\label{complic}
    A_3 = A_3\, (\mathcal{A}, \mathcal{A}^{\prime}, a_1, a_2, a_4, a_5, c_4, c_5, c_6, \mathcal{C}_2, \mathcal{C}_4) \,, \hspace{2 cm}  A_4 = A_4\, (\mathcal{A},\mathcal{A}^{\prime},a_1, a_2, a_4, a_5, c_4, c_5, c_6, \mathcal{C}_2, \mathcal{C}_4)\,.
\end{equation}
The remaining equations $\mathcal{F}_{16}=\mathcal{F}_{17}=0$ provide two consistency relations among the couplings.

Let us now discuss the metric equation \eqref{generalmetriceq} in the torsionless case (setting $A_1=A_2=0$, where these are the functions that represent directly torsion as from \eqref{torphi}),
\begin{equation}
\begin{aligned}\label{Sno}
   & \mathcal{A} \,\tilde{G}_{\mu \nu } + \kappa g_{\mu \nu } \mathcal{V} +\mathcal{A}^{\prime} \,g_{\mu \nu } \widetilde{\square} \phi -\mathcal{A}^{\prime} \tilde{\nabla}_\mu \tilde{\nabla}_\nu \phi + \bigl(\mathcal{F}_{18} + \frac{1}{2} \kappa \mathcal{B}\bigr) g_{\mu \nu } \partial_\rho \phi \partial^\rho \phi + \bigl(\mathcal{F}_{19} -  \kappa \mathcal{B}\bigr) \partial_\mu \phi \partial_\nu \phi = 0\,,
    \end{aligned}
\end{equation}
where $\mathcal{F}_{18}(\mathcal{A},A_i,A_i^{\prime},a_i^{\prime}), \mathcal{F}_{19}(\mathcal{A},A_i,A_i^{\prime},a_i^{\prime})$ are defined in Appendix \ref{AiziPart3}.
One may substitute the explicit torsionless-branch solutions for $A_3$ and $A_4$ in \eqref{complic} obtained from $\mathcal{F}_{14} = \mathcal{F}_{15} = 0$ and also impose the consistency conditions coming from $\mathcal{F}_{16} = \mathcal{F}_{17} = 0$. However, the result is very lengthy and will not be reported here.

We can also consider the scalar field equation \eqref{scalareqlevi} in the case of vanishing torsion, $A_1=A_2=0$, and obtain
\begin{equation}
   \Big(\frac{3 (\mathcal{A}^{\prime})^2}{2 \kappa \mathcal{A}} + \mathcal{B} + \mathcal{F}_{20}\Big) \widetilde{\square} \phi + \partial_\rho \phi \partial^\rho \phi \Big( \frac{\mathcal{B} \mathcal{A}^{\prime}}{2 \mathcal{A}}+\frac{1}{2}\mathcal{B}^{\prime} + \mathcal{F}_{21}\Big) + \frac{2 \mathcal{V} \mathcal{A}^{\prime}}{\mathcal{A}} -  \mathcal{V}^{\prime} = 0\,,
\end{equation}
where $\mathcal{F}_{20} (\mathcal{A}, \mathcal{A}^{\prime}, A_i, \mathcal{C}_i), \mathcal{F}_{21}(\mathcal{A},\mathcal{A}^{\prime},a_i, a_i^{\prime}, A_i, A_i^{\prime}, \mathcal{C}_i, \mathcal{C}_i^{\prime})$ are defined in Appendix \ref{AiziPart3}. One may substitute into $\mathcal{F}_{21}$ and $\mathcal{F}_{20}$ the explicit torsionless-branch solutions for $A_3$ and $A_4$ in \eqref{complic} obtained from $\mathcal{F}_{14} = \mathcal{F}_{15} = 0$ and also impose the consistency conditions coming from $\mathcal{F}_{16} = \mathcal{F}_{17} = 0$. However, the result is very lengthy and will not be reported here.

This torsionless branch of the full ABVabcC theory, obtained by imposing $S^\alpha{}_{\mu\nu}=0$ or equivalently $A_1=A_2=0$ on the connection equations, differs from the torsionless branch of the restricted \ABV\, model discussed in Sec.\ \ref{Palatini}. In the \ABV\, model, the additional couplings are absent, $a_i=b_i=c_i=\mathcal{C}_i=0$, and the torsionless connection solution is simply given by \eqref{Aiex1g2}. In the full theory, instead, the additional quadratic and derivative couplings remain present and modify the nonmetricity solution. The functions $A_3$ and $A_4$ are determined by $\mathcal{F}_{14}=\mathcal{F}_{15}=0$, while $\mathcal{F}_{16}=\mathcal{F}_{17}=0$ provide additional consistency conditions among the couplings.

\subsection{Metric-compatible branch}

We can also consider the metric-compatible branch, obtained by imposing $Q^\alpha{}_{\mu\nu}=0$, or equivalently $A_3=A_4=0$, at the level of the equations of motion. In contrast to the \ABV\, model of Sec.~\ref{metriccompatible}, the full ABVabcC theory is generically not projectively invariant, so metric compatibility is not in general a projective gauge choice but selects a genuine branch of the connection equations. The connection equation \eqref{Gfieldeqs} becomes
\begin{equation}\label{Qno}
    \begin{aligned}
  -2 \bigl(\mathcal{A} -  \frac{1}{2} b_2\bigr) S_{\lambda }{}^{\mu \nu } -  b_2 S_{\lambda }{}^{\nu \mu } -  c_1 S_{\lambda }{}^{\nu \mu } - 2 \bigl(- b_1 + \frac{1}{2} c_1\bigr) S^{\mu \nu }{}_{\lambda } + 2 b_5 \varepsilon^{\mu \nu \rho \sigma} S_{\rho \sigma\lambda } -  c_6 \varepsilon^{\mu \nu \rho \sigma} S_{\rho \sigma\lambda } -  c_6 \varepsilon_{\lambda }{}^{\nu \rho \sigma} S_{\rho \sigma}{}^{\mu } &\\
  - 2 \bigl(- \mathcal{A} -  \frac{1}{2} c_3\bigr) g^{\mu \nu } S_{\lambda } - 2 \bigl(\mathcal{A} -  \frac{1}{2} b_3 -  \frac{1}{2} c_3\bigr) \delta_{\lambda }{}^{\nu } S^{\mu } - 2 \bigl(\frac{1}{2} b_3 -  c_2\bigr) \delta_{\lambda }{}^{\mu } S^{\nu } -  b_4 \varepsilon_{\lambda }{}^{\mu \nu \rho } S_{\rho } + c_5 g^{\mu \nu } t_{\lambda } &\\
  - 2 \bigl(- \frac{1}{4} b_4 -  \frac{1}{2} c_5\bigr) \delta_{\lambda }{}^{\nu } t^{\mu } - 2 \bigl(\frac{1}{4} b_4 -  c_4\bigr) \delta_{\lambda }{}^{\mu } t^{\nu } + \mathcal{C}_2 \kappa g^{\mu \nu } \partial_\lambda \phi + \bigl(2 \mathcal{C}_1 -  \frac{1}{2} \mathcal{C}_3\bigr) \kappa \delta_{\lambda }{}^{\mu } \partial^\nu \phi &\\
  -  \mathcal{C}_4 \kappa\varepsilon_{\lambda }{}^{\mu \nu \rho } \partial_\rho \phi -  g^{\mu \nu } \partial_\lambda \phi \mathcal{A}^{\prime} + \delta_{\lambda }{}^{\nu } \partial^\mu \phi \bigl(\mathcal{C}_2 \kappa + \frac{1}{2} \mathcal{C}_3 \kappa+ \mathcal{A}^{\prime}\bigr) &= 0\,.
    \end{aligned}
\end{equation}

As expected, the pure nonmetricity coefficients $a_i$ drop out of \eqref{Qno}, since the connection variation of quadratic terms in nonmetricity is linear in nonmetricity and therefore vanishes when $Q^\alpha{}_{\mu\nu}=0$. The pure torsion coefficients $b_i$ remain, while the mixed coefficients $c_i$ can also survive because the variation of a mixed torsion--nonmetricity term may remain nonzero when only the nonmetricity is set to zero.

Substituting the torsion ansatz \eqref{torphi}, with $A_1$ and $A_2$ left undetermined, into \eqref{Qno}, we obtain
\begin{equation}\label{rep2}
    \begin{aligned}
      \mathcal{F}_{22} \varepsilon_{\lambda }{}^{\mu \nu \rho} \partial_\rho \phi + \mathcal{F}_{23} g^{\mu \nu } \partial_\lambda \phi + \mathcal{F}_{24} \delta_{\lambda }{}^{\nu } \partial^\mu \phi + \mathcal{F}_{25} \delta_{\lambda }{}^{\mu } \partial^\nu \phi = 0\,,
    \end{aligned}
\end{equation}
where $\mathcal{F}_{22}(\mathcal{A},\mathcal{C}_4,b_i,A_i), \mathcal{F}_{23}(\mathcal{A},\mathcal{A}^{\prime},\mathcal{C}_2,c_i,A_i), \mathcal{F}_{24}(\mathcal{A},\mathcal{A}^{\prime},\mathcal{C}_i,b_i,c_i,A_i), \mathcal{F}_{25}(\mathcal{C}_i,b_i,c_i,A_i)$ are defined in Appendix \ref{AiziPart3}. The functions $A_1$ and $A_2$ must be determined directly from the metric-compatible connection equations rather than simply taken from the generic expressions in Appendix~\ref{AiziPart1}. In fact, the latter were obtained by inverting the entire algebraic system for $A_1,A_2,A_3,A_4$ under the assumption that its determinant is nonzero. Imposing $A_3=A_4=0$ reduces the connection system before this inversion and introduces a different nondegeneracy condition, together with branch-specific consistency relations. Therefore, the reduced system has to be solved independently. On the overlap where the generic full solution is nondegenerate and satisfies $A_3=A_4=0$, the two procedures are equivalent.

For a generic scalar configuration with $\partial_\mu\phi\neq0$, taking traces of \eqref{rep2} in $\nu$ and $\lambda$, in $\mu$ and $\lambda$ and the contractions with the metric $g_{\mu\nu}$ and $\varepsilon^\lambda{}_{\mu\nu \sigma}$ we obtain $\mathcal{F}_{22}=\mathcal{F}_{23}=\mathcal{F}_{24}= \mathcal{F}_{25}=0$. 
For the metric-compatible nondegenerate branch in which the determinant of the linear system defined by $\mathcal{F}_{22}=0$ and $\mathcal{F}_{23}=0$ is nonzero, these equations can be solved for $A_1$ and $A_2$. The expressions obtained are rational functions of the Lagrangian's coefficients:
\begin{equation}\label{complic2}
    A_1 = A_1\, (\mathcal{A}, \mathcal{A}^{\prime}, b_1, b_2, b_4, b_5, c_1, c_3, c_5, c_6, \mathcal{C}_2, \mathcal{C}_4) \,, \hspace{2 cm}  A_2 = A_2\, (\mathcal{A}, \mathcal{A}^{\prime}, b_1, b_2, b_4, b_5, c_1, c_3, c_5, c_6, \mathcal{C}_2, \mathcal{C}_4)\,.
\end{equation}
The remaining equations $\mathcal{F}_{24}=\mathcal{F}_{25}=0$ provide two consistency relations among the couplings.

The general metric equation \eqref{generalmetriceq} on the metric-compatible branch, $A_3=A_4=0$, becomes
\begin{equation}
\begin{aligned}\label{Qno2}
   & \mathcal{A} \,\tilde{G}_{\mu \nu } + \kappa g_{\mu \nu } \mathcal{V} + \mathcal{A}^{\prime} \,g_{\mu \nu } \widetilde{\square} \phi -\mathcal{A}^{\prime}\tilde{\nabla}_\mu \tilde{\nabla}_\nu \phi + \bigl(\mathcal{F}_{26} + \frac{1}{2} \kappa \mathcal{B}\bigr) g_{\mu \nu } \partial_\rho \phi \partial^\rho \phi + \bigl(\mathcal{F}_{27} -  \kappa \mathcal{B}\bigr) \partial_\mu \phi \partial_\nu \phi = 0\,,
    \end{aligned}
\end{equation}
where $\mathcal{F}_{26}(\mathcal{A},\mathcal{C}_i,\mathcal{C}_i^{\prime},b_i,c_i,c_i^{\prime},A_i,A_i^{\prime}), \mathcal{F}_{27}(\mathcal{A},\mathcal{C}_i,\mathcal{C}_i^{\prime},b_i,c_i,c_i^{\prime},A_i,A_i^{\prime})$ are defined in Appendix \ref{AiziPart3}. 
One could also substitute the constraints from the zero-nonmetricity connection equations \eqref{complic2} into $\mathcal{F}_{26}$ and $\mathcal{F}_{27}$; however, the resulting expressions are very lengthy.
In the metric equations \eqref{Qno2}, the pure nonmetricity sector $a_i$ drops out, as expected. The torsion coefficients $b_i$, the mixed coefficients $c_i$, and the derivative couplings $\mathcal{C}_i$ remain explicitly present in $\mathcal{F}_{26}$ and $\mathcal{F}_{27}$.

Let us now consider the metric compatible branch of the scalar equation \eqref{scalareqlevi}, setting $A_3=A_4=0$, which corresponds to vanishing nonmetricity according to \eqref{nonmetphi},
\begin{equation}
    \begin{aligned}
      \Big(\frac{3 (\mathcal{A}^{\prime})^2}{2 \kappa \mathcal{A}} + \mathcal{B} + \mathcal{F}_{28}\Big) \widetilde{\square} \phi + \partial_\rho \phi \partial^\rho \phi \Big( \frac{\mathcal{B} \mathcal{A}^{\prime}}{2 \mathcal{A}}+\frac{1}{2}\mathcal{B}^{\prime} + \mathcal{F}_{29}\Big) + \frac{2 \mathcal{V} \mathcal{A}^{\prime}}{\mathcal{A}} -  \mathcal{V}^{\prime} = 0\,,
    \end{aligned}
\end{equation}
where $\mathcal{F}_{28}(\mathcal{A},\mathcal{A}^{\prime},\mathcal{C}_i,c_i,A_i), \mathcal{F}_{29}(\mathcal{A},\mathcal{A}^{\prime},\mathcal{C}_i,\mathcal{C}_i^{\prime},b_i,b_i^{\prime},c_i,c_i^{\prime},A_i,A_i^{\prime})$ are defined in Appendix \ref{AiziPart3}. We could also substitute into $\mathcal{F}_{28}$ and $\mathcal{F}_{29}$ the explicit expressions for $A_1$, $A_2$ coming from the connection field equation with zero nonmetricity taken at the level of the equations of motion \eqref{complic2}, $\mathcal{F}_{22}=0$, $\mathcal{F}_{23}=0$ and the consistency relations coming from $\mathcal{F}_{24}=0$ and $\mathcal{F}_{25}=0$.

This metric-compatible branch of the full ABVabcC theory, obtained by imposing $Q^\alpha{}_{\mu\nu}=0$ or equivalently $A_3=A_4=0$ on the connection equations, differs from the metric-compatible branch of the restricted \ABV\, model discussed in Sec.\ \ref{metriccompatible}. In the \ABV\, model, the additional couplings are absent, $a_i=b_i=c_i=\mathcal{C}_i=0$, and the metric-compatible connection solution is simply given by \eqref{Aiex1g3}. In the full theory, instead, the additional quadratic and derivative couplings remain present and modify the torsion solution. The functions $A_1$ and $A_2$ are determined by $\mathcal{F}_{22}=\mathcal{F}_{23}=0$, while $\mathcal{F}_{24}=\mathcal{F}_{25}=0$ provide additional consistency conditions among the couplings.

\section{Classification of the quadratic sectors}\label{subcase2}

In this section, we classify representative subcases of the full theory on the generic nondegenerate connection solution described in Sec.~\ref{Alltheothercases}. The \ABV\, and \ABVC \, sectors, for which the connection equations become projectively degenerate, have been treated separately in Secs.~\ref{ABVsection} and \ref{ABVCsection}. The labels of the following sectors refer to the quadratic invariants retained in the action, rather than to which non-Riemannian components survive in the resulting connection solution.

\subsection{Pure nonmetricity sector (\ABVa \,)}\label{KABVa}
In the \ABVa \, sector, we retain, in addition to $\mathcal{A}$, $\mathcal{B}$, and $\mathcal{V}$, only the quadratic nonmetricity terms proportional to $a_i$. On the generic nondegenerate solution, the connection reported in Appendix \ref{AiziPart1} for torsion and nonmetricity in \eqref{nonmetphi}, \eqref{torphi} reduces to
\begin{equation}\label{ABVAi}
    \begin{aligned}
    A_1 &= \frac{1}{4}\frac{\mathcal{A}^{\prime}}{\mathcal{A}}, \qquad & A_2 &=0,\\
    A_3 &= 0,  \qquad &  A_4&=0\,.
    \end{aligned}
\end{equation}
The nonmetricity vanishes, the reason for this somewhat counterintuitive result can be seen directly from the connection equation. The term $\mathcal{A}R$ is projectively invariant, whereas the quadratic nonmetricity terms generically break this invariance. The projective trace of the connection equation therefore gives a relation between the two nonmetricity vectors $Q_\mu$ and $q_\mu$. Combining this relation with the remaining independent traces yields, on the generic nondegenerate branch, $Q_\mu=0$, and the first relation then also gives $q_\mu=0$. For a generic scalar configuration, the connection solution \eqref{nonmetphi} then implies $A_3=A_4=0$, so the full nonmetricity vanishes. The remaining connection equations fix $A_1=\mathcal{A}'/(4\mathcal{A})$ and $A_2=0$.

Substitution of the solution \eqref{ABVAi} into \eqref{kinetic} gives
\begin{equation}\label{gia}
\mathcal{K}(\phi) = \mathcal{B} - \frac{3}{2} \frac{\bigl( \mathcal{A}^{\prime}\bigr)^2}{\kappa\,\mathcal{A}}, \hspace{2 cm} \mathcal{K}_{_{EF}}(\phi)=\frac{\mathcal{B}}{\mathcal{A}}
\end{equation}
Thus, the quadratic nonmetricity sector leaves both Jordan- and Einstein-frame kinetic functions unchanged with respect to the \ABV\, result \eqref{K1}. Since there are no derivative couplings, the hypermomentum also remains equal to \ABV\, expression \eqref{hyper1}.

\subsection{Pure torsion sector (\ABVb \,)}\label{subcase3}
In the \ABVb \, sector, we retain, in addition to $\mathcal{A}$, $\mathcal{B}$, and $\mathcal{V}$, only the quadratic torsion terms proportional to $b_i$. In the generic nondegenerate branch, the connection solution in Appendix \ref{AiziPart1} for torsion and nonmetricity in \eqref{nonmetphi}, \eqref{torphi} reduces to
\begin{equation}\label{ABVBi}
    \begin{aligned}
   & A_1 = 0,   \hspace{2.1 cm} A_2 =0,\\
    & A_3 =  \frac{\mathcal{A}^{\prime}}{\mathcal{A}},  \hspace{1.8 cm}   A_4=0\,.
    \end{aligned}
\end{equation}
Hence, the torsion vanishes, whereas the nonmetricity remains. As from \ref{KABVa}, the reason for this can be understood directly from the structure of the \ABVb \, connection equations. The term $\mathcal{A}R$ is projectively invariant, so the corresponding connection equation has vanishing projective trace. The quadratic torsion terms generically break this invariance, and their contribution to the same trace gives a linear relation between the torsion vector $S_\mu$ and the axial vector $t_\mu$. An independent contraction of the connection equation with the Levi-Civita tensor provides a second relation. On the generic nondegenerate branch, the two relations imply $S_\mu=t_\mu=0$. For the scalar-field connection solution \eqref{torphi}, this is equivalent to $A_1=A_2=0$, and hence the full torsion vanishes.
Moreover, although the \ABVb \, connection equations differ from those of the \ABV\, theory, their generic nondegenerate solution coincides with the Palatini representative \eqref{Aiex1g2} of the \ABV\, connection.

Substituting the solution \eqref{ABVBi} into \eqref{kinetic} gives
\begin{equation}\label{k11}
\mathcal{K}(\phi) = \mathcal{B} - \frac{3}{2} \frac{\bigl( \mathcal{A}^{\prime}\bigr)^2}{\kappa\,\mathcal{A}}, \hspace{2 cm} \mathcal{K}_{_{EF}}(\phi)=\frac{\mathcal{B}}{\mathcal{A}}\,.
\end{equation}
Thus, the quadratic torsion sector leaves both Jordan-and Einstein-frame kinetic functions unchanged with respect to the \ABV\, result \eqref{K1}. Since no derivative couplings are present, the hypermomentum also remains equal to the \ABV\, expression \eqref{hyper1}.

\subsection{Combined pure sectors (\ABVab \,)}\label{subcase6}

Having considered the pure nonmetricity and pure torsion sectors separately, we now turn to the nondegenerate \ABVab \, sector in which both sets of quadratic terms are present.

\subsubsection{Kinetic term}
In our action \eqref{action} we now keep the \ABV\, couplings $\mathcal{A}$, $\mathcal{B}$, and $\mathcal{V}$ together with the quadratic nonmetricity and torsion terms $a_i$ and $b_i$. The general modified kinetic term \eqref{kinetic} becomes
\begin{equation}\label{sub6ab}
\begin{aligned}
   \mathcal{K}(\phi)
   =&\; \mathcal{B}
   - \frac{1}{\kappa} \Big( \left( -24\mathcal{A} +6b_1 -3b_2 +9b_3 \right)A_1^2 +\left( 18 b_4 +24 b_5 \right)A_1A_2 +\left( 6\mathcal{A} -6b_1 -6b_2 \right)A_2^2 \\ &\quad -12\mathcal{A}A_1A_3 +6\mathcal{A}A_1A_4 +\left( -\tfrac{3}{2}\mathcal{A} +4a_1 +a_2 +16a_3 +a_4 +4a_5 \right)A_3^2 \\ &\quad +\left( \tfrac{3}{2}\mathcal{A} +2a_1 +5a_2 +8a_3 +5a_4 +11a_5 \right)A_3A_4 \\ &\quad +\left( \tfrac{3}{4}\mathcal{A} +\tfrac{5}{2}a_1 +\tfrac{7}{4}a_2 +a_3 +\tfrac{25}{4}a_4 +\tfrac{5}{2}a_5 \right)A_4^2 +3\mathcal{A}^{\prime} \left( 4A_1+A_3-\tfrac{1}{2}A_4 \right) \Big)\,,
\end{aligned}
\end{equation}
with $A_1$, $A_2$, $A_3$, $A_4$ the expressions reported in Appendix \ref{Aizi}. When both sectors are present, the kinetic term generically differs from the \ABV\, result \eqref{K1}. This contrasts with the \ABVa \, and \ABVb \, sectors, where each pure quadratic sector separately leaves the \ABV\, kinetic function unchanged. This difference can be understood from the corresponding connection solutions. In the \ABVa \, sector, the connection equations force the nonmetricity to vanish, so the quadratic terms proportional to $a_i$ vanish on shell, while in the \ABVb \, sector they analogously force the torsion to vanish, so the terms proportional to $b_i$ vanish on shell. When both $a_i$ and $b_i$ are present, however, the connection equations form a coupled system for torsion and nonmetricity, already due to the mixed torsion--nonmetricity terms contained in the Ricci scalar \eqref{Rnonriem}. Consequently, neither torsion nor nonmetricity is generically forced to vanish, and both quadratic sectors contribute after the connection is eliminated, leading to a kinetic function that differs from the \ABV\, result.

The kinetic term \eqref{sub6ab} can be illustrated by considering the simplifying assumption $a_i = b_i = \mathcal{A}$:
\begin{equation}
        \mathcal{K}(\phi) = \mathcal{B} +\frac{1}{\kappa} \frac{29985}{11018}\frac{(\mathcal{A}^{\prime})^2}{\mathcal{A}},
        \hspace{2 cm} \mathcal{K}_{_{EF}}(\phi)=\frac{\mathcal{B}}{\mathcal{A}} + \frac{1}{\kappa} \frac{23256}{5509}\frac{(\mathcal{A}^{\prime})^2}{\mathcal{A}^2}\,.
\end{equation}
Even if the kinetic term \eqref{sub6ab} now differs from the \ABV\, case \eqref{K1}, we have that the hypermomentum tensor does not undergo modifications with respect to \eqref{hyper1}.

\subsection{Mixed sector (\ABVC \,)}\label{subcase5}

Another subcase of the general action \eqref{action} and of the equations of motion mentioned above in Sec.\ \ref{Alltheothercases} that we can consider is the case where we have the nonminimally coupled terms of the \ABV\, model and, in addition, the mixed quadratic terms in the torsion and nonmetricity tensors, i.e.\ the terms with coefficients $c_i$. 

\subsubsection{Kinetic term}
In this \ABVC \, subcase, keeping only $\mathcal{A}$, $\mathcal{B}$, $\mathcal{V}$ and $c_i$ different from zero, the general kinetic term \eqref{kinetic} simplifies to
\begin{equation}\label{ci}
  \begin{aligned}
\mathcal{K}(\phi) = &
\mathcal{B} - \frac{3}{4 \kappa}\Big[
\mathcal{A}\left(-32A_1^2+8A_2^2-2A_3^2+2A_3A_4+A_4^2\right)+4\left(-4\mathcal{A}+c_1+4c_2+c_3\right)A_1A_3 +2\left(4\mathcal{A}-c_1+2c_2+5c_3\right)A_1A_4\\
& +8\left(4c_4+c_5+c_6\right)A_2A_3+4\left(2c_4+5c_5-c_6\right)A_2A_4+2\mathcal{A}'\left(8A_1+2A_3-A_4\right)\Big]\,,
   \end{aligned}
\end{equation}
with $A_1$, $A_2$, $A_3$, $A_4$ as in Appendix \ref{Aizi} (keeping only $c_i$ terms therein as well). And the kinetic term in the Einstein-frame follows from \eqref{EFkinetic}. The mixed quadratic sector generically modifies the Einstein-frame kinetic function. However, as discussed below, the parity-odd coefficients $c_4, c_5, c_6$, when present alone or only in combination with one another, leave the \ABV\, kinetic function unchanged on the generic nondegenerate branch.

In this context, no modifications appear in the hypermomentum tensor \eqref{hypedef}. This is because in this \ABVC \, subcase we are modifying the geometric sector only, while the matter sector is unchanged with respect to the \ABV\, model.

\subsubsection{Polynomial coupling functions}
In this paragraph, we now consider for our \ABVC \, functions the specific expressions defined in Sec.\ \ref{dim}. These at the lowest order are $\mathcal{A} = \xi_{\mathcal{A}_0} = 1$, $\mathcal{B} = 1$ and $c_{i}(\phi) =\xi_{c_{i_{0}}}$, and the kinetic term is
\begin{equation}\label{csmall}
  \begin{aligned}
    \mathcal{K}(\phi) = 1\,.
   \end{aligned}
\end{equation}
We could then consider the functions in Sec.\ \ref{dim} \eqref{expansionABV} and \eqref{expansionc} up to the second order.
For the full second-order polynomial ansatz, the resulting kinetic functions are lengthy and not particularly illuminating. Their large-field behavior can nevertheless be read off from the general expression \eqref{ci} together with \eqref{expansion}:
\begin{equation}\label{csmalllim}
\mathcal{K}(\phi)=\mathcal{O}(\phi^2)\,,\qquad \phi\to\infty\,,
\end{equation}
\begin{equation}\label{csmallEFlim}
\mathcal{K}_{_{EF}}(\phi)=\text{constant}+\mathcal{O}(\phi^{-1})\,,\qquad \phi\to\infty\,.
\end{equation}
To give concrete examples of some kinetic terms that may be interesting for inflation scenarios, we here continue with two illustrative subcases of the \ABVC \, case \eqref{csmall}.

\vspace{1 cm}\noindent\paragraph{Only $c_1$}~\\
When we keep only $c_1 \neq 0 $ and let other $c_i$ to vanish, the functions $A_i$ from Appendix \ref{Aizi} become
\begin{equation}
    \begin{aligned}
   A_1 &= \frac{\mathcal{A}^{\prime}}{8\mathcal{A}-c_1},   \qquad & A_2 &=0,\\
    A_3 &=  \frac{4 \mathcal{A}^{\prime}}{8 \mathcal{A} - c_1},  \qquad &  A_4&=0\,.
    \end{aligned}
\end{equation}
In the formal limit $c_1\to0$, these coefficients satisfy the \ABV\, relation \eqref{Aiex1g1}, selecting a particular representative of the projectively degenerate \ABV\, connection solution in Sec.\ \ref{ABVsection}.

The corresponding modified kinetic term is
\begin{equation}
    \mathcal{K}(\phi) =  \mathcal{B} - \frac{12 (\mathcal{A}^{\prime})^2}{\kappa \Bigl(8 \mathcal{A}- c_1 \Bigr)}\,.
\end{equation}
Again, it is clear that taking $c_1 = 0$ leads back to the usual \ABV\, model kinetic term \eqref{K1}. We see that this solution requires $8 \mathcal{A} - c_1 \neq 0$.

For the polynomial parametrization introduced in Sec.\ \ref{dim},
$c_1(\phi)= \xi_{c_{1_{0}}}+\sqrt{\kappa}\, \xi_{c_{1_{1}}} \,\phi + \kappa\,\xi_{c_{1_{2}}}\, \phi^2$, the kinetic function becomes
\begin{equation}\label{subcase5n2}
    \begin{aligned}
      \mathcal{K}(\phi) = 1 + \kappa \, \xi_{\mathcal{B}_2} \phi^2 + \frac{12 ( \, \xi_{\mathcal{A}_1} + 2 \sqrt{\kappa}  \, \xi_{\mathcal{A}_2} \phi)^2}{-8 + \xi_{c_{1_{0}}} + \sqrt{\kappa} (-8  \, \xi_{\mathcal{A}_1} + \xi_{c_{1_{1}}}) \phi + \kappa (-8  \, \xi_{\mathcal{A}_2} + \xi_{c_{1_{2}}}) \phi^2}\,.
    \end{aligned}
\end{equation}
At large $\phi$, the kinetic function behaves as $\mathcal{K}(\phi)=\mathcal{O}(\phi^2)$, as in the general case \eqref{csmalllim}.

In the Einstein-frame, this gives
\begin{equation}\label{kefc1}
\begin{aligned}
\mathcal{K}_{_{EF}}(\phi)=&  \frac{\mathcal{K}(\phi)}{\mathcal{A}}+\frac{3}{2} \frac{(\mathcal{A}^{\prime})^2}{\kappa \mathcal{A}^2}=\\& \frac{1 + \kappa \xi_{\mathcal{B}_2} \phi^2}{1 + \sqrt{\kappa} \xi_{\mathcal{A}_1} \phi + \kappa \,\xi_{\mathcal{A}_2} \phi^2} + \frac{3 (\xi_{\mathcal{A}_1} + 2 \sqrt{\kappa} \xi_{\mathcal{A}_2} \phi)^2 (\xi_{c_{1_{0}}} + \sqrt{\kappa} \xi_{c_{1_{1}}} \phi + \kappa \xi_{c_{1_{2}}} \phi^2)}{2 (1 + \sqrt{\kappa} \xi_{\mathcal{A}_1} \phi + \kappa \,\xi_{\mathcal{A}_2} \phi^2)^2 \bigl(-8 + \xi_{c_{1_{0}}} + \sqrt{\kappa} (-8 \xi_{\mathcal{A}_1} + \xi_{c_{1_{1}}}) \phi + \kappa (-8 \xi_{\mathcal{A}_2} + \xi_{c_{1_{2}}}) \phi^2\bigr)}\,.
\end{aligned}
\end{equation}
We see how if the $c_i$ vanish, the second term goes to zero, and we recover the standard \ABV\, case \eqref{K1}. Again, at large $\phi$ the Einstein-frame kinetic function approaches a constant, as in the general case \eqref{csmallEFlim}.

\vspace{1 cm}\noindent\paragraph{Other $c_i$ couplings}\label{cicomment}
It can be shown that each of the couplings $c_1$, $c_2$, and $c_3$ in the action \eqref{action}, when present individually, modifies the \ABV\, kinetic function \eqref{K1} (as shown explicitly for $c_1$ in \eqref{kefc1}). By contrast, $c_4$, $c_5$, and $c_6$, which correspond to the parity-odd mixed invariants, leave the \ABV\, kinetic function \eqref{K1} unchanged when present individually or in combinations among themselves. When combined with $c_1$, $c_2$, or $c_3$, however, they can contribute to the modified kinetic function.

\subsection{Universal coupling example}

As a final illustrative subcase, let us consider the case where, aside from the \ABV\, model couplings, all the other terms in the full action \eqref{action} are present but are set equal or proportional to $\mathcal{A}$:
\begin{equation}\label{no5}
a_i = b_i = c_i = \mathcal{A},  \hspace{1 cm}  \mathcal{C}_i =\frac{1}{\sqrt{\kappa}} \mathcal{A}\,.
\end{equation}
This choice defines a nondegenerate, projectively noninvariant solution of the connection equations for $\mathcal{A} \neq 0$.

The functions $A_i$ of the Appendix \ref{Aizi} become
\begin{equation}
    \begin{aligned}
    A_1&=-\frac{196 \sqrt{\kappa }}{365}-\frac{32 \mathcal{A}'}{25 \mathcal{A}},   \qquad & A_2&=-\frac{377 \sqrt{\kappa }}{730}-\frac{37 \mathcal{A}'}{25 \mathcal{A}}, \\\\
     A_3&=\frac{43 \sqrt{\kappa }}{365}+\frac{11 \mathcal{A}'}{25 \mathcal{A}},  \qquad &   A_4&=\frac{124 \sqrt{\kappa }}{365}+\frac{28 \mathcal{A}'}{25 \mathcal{A}} \,.
    \end{aligned}
\end{equation}
We see that both torsion and nonmetricity, as expected, are non-vanishing.

With the choice \eqref{no5}, the Jordan-frame kinetic term becomes
\begin{equation}
\begin{aligned}\label{alleqA}
\mathcal{K}(\phi) =& \mathcal{B}+\frac{107}{73} \mathcal{A} + \frac{33  \mathcal{A}^{\prime}}{5 \sqrt{\kappa}} + \frac{393 \bigl( \mathcal{A}^{\prime}\bigr)^2}{50 \kappa \mathcal{A}}\,.
    \end{aligned}
\end{equation}
We see that the kinetic term differs, in general, from the \ABV\, case \eqref{K1}. The expression in the Einstein-frame follows as usual from \eqref{EFkinetic}:
\begin{equation}
\begin{aligned}\label{alleqAEF}
\mathcal{K}_{_{EF}}(\phi) =& \frac{\mathcal{B}}{\mathcal{A}}+ \frac{107}{73}+\frac{33 \mathcal{A}'}{5 \sqrt{\kappa } \mathcal{A}}+\frac{234\left(\mathcal{A}'\right)^2}{25 \kappa  \mathcal{A}^2}\,.
    \end{aligned}
\end{equation}

\section{Canonical field redefinition and reshaping of the potential}\label{canonicalredefinition}

In the previous sections, we determined the effective kinetic function in several subcases of the general action \eqref{action}. We now study how these modifications affect the relation between the original scalar field $\phi$ and the canonically normalized field $\varphi$, and consequently reshape the Einstein-frame potential when expressed as a function of $\varphi$.

Starting from the Einstein-frame action \eqref{actionEF}, the canonical field is defined by $d\varphi/d\phi=\pm\sqrt{\mathcal{K}_{_{EF}}(\phi)}$ \cite{Jarv:2014hma,Burns:2016ric}. The potential itself is not changed by this field redefinition, but its functional form becomes $\mathcal{V}_{_{EF}}(\phi(\varphi))$.

\subsection{General}
For generic values of the coupling coefficients in the expansions in Sec.\ \ref{dim} and in the absence of cancellations, the Jordan-frame and Einstein-frame kinetic functions, \eqref{kinetic} and \eqref{EFkinetic}, take the rational forms
\begin{equation}
    \mathcal{K}(\phi) = \frac{\sum\limits_{i=0}^{10} f_i \,\phi^i}{\sum\limits_{j=0}^8 g_j\, \phi^j}, \hspace{2 cm}  \mathcal{K}_{_{EF}}(\phi) = \frac{\sum\limits_{i=0}^{12} r_i \,\phi^i}{\sum\limits_{j=0}^{12} s_j\, \phi^j}\,.
\end{equation}
Here, $f_i,g_j, r_i$ and $s_j$ are combinations of the coupling constants and appropriate powers of $\kappa$, independent of $\phi$. For particular relations among the couplings, cancellations may reduce the polynomial degrees.

However, to obtain more concrete results, we now study the modifications of the canonical field and of the corresponding Einstein-frame potential in representative subcases of \eqref{kinetic}. The \ABVa \, and \ABVb \, models have the same Einstein-frame kinetic function as the \ABV\, model and therefore do not produce any additional modification of the canonical field or potential. We first illustrate how derivative and mixed torsion--nonmetricity couplings reshape the potential through the modified canonical field redefinition. We then consider the inverse problem and show, in a particular \ABVC \, example, how the couplings can be chosen to reproduce a prescribed target potential.

\subsection{Potential reshaping from non-Riemannian couplings}

\subsubsection{Derivative couplings}

We first consider the effect of a given modified kinetic function on the potential expressed in terms of the canonically normalized field $\mathcal{V}_{_{EF}}(\phi(\varphi))$. We start from the condition for canonical normalization,
\begin{equation}\label{eqek3}
  - \frac{1}{2}\mathcal{K}_{_{EF}}(\phi) \,\partial_\mu \phi \,\partial^\mu \phi \equiv  - \frac{1}{2}\,\partial_\mu \varphi \,\partial^\mu \varphi  = - \frac{1}{2} \Big(\frac{d \varphi (\phi)}{d \phi} \frac{d \varphi (\phi)}{d \phi}\Big) \partial_\mu \phi \,\partial^\mu \phi
\end{equation}
where $\varphi$ is the canonically normalized scalar field ($\partial_\mu \varphi= \tfrac{d \varphi (\phi)}{d \phi} \partial_\mu \phi$). Therefore
\begin{equation}\label{expr1}
 \mathcal{K}_{_{EF}}(\phi)= \Big(\frac{d \varphi (\phi)}{d \phi}\Big)^2 \,.
\end{equation}
The procedure is as follows. First, we choose a specific kinetic term. On any field interval where $\mathcal{K}_{_{EF}}(\phi)>0$, we take the positive square root of \eqref{expr1}. We integrate it numerically, invert the integral result and obtain $\phi(\varphi)$ (that is our scalar field $\phi$ expressed in terms of the canonically normalized one $\varphi$). Finally, we use the $\phi(\varphi)$ inside the Einstein-frame potentials corresponding to the quadratic and the quartic Jordan-frame potentials to visualize the effect of including the derivative couplings $\mathcal{C}_i$ in our theory.

In general, the Einstein-frame potentials in terms of $\phi(\varphi)$ are
\begin{equation}\label{quadrV}
 \mathcal{V}_{\text{quadratic}} \equiv \mathcal{V}_{_{EF}}(\phi(\varphi)) = \frac{1}{(\xi_{\mathcal{A}_0} + \sqrt{\kappa}\,\, \xi_{\mathcal{A}_1}\,\phi(\varphi)  + \kappa\,\, \xi_{\mathcal{A}_2}\, \phi(\varphi)^2)^2} \frac{m^2 \phi(\varphi)^2}{2}\,,
\end{equation}
and
\begin{equation}\label{quarticV}
  \mathcal{V}_{\text{quartic}} \equiv\mathcal{V}_{_{EF}}(\phi(\varphi)) =\frac{1}{ (\xi_{\mathcal{A}_0} + \sqrt{\kappa} \xi_{\mathcal{A}_1} \phi(\varphi) + \kappa \xi_{\mathcal{A}_2} \phi(\varphi)^2)^{2}} \frac{1}{4} \lambda\, \phi(\varphi)^4\,.
\end{equation}

We can now apply the procedure described above to the derivative coupling case. The kinetic term to consider therefore is \eqref{kefC}, that here we report for clarity
\begin{equation}
 \begin{aligned}
&\mathcal{K}_{_{EF}}(\phi) = \frac{1}{(1 + \sqrt{\kappa}\, \xi_{\mathcal{A}_{1}} \phi + \kappa \,\xi_{\mathcal{A}_{2}} \phi^2)}+  \frac{ ( \sqrt{\kappa} \,\xi_{w^{\prime}} \phi + \kappa \, \xi_{y^{\prime}} \phi^2 + \kappa^{3/2} \,\xi_{z^{\prime}} \phi^3 + \kappa^2 \,\xi_{s^{\prime}} \phi^4)}{ (1 + \sqrt{\kappa}\, \xi_{\mathcal{A}_{1}} \phi + \kappa \,\xi_{\mathcal{A}_{2}} \phi^2)^2}
 \end{aligned}
\end{equation}
with
\begin{equation*}
 \begin{aligned}
 \xi_{w^{\prime}} &=- \frac{3}{128} \xi_w + 6 \xi_{\mathcal{A}_{1}} \xi_{\mathcal{A}_{2}}, \qquad &  \xi_{y^{\prime}}&=- \frac{3}{128} \xi_y + 6 \xi_{\mathcal{A}_{2}}^2 +\xi_{\mathcal{B}_2},\\
\xi_{z^{\prime}} &= - \frac{3}{128} \xi_z + \xi_{\mathcal{A}_{1}}\xi_{\mathcal{B}_2}, \qquad & \xi_{s^{\prime}} &=- \frac{3}{128} \xi_s + \xi_{\mathcal{A}_{2}} \xi_{\mathcal{B}_2}\,.
 \end{aligned}
\end{equation*}

The procedure explained above leads to the following plots for the $\phi(\varphi)$ and for the quadratic, \eqref{quadrV}, and quartic, \eqref{quarticV} potentials:
\begin{figure}[t]
\centering
\begin{subfigure}[b]{0.48\textwidth}
   \includegraphics[width=0.9\textwidth]{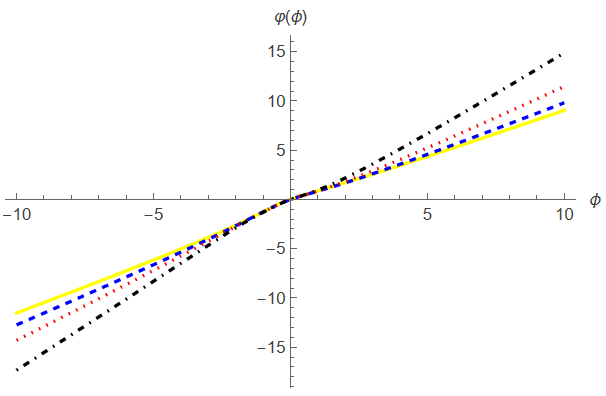}
   \caption{} \label{fig:CAll-varphi}
\end{subfigure}%
\hfill
\begin{subfigure}[b]{0.48\textwidth}
   
   \includegraphics[width=0.9\textwidth]{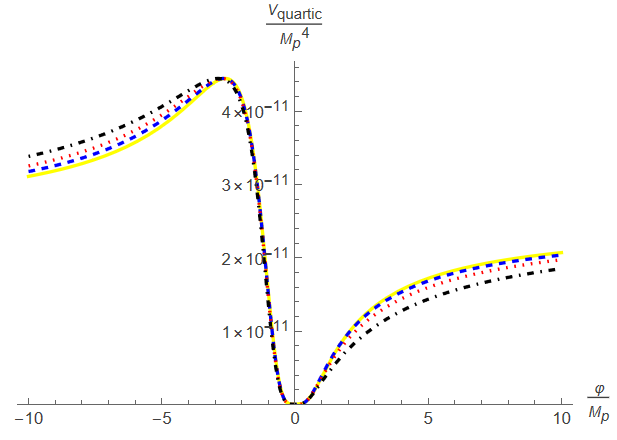}
   \caption{}\label{fig:CAll-quartic}

\end{subfigure}

\vspace{0.3cm}

\begin{subfigure}[b]{0.48\textwidth}
 \includegraphics[width=0.9\textwidth]{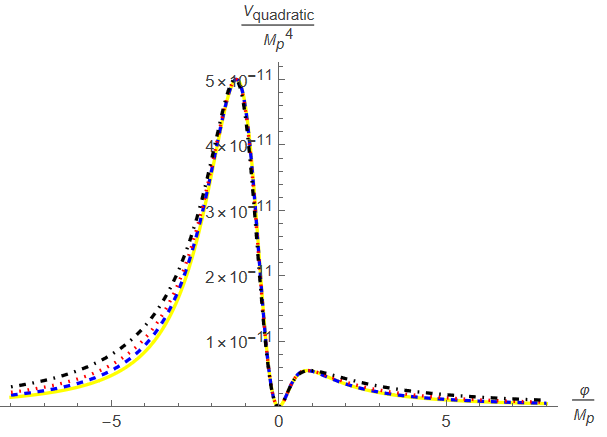}
   \caption{}\label{fig:CAll-quadratic}
\end{subfigure}%
\hfill
\begin{subfigure}[b]{0.48\textwidth}
\includegraphics[width=0.9\textwidth]{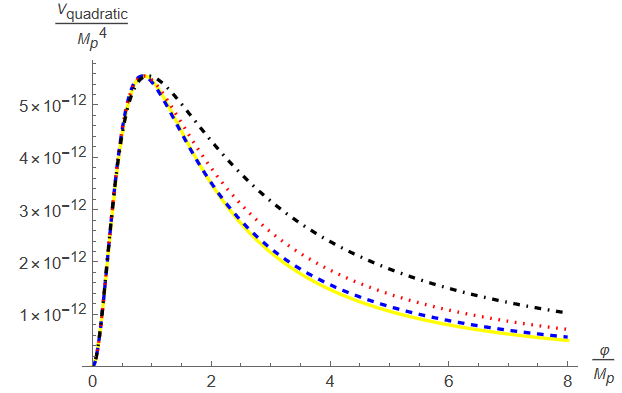}
  \caption{}\label{fig:CAll-quadratic-positive}
\end{subfigure}
\caption{\justifying
Canonical field redefinition and the corresponding Einstein-frame
potentials in the presence of all the derivative couplings
$\mathcal{C}_2$, $\mathcal{C}_3$, and $\mathcal{C}_4$.
Panel (a) shows the canonically normalized field $\varphi(\phi)$. Panel (b) shows the quartic potential \eqref{quarticV} expressed in
terms of the canonical field, while panels (c) and (d) show the
quadratic potential \eqref{quadrV} over the full displayed field
range and on its positive-field branch, respectively.
We fix $\xi_{\mathcal{A}_1}=\xi_{\mathcal{A}_2}=1$, $\xi_{\mathcal{B}_2}=1$
and take the derivative-coupling parameters to be equal, $\xi_{\mathcal{C}_{i_j}}=\xi_{\mathcal C}$ for $i=2,3,4$ and $j=1,2$. The corresponding coefficients of $\mathcal{C}_1$ are fixed by the projective condition \eqref{constr}, giving $\xi_{\mathcal{C}_{1_1}}=\xi_{\mathcal{C}_{1_2}}=-\xi_{\mathcal{C}}/16$. The solid yellow, dashed blue, dotted red, and dot-dashed black curves
correspond to
$\xi_{\mathcal{C}}=0$, $3\times10^{-1}$,
$5\times10^{-1}$, and $8\times10^{-1}$, respectively.
The scalar mass and quartic coupling are fixed to
$m=10^{-5}M_{\mathrm{Pl}}$ and $\lambda=10^{-10}$.}
   \label{fig:CAll}
\end{figure}

Figure~\ref{fig:CAll} illustrates the effect of the simultaneous presence of the derivative couplings $\mathcal{C}_2$, $\mathcal{C}_3$, and $\mathcal{C}_4$ on the canonical normalization and on the potentials written in terms of the canonical field. The derivative couplings do not modify $\mathcal V_{\mathrm{EF}}(\phi)$ directly. Instead, they modify the kinetic function $\mathcal{K}_{_{EF}}(\phi)$ and hence the field redefinition. Their effect on the potential therefore arises through the replacement $\mathcal V_{\mathrm{EF}}(\phi)\rightarrow\mathcal V_{\mathrm{EF}}(\phi(\varphi))$. In figure \ref{fig:CAll-varphi} all curves satisfy $\varphi(0)=0$ and $\left.d\varphi /d\phi\right|_{\phi=0}=1$, so they coincide to leading order near the origin. The effect of the derivative couplings is asymmetric at intermediate field values because of the linear term in $\mathcal{A}(\phi)$. For the parameter choices shown, increasing $\xi_{\mathcal{C}}$ increases the asymptotic slope of $\varphi(\phi)$ producing a stronger horizontal stretching of the canonically normalized potentials. The canonically normalized quartic potential $\mathcal{V}_{\mathrm{EF}}(\phi(\varphi))$ in Fig.~\ref{fig:CAll-quartic} depends on the derivative couplings through the inverse field redefinition $\phi(\varphi)$. Therefore, different coupling choices generally give different potential values at fixed $\varphi$. However, the derivative couplings do not change the large-field plateau value, but only the way in which it is approached in canonical-field space. The canonically normalized quadratic potential $\mathcal{V}_{\mathrm{EF}}(\phi(\varphi))$ in Fig.~\ref{fig:CAll-quadratic} depends on the derivative couplings through the inverse field redefinition $\phi(\varphi)$. Therefore, different coupling choices generally give different potential values at fixed $\varphi$. The derivative couplings shift the positions and modify the widths of the extrema in canonical-field space, while leaving their heights and the asymptotic value $\mathcal{V}_{\mathrm{EF}}(\phi(\varphi))\to 0$ unchanged.

In summary, for the plotted values of the couplings, the effect of the extra Nieh-Yan-like couplings is mainly to stretch or squeeze the canonically normalized scalar-field potential along the horizontal $\varphi$ axis, compared to the case without these couplings. It is well known that the quartic potential develops an asymptotic plateau in the Einstein frame, and this feature remains qualitatively unchanged in the presence of the extra couplings, although the way in which the plateau is approached in canonical-field space is modified. Similarly, the quadratic potential in the Einstein frame retains its local maximum, while its position and the steepness of the potential around it are modified. Since inflationary observables depend on the derivatives of the potential with respect to the canonically normalized field \cite{Jarv:2016sow}, these modifications can nevertheless have observationally significant consequences.

\subsubsection{Mixed torsion--nonmetricity couplings}

\vspace{0.5 cm}\paragraph{$c_4,c_5,c_6$}
When $c_4$, $c_5$, and $c_6$ are present individually or only in combinations among themselves, the Einstein-frame kinetic function remains equal to the \ABV\, result, as discussed in Sec.\ \ref{cicomment}. Consequently, the canonical field redefinition and the corresponding potential plots are unchanged with respect to the \ABV\, reference case.

\vspace{0.5 cm}\paragraph{Only $c_1$}
Considering only the coupling $c_1$ in addition to the \ABV\, action gives the modified Einstein-frame kinetic function \eqref{kefc1}
\begin{equation}
\begin{aligned}
\mathcal{K}_{_{EF}}(\phi)=& \frac{1 + \kappa \xi_{\mathcal{B}_2} \phi^2}{1 + \sqrt{\kappa} \xi_{\mathcal{A}_1} \phi + \kappa \,\xi_{\mathcal{A}_2} \phi^2} + \frac{3 (\xi_{\mathcal{A}_1} + 2 \sqrt{\kappa} \xi_{\mathcal{A}_2} \phi)^2 (\xi_{c_{1_{0}}} + \sqrt{\kappa} \xi_{c_{1_{1}}} \phi + \kappa \xi_{c_{1_{2}}} \phi^2)}{2 (1 + \sqrt{\kappa} \xi_{\mathcal{A}_1} \phi + \kappa \,\xi_{\mathcal{A}_2} \phi^2)^2 \bigl(-8 + \xi_{c_{1_{0}}} + \sqrt{\kappa} (-8 \xi_{\mathcal{A}_1} + \xi_{c_{1_{1}}}) \phi + \kappa (-8 \xi_{\mathcal{A}_2} + \xi_{c_{1_{2}}}) \phi^2\bigr)}\,.
\end{aligned}
\end{equation}

\begin{figure}[]
\centering
\begin{subfigure}[b]{0.48\textwidth}
   \includegraphics[width=0.9\textwidth]{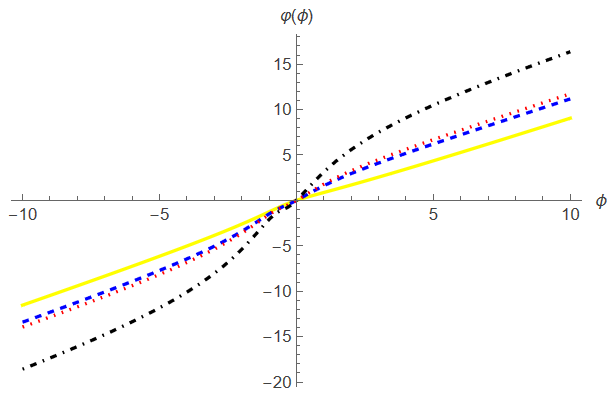}
   \caption{}\label{a33}
\end{subfigure}%
\hfill
\begin{subfigure}[b]{0.48\textwidth}
 \includegraphics[width=0.9\textwidth]{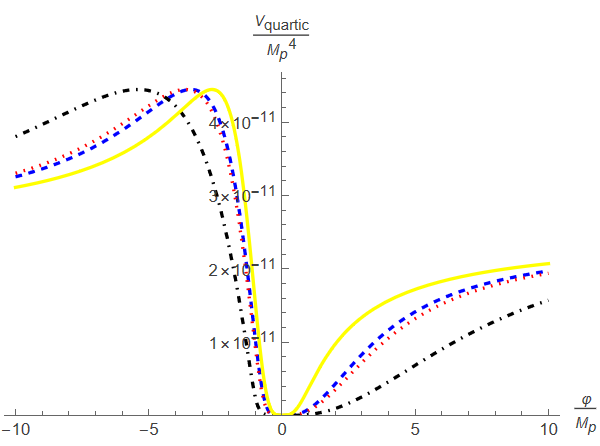}
   \caption{}\label{b33}
\end{subfigure}

\vspace{0.3cm}

\begin{subfigure}[b]{0.48\textwidth}
 \includegraphics[width=0.9\textwidth]{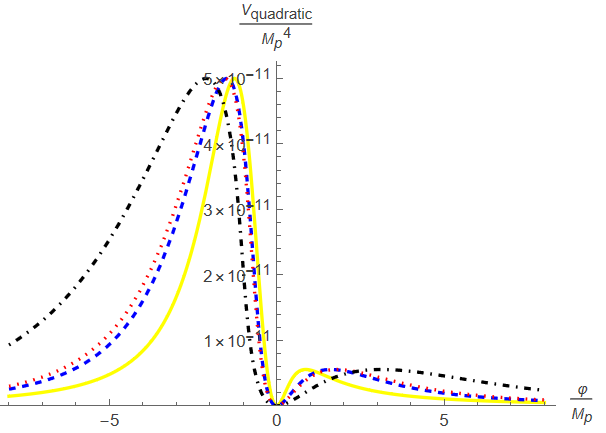}
   \caption{}\label{c33}
\end{subfigure}%
\hfill
\begin{subfigure}[b]{0.48\textwidth}
 \includegraphics[width=0.9\textwidth]{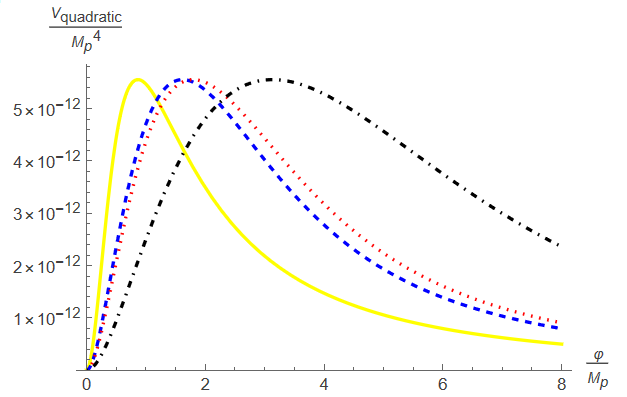}
  \caption{}\label{d33}
\end{subfigure}
\caption{\justifying
Canonical field $\varphi(\phi)$ as a function of the original scalar field $\phi$ in (a). The quartic potential \eqref{quarticV} in (b). The quadratic potential \eqref{quadrV} for both positive and negative values of the field in (c) and zoomed for the positive values only in (d). We fix $\xi_{\mathcal{A}_1}=\xi_{\mathcal{A}_2}=\xi_{\mathcal{B}_2}=1$ and vary the coefficients $\xi_{c_{1_0}}$, $\xi_{c_{1_1}}$ and $\xi_{c_{1_2}}$. The yellow continuous curve corresponds to the reference case $(\xi_{c_{1_0}},\xi_{c_{1_1}},\xi_{c_{1_2}})=(0,0,0)$, while the black, red and blue dashed curves correspond respectively to $(10,10,10)$, $(3\times10^{1}, 3\times10^{1},3\times10^{1})$ and $(3\times10^{2},3\times10^{2}, 3\times10^{2})$.}
   \label{fig:c1L}
\end{figure}

In Fig.~\ref{fig:c1L} we show the effect of the $c_1$-sector couplings on the canonical field redefinition and on the Einstein-frame potentials. For the choice $\xi_{c_{1_0}}=\xi_{c_{1_1}}=\xi_{c_{1_2}}\equiv\xi_{c_1}$, the additional contribution to the Einstein-frame kinetic function is controlled by the factor $\xi_{c_1}/(\xi_{c_1}-8)$. The value $\xi_{c_1}=8$ is singular and must be excluded. For $\xi_{c_1}>8$, this factor decreases monotonically toward $1$ as $\xi_{c_1}\to\infty$. Therefore, among the values considered, $\xi_{c_1}=10$ produces the strongest modification of the kinetic function and of the field redefinition, since it is closest to the singular value, whereas the curve for $\xi_{c_1}=300$ is already close to the large-$\xi_{c_1}$ limiting behavior. This saturation of the field redefinition is reflected in both the quartic and quadratic potentials expressed in terms of the canonically normalized field. The $c_1$ coupling modifies the potentials indirectly through the field redefinition. As shown in panel (\subref{a33}), the relation between the canonical field $\varphi$ and the original field $\phi$ becomes nonlinear and asymmetric. For the quartic potential in panel (\subref{b33}), the positions of the maxima and the approach to the large-field asymptotic regions are shifted in terms of the canonical field. For the quadratic potential, shown over the full field range in panel (\subref{c33}) and on the positive-field branch in panel (\subref{d33}), the extrema are displaced in canonical-field space while their heights remain unchanged. The positive-field maximum moves towards larger values of $\varphi$, whereas the negative-field maximum moves towards more negative values of $\varphi$. Therefore, the main effect of the $c_1$ sector is to stretch and asymmetrically reshape the potentials when they are expressed in terms of the canonically normalized field.

\subsection{Reconstructing a kinetic function from a target potential}

Having illustrated how non-Riemannian couplings can reshape a given potential through the canonical field redefinition, we now consider the inverse problem. As an example, we ask whether the Einstein-frame potential $\mathcal{V}_{_{EF}}(\phi)$ in our action \eqref{actionEF}, obtained from the quadratic Jordan-frame potential $\mathcal{V}(\phi)=m^2\phi^2/2$, can be mapped by canonical normalization to a natural-inflation potential $\hat{\mathcal{V}}(\varphi)$ \cite{Freese:1990rb}. The quadratic potential for a generic scalar field $\phi$ becomes in the Einstein frame
\begin{equation}\label{quadraticpot1}
 \mathcal{V}_{_{EF}}(\phi) = \frac{1}{\mathcal{A}^2} \frac{m^2\phi^2}{2}  = \frac{1}{(\xi_{\mathcal{A}_0} + \sqrt{\kappa}\,\, \xi_{\mathcal{A}_1}\, \phi  + \kappa\,\, \xi_{\mathcal{A}_2}\, \phi^2)^2} \frac{m^2\phi^2}{2}\,.
\end{equation}
The natural-inflation potential in terms of a canonically normalized scalar field $\varphi$ is
\begin{equation}\label{naturalpotential01}
 \hat{\mathcal{V}}(\varphi) =  V_0 \Big(1- \cos{\Big(\frac{\varphi}{\mu}\Big)}\Big)\,,
\end{equation}
where $V_0$ and $\mu$ are some constants. (A pseudo Nambu-Goldstone boson, with a potential of the form \eqref{naturalpotential01}, can give rise to an epoch of inflation in the early Universe \cite{Freese:1990rb}.) We proceed by setting the potentials \eqref{naturalpotential01} and \eqref{quadraticpot1} equal, $\mathcal{V}_{_{EF}}(\phi)=\hat{\mathcal{V}}(\varphi)$, that is:
\begin{equation}\label{re}
\frac{1}{(\xi_{\mathcal{A}_0} + \sqrt{\kappa}\,\, \xi_{\mathcal{A}_1}\, \phi  + \kappa\,\, \xi_{\mathcal{A}_2}\, \phi^2)^2} \frac{m^2\phi^2}{2}  = V_0 \Big(1- \cos{\Big(\frac{\varphi}{\mu}\Big)}\Big)\,.
\end{equation}
We solve this equality for the canonically normalized scalar field $\varphi(\phi)$:
\begin{equation}\label{varphi}
  \varphi [\phi]= \pm \mu \,\,\text{Arccos} \Big(1 - \frac{\mathcal{V}_{_{EF}}(\phi)}{V_0}\Big) = \pm \mu \,\,\text{Arccos} \bigl(1 - \frac{m^2 \phi^2}{2 V_0 (\xi_{\mathcal{A}_0} + \sqrt{\kappa} \xi_{\mathcal{A}_1} \,\phi + \kappa \xi_{\mathcal{A}_2} \,\phi^2)^2}\bigr)\,,
\end{equation}
where this expression is understood locally on a chosen branch of the inverse cosine. The different branches give the same target kinetic function, since it depends on $(d\varphi/d\phi)^2$.

Choosing the positive branch, the vanishing of the Einstein-frame potential at $\phi=0$ corresponds to $\varphi(0)=2\pi n\mu$, which is the minimum of the potential $\hat{\mathcal{V}}(\varphi)$. We can choose $\varphi(0)=0$. We then differentiate \eqref{varphi} to obtain the required Einstein-frame kinetic function.

Differentiating \eqref{varphi}, the target kinetic function is fixed by
\begin{equation}\label{eqek}
  - \frac{1}{2}\mathcal{K}_{_{EF}}(\phi)^{\text{target}} \,\partial_\mu \phi \,\partial^\mu \phi \equiv  - \frac{1}{2}\,\partial_\mu \varphi \,\partial^\mu \varphi  = - \frac{1}{2} \Big(\frac{d \varphi (\phi)}{d \phi} \frac{d \varphi (\phi)}{d \phi}\Big) \partial_\mu \phi \,\partial^\mu \phi\,.
\end{equation}
Then we solve \eqref{eqek} for $\mathcal{K}_{_{EF}}(\phi)^{\text{target}}$:
\begin{equation}\label{re2}
\begin{aligned}
  &\mathcal{K}_{_{EF}}(\phi)^{\text{target}} =  \frac{4 m^2 \mu^2 (\xi_{\mathcal{A}_0} -  \kappa\xi_{\mathcal{A}_2} \phi^2)^2}{(\xi_{\mathcal{A}_0} + \sqrt{\kappa}\xi_{\mathcal{A}_1} \phi + \kappa\xi_{\mathcal{A}_2} \phi^2)^2 \bigl(- m^2 \phi^2 + 4 V_0 (\xi_{\mathcal{A}_0} + \sqrt{\kappa}\xi_{\mathcal{A}_1} \phi + \kappa\xi_{\mathcal{A}_2} \phi^2)^2\bigr)}
   \end{aligned}
\end{equation}
that can be simplified considering $\xi_{\mathcal{A}_{0}} = 1, \,\xi_{\mathcal{A}_{1}} = 0, \,\xi_{\mathcal{B}_2} = 0$ as \begin{equation}\label{uno}
  \mathcal{K}_{_{EF}}(\phi)^{\text{target}} = \frac{4 m^2 \mu^2 \bigl(-1 + \kappa \xi_{\mathcal{A}_{2}} \phi^2\bigr)^2}{\bigl(1 + \kappa \xi_{\mathcal{A}_{2}} \phi^2\bigr)^2 \Bigl(- m^2 \phi^2 + 4 V_0 \bigl(1 + \kappa \xi_{\mathcal{A}_{2}} \phi^2\bigr)^2\Bigr)}\,.
\end{equation}
Note that the Einstein-frame kinetic term can also be written, thanks to the \eqref{naturalpotential01} and \eqref{eqek}, directly in terms of the Einstein-frame potential as:
\begin{equation}
   \mathcal{K}_{_{EF}}(\phi)^{\text{target}} = \mu^2 \frac{\big(\mathcal{V}_{_{EF}}(\phi)^{\prime}\big)^2}{\mathcal{V}_{_{EF}}(\phi) (2 V_0 - \mathcal{V}_{_{EF}}(\phi))}\,.
\end{equation}
Either way, the resulting expression in this context is \eqref{uno}. At this point, we want to compare \eqref{uno} with the kinetic term for the derivative couplings $\mathcal{C}_i$ \eqref{kefC} in the same assumptions, and setting $\xi_{\mathcal{B}_2} = 0$ for simplicity, that is
\begin{equation}\label{due}
\begin{aligned}
   \mathcal{K}_{_{EF}}(\phi)^{\text{\ABVC \,}} = & \Big(128 + \kappa \bigl(-64 \xi_{\mathcal{A}_{2}} (-2 + 3 \xi_{\mathcal{C}_{3_{1}}}) + 3 (48 \xi_{\mathcal{C}_{2_{1}}}^2 + 24 \xi_{\mathcal{C}_{2_{1}}} \xi_{\mathcal{C}_{3_{1}}} -  \xi_{\mathcal{C}_{3_{1}}}^2 + 64 \xi_{\mathcal{C}_{4_{1}}}^2)\bigr) \phi^2 \\
   &+ 6 \kappa^{3/2} \bigl(12 \xi_{\mathcal{C}_{2_{2}}} \xi_{\mathcal{C}_{3_{1}}} - 32 \xi_{\mathcal{A}_{2}} \xi_{\mathcal{C}_{3_{2}}} -  \xi_{\mathcal{C}_{3_{1}}} \xi_{\mathcal{C}_{3_{2}}} + 12 \xi_{\mathcal{C}_{2_{1}}} (4 \xi_{\mathcal{C}_{2_{2}}} + \xi_{\mathcal{C}_{3_{2}}}) + 64 \xi_{\mathcal{C}_{4_{1}}} \xi_{\mathcal{C}_{4_{2}}}\bigr) \phi^3\\
   &+ 3 \kappa^2 (48 \xi_{\mathcal{C}_{2_{2}}}^2 + 24 \xi_{\mathcal{C}_{2_{2}}} \xi_{\mathcal{C}_{3_{2}}} -  \xi_{\mathcal{C}_{3_{2}}}^2 + 64 \xi_{\mathcal{C}_{4_{2}}}^2) \phi^4\Big) / \Big(128 (1 + \kappa \xi_{\mathcal{A}_{2}} \phi^2)^2\Big)\,.
    \end{aligned}
\end{equation}
To obtain an explicit matching, we impose the simplifying conditions $\xi_{\mathcal{C}_{2_{1}}}=\xi_{\mathcal{C}_{3_{1}}}=\xi_{\mathcal{C}_{4_{1}}}=\xi$, $\xi_{\mathcal{C}_{2_{2}}}=\xi_{\mathcal{C}_{3_{2}}}=\xi_{\mathcal{C}_{4_{2}}}=0$. And the projective consistency \eqref{constr} implies $\xi_{\mathcal{C}_{1_{1}}}= - (\xi / 16)$ and $\xi_{\mathcal{C}_{1_{2}}} = 0$. With these assumptions \eqref{due} becomes
\begin{equation}\label{due2}
\begin{aligned}
   \mathcal{K}_{_{EF}}(\phi)^{\text{\ABVC \,}} = &
    \frac{128 + \kappa \bigl(405 \xi^2 - 64 (-2 + 3 \xi) \xi_{\mathcal{A}_{2}}\bigr) \phi^2}{128 (1 + \kappa \xi_{\mathcal{A}_{2}} \phi^2)^2}\,.
    \end{aligned}
\end{equation}
Matching \eqref{uno} and \eqref{due2} gives, for $\mu>0, V_0>0, \kappa \mu^2 \leq(1/90)$,
\begin{equation}
    \begin{aligned}\label{eqw}
         m &= \frac{\sqrt{V_0}}{\mu}, \qquad &\xi_{\mathcal{A}_2} &= \frac{1}{16 \kappa \mu^2}, \qquad & \xi &= 2 \frac{1 \pm \sqrt{1-90 \kappa \mu^2}}{135 \kappa \mu^2}\,.
    \end{aligned}
\end{equation}
Substituting the conditions \eqref{eqw} into either \eqref{uno} or \eqref{due2}, the Einstein-frame kinetic function becomes
\begin{equation}\label{solj}
\mathcal{K}_{_{EF}}^{\mathrm{\ABVC \,}}(\phi)=\frac{256\mu^4}{(16\mu^2+\phi^2)^2}.
\end{equation}
Although the final kinetic function no longer depends explicitly on the derivative-coupling parameter $\xi$, the derivative couplings are essential for the reconstruction. Their values must satisfy the tuning condition above. Setting them to zero from the outset does not reproduce the target kinetic function.

As a consistency check, we now verify the reconstruction explicitly. Choosing the positive orientation of the canonical field in \eqref{solj}, the field redefinition therefore satisfies
\begin{equation}
\frac{d\varphi}{d\phi}=\frac{16\mu^2}{16\mu^2+\phi^2}.
\end{equation}
With the choice $\varphi(0)=0$, this integrates to
\begin{equation}\label{sd}
\varphi(\phi)=4\mu\,\arctan\left(\frac{\phi}{4\mu}\right).
\end{equation}
Inverting \eqref{sd}, using $m=\sqrt{V_0}/\mu$, $\xi_{\mathcal A_2}=1/(16\kappa\mu^2)$ and our choices $\xi_{\mathcal{A}_{0}} = 1, \,\xi_{\mathcal{A}_{1}} = 0$, the Einstein-frame quadratic potential becomes
\begin{equation}
\mathcal{V}_{_{EF}}(\phi)=\frac{V_0\phi^2}{2\mu^2\left(1+\phi^2/(16\mu^2)\right)^2}.
\end{equation}
Substituting the canonical field redefinition above, one verifies that
\begin{equation}
\mathcal{V}_{_{EF}}(\phi)=V_0\left(1-\cos\left(\frac{\varphi(\phi)}{\mu}\right)\right),
\end{equation}
which reproduces the target natural-inflation potential.

\section{Cosmological specialization}\label{CPgeneral}

The cosmological principle consists of the assumptions of homogeneity and isotropy of the spatial hypersurfaces. This gives the Friedmann-Lema\^itre-Robertson-Walker metric,
\begin{equation}
    \mathrm{d}s^2 = - \mathrm{d}t^2 + a(t)^2 \mathrm{d}\vec{x}^2 \,,
    \label{FLRW metric}
\end{equation}
for vanishing spatial curvature, where $a(t)$ is the scale factor. In the previous sections, like in the bulk of the literature, including Ref.\ \cite{Rigouzzo:2022yan}, we kept the connection arbitrary. However, imposing these symmetries on the cosmological background completely, i.e.\ on both the metric and the independent connection, restricts the allowed tensorial forms of torsion and nonmetricity \cite{Iosifidis:2021fnq,Iosifidis:2024bsq}. As a consequence, the quadratic invariants of the full covariant theory enter the cosmological equations only through a reduced number of independent combinations. We use the relations derived in \cite{Iosifidis:2024bsq} and translate them into the notation of the action \eqref{action}:
\begin{equation}
\begin{array}{l@{\hspace{2cm}}l@{\hspace{2cm}}l}
\alpha_1=-a_1-a_2, & \beta_1=b_1+b_2,  & \gamma_1=-c_1-3 c_2, \\
\alpha_2=-\frac{1}{3} a_1-a_3,& \beta_2=-b_1-b_3, & \gamma_2=c_2+c_3, \\
\alpha_3=-\frac{1}{3} a_1-a_4, & \beta_3=-\frac{3}{4} b_4-b_5,& \gamma_3=c_4+c_5, \\
\alpha_4=\frac{2}{3} a_1-a_5, & & \gamma_4=-3 c_4-c_6,\\
a_6 =0\,.
\end{array}
\end{equation}
Equivalently, at the level of the homogeneous and isotropic cosmological background, we may choose a convenient representative parametrization of these combinations. These relations should not be understood as constraints on the coefficient functions of the full covariant action. Rather, once torsion and nonmetricity are restricted to their homogeneous and isotropic forms, different quadratic invariants become degenerate on the cosmological background, so that the background equations depend on the original coefficients only through the combinations $\alpha_i$, $\beta_i$, and $\gamma_i$. The map from the original coefficients to these combinations is therefore not one-to-one. For fixed $\alpha_i$, one function among $a_1,\ldots,a_5$ remains arbitrary; for fixed $\beta_i$, two functions among $b_1,\ldots,b_5$ remain arbitrary; and for fixed $\gamma_i$, two functions among $c_1,\ldots,c_6$ remain arbitrary. We may therefore choose a convenient representative parametrization of each set of coefficients. In the following, we use
\begin{equation}\label{condit}
\begin{array}{l@{\hspace{2cm}}l@{\hspace{2cm}}l}
a_1=0, & b_1=0, & c_1 = - \gamma_1, \\
a_2 = - \alpha_1, & b_2 = \beta_1, & c_2 = 0, \\
a_3 = - \alpha_2, & b_3 = -  \beta_2, & c_3 = \gamma_2, \\
a_4 = - \alpha_3, & b_4 =0, & c_4=0,\\
a_5 = -\alpha_4, & b_5 = - \beta_3,  &c_5 = \gamma_3,\\
a_6 =0,&    &c_6 = - \gamma_4\,.
\end{array}
\end{equation}
This choice is only a convenient parametrization of the combinations that remain independent after imposing the cosmological principle and does not constitute an additional restriction on the coefficients of the full covariant theory. The invariant proportional to $a_6$ does not contribute to the homogeneous and isotropic background and can therefore be omitted in the representative parametrization used below \cite{Iosifidis:2024bsq}. We see how, once we impose the cosmological principle, not all the quadratic tensors in torsion and nonmetricity are independent. Different sets of $a_i$, $b_i$, and $c_i$ giving the same $\alpha_i$, $\beta_i$, and $\gamma_i$ are indistinguishable at the homogeneous and isotropic background level.

The cosmological principle also restricts the tensorial form of the torsion and nonmetricity solutions \eqref{nonmetphi}, \eqref{torphi}. Since the scalar field is homogeneous, its derivative is aligned with the cosmological four-velocity, $\partial_\mu\phi=-\dot{\phi} \,u_\mu$, with our metric-signature convention. The torsion and nonmetricity therefore take the forms
\beq
S_{\mu \nu \alpha } = D_1 (- \dot{\phi} g_{\nu \alpha } u_{\mu } + \dot{\phi} g_{\mu \alpha } u_{\nu }) -   D_2 \,\dot{\phi} \varepsilon_{\mu \nu \alpha \rho } u^{\rho }
\label{torphiCP}
\eeq
and 
\beq 
\label{nonmetphiCP}
Q_{\alpha \mu \nu } = - D_3 \,\dot{\phi}  g_{\mu \nu } u_{\alpha } + \frac{1}{2} D_4 (- \dot{\phi} g_{\alpha \nu } u_{\mu } -  \dot{\phi} g_{\alpha \mu } u_{\nu })\,.
\eeq
The functions $D_i$ are obtained from the general functions $A_i$ in Appendix \ref{Aizi} after imposing the cosmological principle relations \eqref{condit} ($[D_i] = M^{-1})$).

In the following sections, when suitable, we will compare our hypermomentum tensor as defined in \eqref{hyper} with its general shape when the cosmological principle is assumed \cite{Iosifidis:2020gth}:
\begin{equation}\label{hyp}
    \Delta_{\alpha \mu \nu } = \psi\, h_{\mu \nu } u_{\alpha } + \chi\, h_{\nu \alpha } u_{\mu } + \Phi\,h_{\mu \alpha } u_{\nu } + \omega \,u_{\alpha } u_{\mu } u_{\nu } + \zeta \,\varepsilon_{\alpha \mu \nu \rho} u^{\rho}\,,
\end{equation}
where $h_{\mu\nu}\equiv g_{\mu\nu}+u_\mu u_\nu$ is the spatial projector.

The modified Friedmann equations for the metric \eqref{FLRW metric} and in terms of some functions $p_h$ and $\rho_h$ of $\psi,\chi,\Phi,\omega,\zeta$ were found in \cite{Andrei:2024vvy}
for the minimal coupling ($\mathcal{A}=1$) to be
\begin{subequations}
\label{eq: FLRW equations general}
\begin{align}
\label{eq: FR1}
    3 H^2 &= \kappa \rho + \kappa \rho_h \,, \\
\label{eq: FR2}
    2 \dot{H} + 3 H^2 &= - \kappa p - \kappa p_h \,, \\
\label{eq: continuity eq}
    \dot{\rho} + 3 H (\rho + p ) &= -\dot{\rho}_h - 3 H (\rho_h + p_h ) \,.
\end{align}
\end{subequations}
Here $H(t)=\frac{\dot{a}}{a}$ denotes the Hubble parameter. Comparing \eqref{hyper} with the scalar field hypermomentum \eqref{hyp} allows us to identify the functions $\psi,\chi,\Phi,\omega,\zeta$. For the \ABV\, and \ABVC \, sectors, the resulting setup can be directly matched to the spatially flat Einstein-Hilbert MAG cosmology studied in Ref.~\cite{Andrei:2024vvy}. We can therefore use the results of that work to determine the corresponding effective hypermomentum density $\rho_h$ and pressure $p_h$.
For the more general quadratic sectors, this direct identification is no longer available because the gravitational sector contains additional geometric terms beyond the Einstein-Hilbert contribution. In those cases, we restrict the discussion to the reduction of the independent coupling combinations and to the corresponding effective kinetic functions.

\subsection{\ABV\, cosmology}
Imposing the cosmological principle on the \ABV\, hypermomentum \eqref{hyper1}, and comparing with \eqref{hyp}, gives
\begin{equation}\label{hypfun}
    \psi = - \dot{\phi} \frac{\mathcal{A}^{\prime}}{\kappa\,\mathcal{A}} \,, \qquad 
    \chi = + \dot{\phi} \frac{\mathcal{A}^{\prime}}{\kappa\,\mathcal{A}} \,, \qquad
    \Phi = 0  \,, \qquad 
    \omega= 0 \,, \quad  
    \zeta =0 \,.
\end{equation}
We can rewrite the \eqref{hypfun} as the spin $\sigma$ and shear $\Sigma_1$, $\Sigma_2$ hypermomentum variables, as defined in \cite{Andrei:2024vvy}:
\begin{equation}\label{spinandshear}
    \sigma \equiv \frac{1}{2} (\psi - \chi) = - \dot{\phi} \frac{\mathcal{A}^{\prime}}{\kappa\,\mathcal{A}}\,, 
    \quad \Sigma_1\equiv \frac{1}{2} (\psi + \chi) = 0\,,
    \quad \Sigma_2 \equiv\frac{1}{4} (\Phi + \omega) = 0 \,, 
    \quad \zeta = 0 \,,\quad
    \Delta\equiv 3 \Phi - \omega =0\,.
\end{equation}
The fact that only the spin component survives follows directly from the tensorial structure of the \ABV\, hypermomentum. Lowering its indices, Eq.~\eqref{hyper1} is antisymmetric in its first two indices. Consequently, the symmetric hypermomentum components associated with shear and dilation vanish identically. For a homogeneous scalar field, $\partial_\mu\phi=-\dot{\phi}u_\mu$, this structure gives $\psi=-\chi$ and $\Phi=\omega=0$, while the absence of a Levi-Civita-tensor contribution implies $\zeta=0$. Therefore only the spin component $\sigma=(\psi-\chi)/2$ remains nonzero.

Then with \eqref{spinandshear} directly from \cite{Andrei:2024vvy} we have the effective hypermomentum density and pressure
\begin{equation}
\begin{aligned}
    & \rho_h = - \frac{1}{\kappa \mathcal{A}}\Big( 3 H \mathcal{A}^{\prime} \dot{\phi} +\frac{3 \big(\mathcal{A}^{\prime}\big)^2 \dot{\phi}^2}{4 \mathcal{A}} \Big),\\
    & p_h = \frac{1}{\kappa \mathcal{A}}\Big(2 H \mathcal{A}^{\prime} \dot{\phi} -  \frac{3 \big(\mathcal{A}^{\prime}\big)^2 \dot{\phi}^2}{4 \mathcal{A}} + \dot{\phi}^2 \mathcal{A}^{\prime\prime} + \mathcal{A}^{\prime} \ddot{\phi}\Big) \,.
\end{aligned}    
\end{equation}
In order to cast the cosmological equations in the form \eqref{eq: FLRW equations general}, we define the effective rescaled scalar-field energy density and pressure after dividing the metric equations by the nonminimal coupling $\mathcal{A}$ as
\begin{equation}\label{e11}
\rho \equiv\frac{1}{\mathcal A}\left(\frac{1}{2}\mathcal B\,\dot{\phi}^2+\mathcal V\right),
\qquad
p\equiv\frac{1}{\mathcal A}\left(\frac{1}{2}\mathcal B\,\dot{\phi}^2-\mathcal V\right)\,.
\end{equation}
Finally, in the \ABV\, model with cosmological principle, the metric equations \eqref{eq: FLRW equations general} are 
\begin{equation}
\begin{aligned}
   3 H^2 &= \kappa \rho -  \frac{1}{\mathcal{A}} \Big(3 H \mathcal{A}^{\prime} \dot{\phi} + \frac{3 \big(\mathcal{A}^{\prime}\big)^2 \dot{\phi}^2}{4 \mathcal{A}}\Big)\,,\\
    3 H^2 + 2 \dot{H} &= - \kappa p -  \frac{1}{\mathcal{A}} \Big(2 H \mathcal{A}^{\prime} \dot{\phi} -  \frac{3 \big(\mathcal{A}^{\prime}\big)^2 \dot{\phi}^2}{4 \mathcal{A}} + \dot{\phi}^2 \mathcal{A}^{\prime\prime} + \mathcal{A}^{\prime} \ddot{\phi}\Big)\,,\\
   \dot{\rho}+3 H \bigl(p + \rho\bigr) &= \frac{3 \mathcal{A}^{\prime} \dot{\phi}}{2 \kappa \mathcal{A}^3}\biggl(2 \mathcal{A}^2 \Bigl(H^2 + \dot{H}\Bigr) -  \big(\mathcal{A}^{\prime}\big)^2 \dot{\phi}^2 + \mathcal{A} \Bigl(H \mathcal{A}^{\prime} \dot{\phi} + \dot{\phi}^2 \mathcal{A}^{\prime\prime} + \mathcal{A}^{\prime} \ddot{\phi}\Bigr)\biggr)\,.
\end{aligned}
\end{equation}
Ref.\ \cite{Andrei:2024vvy} considers the phenomenological ansatz
\begin{align}
\label{eq: sigma(rho) assume}
    \sigma &= \lambda_\sigma \sqrt{\frac{3 \rho}{\kappa}}\,,
\end{align}
where $\rho$ denotes the energy density of matter. Another cosmological application of the same type of spin-density relation to a dark matter component was considered in Ref.\ \cite{Andrei_2026}. Taking $\rho$ to be the scalar-field energy density \eqref{e11}, the spin density \eqref{spinandshear} satisfies the phenomenological ansatz \eqref{eq: sigma(rho) assume}, in the potential-free case $\mathcal{V}=0$, provided $\mathcal{A}$ and $\mathcal{B}$ obey
\begin{equation}
    - \frac{\mathcal{A}^{\prime}}{\kappa\mathcal{A}} = \lambda_\sigma \sqrt{\frac{3 \,\mathcal{B}}{2 \,\kappa\,\mathcal{A}}}
\end{equation}
that solved gives
\begin{equation}
  \mathcal{A}(\phi) = \Big( c_0\,- \lambda_\sigma \sqrt{\frac{3 \kappa}{8}} \int_1^{\phi} \sqrt{\mathcal{B}} \, d\phi \Big)^2\,.
\end{equation}
This relation should be understood as a special compatibility condition, not as a restriction of the general \ABV\, theory.

\subsection{\ABVC \, cosmology}
Imposing the cosmological principle on the \ABVC \, hypermomentum \eqref{hyper} gives
\begin{equation}
\begin{aligned}\label{hypfun2}
   & \psi = \frac{\Bigl( \kappa \,\mathcal{C}_2 - \mathcal{A}^{\prime}\Bigr)}{\kappa\,\mathcal{A}}\dot{\phi}, \hspace{0.7 cm} \chi= \frac{\Bigl(2 \kappa\, \mathcal{C}_2 + \kappa\, \mathcal{C}_3 + 2 \mathcal{A}^{\prime}\Bigr)}{2 \kappa\,\mathcal{A}}\dot{\phi}\,,\quad \Phi=\frac{\Bigl(4 \mathcal{C}_1 - \mathcal{C}_3\Bigr)}{2 \mathcal{A}} \dot{\phi}, \\
   &\omega=- \frac{2 \Bigl( \mathcal{C}_1 + \mathcal{C}_2\Bigr)}{\mathcal{A}} \dot{\phi}\,,\quad \zeta = - \frac{\,\mathcal{C}_4}{\mathcal{A}}\dot{\phi}\,.
\end{aligned}
\end{equation}
These can be rewritten in terms of the spin $\sigma$ and shear $\Sigma_1$, $\Sigma_2$ hypermomentum variables defined in \cite{Andrei:2024vvy}:
\begin{equation}\label{spinandshear2}
\begin{aligned}
   & \sigma \equiv \frac{1}{2} (\psi - \chi) = - \frac{\Bigl(\kappa \mathcal{C}_3 + 4 \mathcal{A}^{\prime}\Bigr) \dot{\phi}}{4 \kappa \mathcal{A}}\,, 
    \quad \Sigma_1\equiv \frac{1}{2} (\psi + \chi) = \frac{\Bigl(4 \mathcal{C}_2 + \mathcal{C}_3\Bigr) \dot{\phi}}{4 \mathcal{A}}\,,
    \quad\\
    &\Sigma_2 \equiv\frac{1}{4} (\Phi + \omega) = - \frac{\Bigl(4 \mathcal{C}_2 + \mathcal{C}_3\Bigr) \dot{\phi}}{8 \mathcal{A}}\,,
    \quad \zeta = - \frac{\mathcal{C}_4 \dot{\phi}}{\mathcal{A}} \, ,\quad
    \Delta\equiv 3 \Phi - \omega =0\,.
    \end{aligned}
\end{equation}
In contrast to the \ABV\, case, the derivative couplings generate additional tensorial structures in the scalar-field hypermomentum, so that it is no longer purely antisymmetric in its first two indices. Consequently, both spin and shear can be nonzero. In particular, the combination $4\mathcal{C}_2+\mathcal{C}_3$ generates the shear components, with $\Sigma_2=-\Sigma_1/2$, while the spin component receives contributions from both $\mathcal{A}^{\prime}$ and $\mathcal{C}_3$. The dilation component still vanishes as a consequence of the projective consistency condition \eqref{constr}, whereas the parity-odd coupling $\mathcal{C}_4$ generates the independent axial component $\zeta$.

Ref.\ \cite{Andrei:2024vvy} considers a spatially flat FLRW background and subsequently restricts to the branch $\zeta=0$, since the alternative solution $\zeta\propto a^{-1}$ contributes to the Friedmann equation as an effective spatial-curvature term. In the present \ABVC \, model, $\zeta=-\mathcal{C}_4\dot{\phi}/\mathcal{A}$, so for a dynamical scalar field the choice $\zeta=0$ implies $\mathcal{C}_4=0$.

Substituting \eqref{spinandshear2} into the results of Ref.~\cite{Andrei:2024vvy} gives the effective hypermomentum density and pressure entering \eqref{eq: FLRW equations general}:
\begin{equation}
\begin{aligned}\label{e13}
\rho_h = \frac{3}{256\kappa\mathcal{A}^2}\Big[&
-16\mathcal{A}\Big(16\mathcal{A}'-4\kappa\mathcal{C}_2+3\kappa\mathcal{C}_3\Big)H\dot{\phi}
-16\kappa\mathcal{A}\Big(4\mathcal{C}_2+\mathcal{C}_3\Big)\ddot{\phi}\\
&+\Big(
\kappa^2\big(16\mathcal{C}_2^2+24\mathcal{C}_2\mathcal{C}_3+\mathcal{C}_3^2\big)
+128\kappa\mathcal{C}_2\mathcal{A}'
-64\big(\mathcal{A}'\big)^2
-16\kappa\mathcal{A}\big(4\mathcal{C}_2'+\mathcal{C}_3'\big)
\Big)\dot{\phi}^2
\Big],
\\[0.3cm]
p_h = \frac{1}{256\kappa\mathcal{A}^2}\Big[&
16\mathcal{A}\Big(32\mathcal{A}'+4\kappa\mathcal{C}_2+9\kappa\mathcal{C}_3\Big)H\dot{\phi}
+16\mathcal{A}\Big(16\mathcal{A}'-4\kappa\mathcal{C}_2+3\kappa\mathcal{C}_3\Big)\ddot{\phi}\\
&+\Big(
\kappa^2\big(-80\mathcal{C}_2^2-24\mathcal{C}_2\mathcal{C}_3+3\mathcal{C}_3^2\big)
+128\kappa\mathcal{C}_2\mathcal{A}'
-192\big(\mathcal{A}'\big)^2\\
&\hspace{1.1cm}
+16\mathcal{A}\big(16\mathcal{A}''-4\kappa\mathcal{C}_2'+3\kappa\mathcal{C}_3'\big)
\Big)\dot{\phi}^2
\Big]\,.
\end{aligned}
\end{equation}
For comparison with \cite{Andrei:2024vvy}, we have to consider the density and pressure defined as
  \begin{equation}\label{e12}
\rho \equiv\frac{1}{\mathcal A}\left(\frac{1}{2}\mathcal B\,\dot{\phi}^2+\mathcal V\right) + \rho_{\text{\ABVC \,}} ,
\qquad
p\equiv\frac{1}{\mathcal A}\left(\frac{1}{2}\mathcal B\,\dot{\phi}^2-\mathcal V\right) + p_{\text{\ABVC \,}}\,,
\end{equation}
with
\begin{equation}
\begin{aligned}
\rho_{\text{\ABVC \,}} =& \frac{1}{128\mathcal{A}^2}\Big[
8\mathcal{A}(4\mathcal{C}_2-3\mathcal{C}_3)\ddot{\phi}
-8\mathcal{A}(4\mathcal{C}_2+9\mathcal{C}_3)H\dot{\phi}\\
&+\Big(
112\kappa\mathcal{C}_2^2
+48\kappa\mathcal{C}_2\mathcal{C}_3
-3\kappa\mathcal{C}_3^2
-16\mathcal{A}'(4\mathcal{C}_2+3\mathcal{C}_3)
+8\mathcal{A}(4\mathcal{C}_2'-3\mathcal{C}_3')
\Big)\dot{\phi}^2
\Big]\,,\\
p_{\text{\ABVC \,}} =& \frac{1}{128\mathcal{A}^2}\Big[
-8\mathcal{A}(4\mathcal{C}_2+9\mathcal{C}_3)H\dot{\phi}
+8\mathcal{A}(4\mathcal{C}_2-3\mathcal{C}_3)\ddot{\phi}\\
&+\Big(
\kappa\left(112\mathcal{C}_2^2+48\mathcal{C}_2\mathcal{C}_3-3\mathcal{C}_3^2\right)
-16\mathcal{A}'(4\mathcal{C}_2+3\mathcal{C}_3)
+8\mathcal{A}(4\mathcal{C}_2'-3\mathcal{C}_3')
\Big)\dot{\phi}^2
\Big]\,.
 \end{aligned}
\end{equation}
The Friedmann equations in the \ABVC \, model with cosmological principle are \eqref{eq: FLRW equations general} with \eqref{e12}, \eqref{e13}.

\subsection{Quadratic sectors}
\subsubsection{\ABVab \,}
For the \ABVab \, sector, imposing the cosmological principle on the general kinetic function \eqref{kinetic} and using the parametrization \eqref{condit} gives
\begin{equation}
\begin{aligned}
     \mathcal{K}(\phi) =\ \mathcal{B}
     +\frac{1}{4 \kappa} &\Bigg[
        3 \mathcal{A}\Bigl(
            32 D_1^2 - 8 D_2^2 + 2 D_3^2
            + 16 D_1 D_3 - 8 D_1 D_4
            - 2 D_3 D_4 - D_4^2
        \Bigr) \\
     &\quad
        + D_1^2 \bigl(12 \beta_1 + 36 \beta_2\bigr)
        + D_2^2 \bigl(24 \beta_1\bigr)
        + D_1 D_2 \bigl(96 \beta_3\bigr) \\
     &\quad
        + D_3^2\,4\bigl(\alpha_1 + 16 \alpha_2 + \alpha_3 + 4 \alpha_4\bigr)
        + D_4^2 \bigl(7 \alpha_1 + 4 \alpha_2 + 25 \alpha_3 + 10 \alpha_4\bigr) \\
     &\quad
        + D_3 D_4\,4\bigl(5 \alpha_1 + 8 \alpha_2 + 5 \alpha_3 + 11 \alpha_4\bigr)
        +\mathcal{A}^{\prime}(- 48 D_1
        - 12 D_3
        + 6 D_4 )
     \Bigg]\,,
\end{aligned}
\end{equation}
where the $D_i$ functions are the $A_i$ ones after setting $c_i = \mathcal{C}_i = 0$ and imposing the cosmological principle parametrization \eqref{condit}. The cosmological restriction reduces the number of independent coupling combinations, although the resulting kinetic function remains too involved for further analytic treatment. The Einstein-frame kinetic function follows from \eqref{EFkinetic}.

\subsubsection{\ABVC \,}
Under the cosmological principle, the kinetic term \eqref{ci} simplifies slightly relative to the general case, remaining however, 
nontrivial:
\begin{equation}
\begin{aligned}
    \mathcal{K}(\phi) =&\ \mathcal{B} - \frac{3}{4 \kappa} 
\Big[
\mathcal{A}\left(-32D_1^2+8D_2^2-2D_3^2+2D_3D_4+D_4^2\right)
+4\left(-4\mathcal{A}-\gamma_1+\gamma_2\right)D_1D_3\\
&+2\left(
4\mathcal{A}+\gamma_1+5\gamma_2\right)D_1D_4
+8\left(\gamma_3-\gamma_4\right)D_2D_3+4\left(5\gamma_3
+\gamma_4\right)D_2D_4+2\mathcal{A}'\left(8D_1+2D_3-D_4
\right)
\Big]\,.
\end{aligned}
\end{equation}
The $D_i$ are obtained from the functions $A_i$ in Appendix \ref{Aizi} after applying the reparametrization \eqref{condit}. The cosmological principle reduces the number of independent combinations entering the \ABVC \, kinetic function, although the resulting expression remains nontrivial. The corresponding Einstein-frame kinetic function follows from \eqref{EFkinetic}.\\

In summary, a consistent homogeneous and isotropic cosmological background requires the cosmological symmetry to be imposed on all geometric and matter quantities. This reduces the number of independent tensorial structures and coupling combinations entering the background equations. However, the generic metric-affine dynamics remains nontrivial: the surviving combinations still enter the connection solution and the effective kinetic function in a complicated way. Thus, the cosmological principle reduces the parameter space relevant for the background evolution, but does not in general drastically simplify the theory.

\section{Conclusions}\label{conclusions}
In this work, we studied a scalar field nonminimally coupled to metric-affine geometry within the class of actions that are linear in the affine curvature and contain all independent parity-even and parity-odd terms quadratic in torsion and nonmetricity, including their mixed contractions. We also included the Nieh-Yan-like terms, derivative couplings between the scalar-field derivative and the four independent torsion and nonmetricity vectors. All the coefficients involved are allowed to depend on the scalar field. Direct couplings between curvature and torsion or nonmetricity, as well as higher-order curvature terms, were not considered. At the level of operator content, the action considered here is equivalent to the general scalar-field metric-affine action introduced in Ref.\ \cite{Rigouzzo:2022yan} (see Appendix \ref{AppendixC}). Our focus has instead been on developing the complete field-equation structure of this class, its different connection branches and quadratic sectors, and the consequences for the effective scalar dynamics and cosmological specialization. Since the connection equation is algebraic and linear in torsion and nonmetricity, the independent connection can be eliminated on a generic nondegenerate branch. Substituting its solution back into the action yields a dynamically equivalent metric scalar-tensor theory in which the effects of the non-Riemannian sectors are encoded in an effective Jordan-frame kinetic function $\mathcal{K}(\phi)$ and, after a conformal transformation, in the corresponding Einstein-frame function $\mathcal{K}_{_{EF}}(\phi)$. We also introduced polynomial expressions for the scalar-dependent coefficient functions. For generic nondegenerate connection branches, nonvanishing quadratic coefficients, and in the absence of cancellations, these lead to a Jordan-frame kinetic function that grows quadratically at large field values, while its Einstein-frame counterpart approaches a constant.

We first applied the general framework to the simplest scalar-tensor case that contains only the nonminimal coupling to curvature $\mathcal{A}$, the kinetic function $\mathcal{B}$, and the potential $\mathcal{V}$ (\ABV\, model). Its connection equation is projectively invariant and leaves one vectorial component of the distortion undetermined. Different projective gauge choices can therefore select torsionless or metric-compatible representatives without changing the metric and scalar-field dynamics. The Einstein-frame kinetic function reduces to the reference result $\mathcal{K}_{_{EF}}=\mathcal{B}/\mathcal{A}$. We also identified directly from the connection equations the special Riemannian branch, which for a dynamical scalar requires $\mathcal{A}^{\prime}=0$, as well as the torsionless Palatini and metric-compatible projective gauges of the generic \ABV\, solution.

Adding the derivative couplings $\mathcal{C}_i$ in the \ABVC \, model modifies this picture. The geometric part of the connection equation remains projectively invariant, whereas the derivative-coupling sector is not projectively invariant for arbitrary $\mathcal{C}_i$. In the absence of additional matter sources, projective consistency requires $\mathcal{C}_1=(-4\mathcal{C}_2+3\mathcal{C}_3)/16$, after which the remaining projective freedom can be gauge fixed. The derivative couplings modify both the connection solution and the effective kinetic function. After eliminating the connection, they do not introduce new tensorial structures into the metric and scalar-field equations with respect to the \ABV\, model, but modify the coefficients multiplying the scalar derivative terms.

Next, we analyzed the full quadratic theory and its Riemannian, torsionless, and metric-compatible branches. These branches are distinct from theories in which torsion or nonmetricity is removed directly from the Lagrangian, since imposing their vanishing at the level of the equations of motion generally leaves the remaining coupling sectors present and produces additional consistency relations. Moreover, the torsionless and metric-compatible conditions reduce the algebraic connection system before it is inverted, so these branches possess their own nondegeneracy conditions and branch-specific connection solutions rather than being obtained in general by simply restricting the generic nondegenerate solution. The classification of the quadratic sectors revealed several simplifications. In the generic nondegenerate branch, the pure quadratic nonmetricity sector \ABVa \, and the pure quadratic torsion sector \ABVb \, separately leave the \ABV\, kinetic functions unchanged, although the surviving connection has, respectively, nonzero torsion and nonzero nonmetricity. Their simultaneous presence in the \ABVab \, model, instead, generically modifies the effective kinetic function. In the mixed torsion--nonmetricity sector, the parity-even coefficients $c_1$, $c_2$, and $c_3$ can individually modify the \ABV\, kinetic function, while the parity-odd coefficients $c_4$, $c_5$, and $c_6$ leave it unchanged when present individually or only in combinations among themselves, although they can contribute when combined with the parity-even mixed terms. Thus, the presence of a non-Riemannian invariant in the original action does not by itself imply an independent modification of the equivalent metric scalar dynamics.

We then investigated how the modified Einstein-frame kinetic functions affect the relation between the original scalar field $\phi$ and the canonically normalized field $\varphi$, and consequently the potentials expressed in canonical-field space. Starting from representative kinetic functions, we showed how derivative and mixed torsion--nonmetricity couplings can reshape quadratic and quartic potentials through the field redefinition $\phi(\varphi)$. For a regular transformation, the potential values at the corresponding extrema remain unchanged, while their positions and the field-space distances between them can be modified. The illustrative $c_1$ example provides a particularly clear case of such a nonlinear and asymmetric reshaping (Fig.\ \ref{fig:c1L}). By contrast, the parity-odd mixed sector containing only $c_4$, $c_5$, and $c_6$ does not produce an additional modification of the canonical field with respect to the \ABV\, case, since the corresponding Einstein-frame kinetic function remains unchanged. We then considered the inverse problem. In the \ABVC \, sector, we showed that, starting from a quadratic Jordan-frame potential, a suitable choice of the derivative couplings and of the nonminimal coupling can reproduce a natural-inflation potential after transformation to the Einstein-frame and canonical normalization. The corresponding field redefinition can be obtained analytically, providing a direct consistency check of the reconstruction.

Finally, we separately investigated the restrictions imposed by the cosmological principle. Homogeneity and isotropy restrict the allowed tensorial forms of torsion and nonmetricity and imply that the quadratic invariants of the full covariant theory enter the cosmological equations only through a reduced number of independent combinations. In the \ABV\, model, the scalar field hypermomentum reduces to a pure spin contribution, with vanishing shear, dilation, and pseudoscalar hypermomentum component $\zeta$. Using the hyperfluid description, we obtained the corresponding effective energy density $\rho_h$ and pressure $p_h$ induced by the hypermomentum and wrote the modified Friedmann and continuity equations. A comparison with a previously considered phenomenological spin-density
for hyperfluids yields, for a vanishing scalar potential, a particular compatibility relation between $\mathcal{A}$ and $\mathcal{B}$, which represents a special subclass rather than a restriction of the general \ABV\, theory. In the \ABVC \, model, the derivative couplings generate both spin and shear, with $\Delta=0$ and $\Sigma_2=-\Sigma_1/2$. The variable $\zeta$ is in general proportional to $\mathcal{C}_4$, while on the $\zeta=0$ branch used for the comparison with spatially flat hyperfluid cosmology, one has $\mathcal{C}_4=0$ for a dynamical scalar field. We determined the corresponding hypermomentum contributions $\rho_h$ and $p_h$. In addition, the derivative-coupling sector contributes to the effective scalar-field density and pressure through $\rho_{\mathrm{\ABVC \,}}$ and $p_{\mathrm{\ABVC \,}}$, which are included in the definitions of $\rho$ and $p$ entering the Friedmann and continuity equations. For more general quadratic sectors, the same hyperfluid identification cannot in general be applied directly, since the gravitational sector itself is modified. Nevertheless, the cosmological principle still reduces the number of independent combinations entering the background dynamics and simplifies the effective kinetic functions, as illustrated for the \ABVab \, and \ABVC \, sectors.

The analysis developed here provides a systematic framework for determining which metric-affine interactions survive after the auxiliary connection is eliminated and which apparently different non-Riemannian actions lead to the same effective scalar dynamics. This provides a basis for the systematic classification of models, allowing us to expand the approach of Refs.\ \cite{Jarv:2016sow,Jarv:2020qqm} to the metric-affine setting. Natural extensions of the work include the identification of phenomenologically viable subclasses and the calculation of their inflationary observables, the study of cosmological perturbations and stability, the analysis of degenerate branches of the connection equations, the inclusion of additional matter sources carrying hypermomentum, and extensions involving direct curvature--torsion or curvature--nonmetricity couplings.


\enlargethispage{\baselineskip}
\begin{acknowledgments}
	IA, LJ, and MS were supported by the Estonian Research Council team grant ``Space - Time - Matter'' (PRG2608) as well as by the Estonian Ministry of Education and Research Centre of Excellence ``Foundations of the Universe'' (TK202U4). DI's work was supported by the Istituto Nazionale di Fisica Nucleare (INFN), Sezioni di Napoli, {\it Iniziative Specifiche} QGSKY. Some of the authors used OpenAI ChatGPT to assist with language editing and manuscript clarity. All scientific content and conclusions were independently checked by the authors.
\end{acknowledgments}


\appendix

\section{Field-equation contributions}\label{Aizi0}
\subsection{Connection field equations}
\begin{equation}
\begin{aligned}
    & P^{(2)}{}_\lambda{}^{\mu\nu}:= \frac{1}{\mathcal{A}}\Big(+4 a_{1}(\phi) Q^{\nu\mu}{}_{\lambda}
    +2 a_{2}(\phi)\bigl(Q^{\mu\nu}{}_{\lambda}+Q_{\lambda}{}^{\mu\nu}\bigr)
    +2 b_{1}(\phi) S^{\mu\nu}{}_{\lambda}  +2 b_{2}(\phi) S_{\lambda}{}^{[\mu\nu]}\\
    &\quad
    +c_{1}(\phi)\Big( S^{\nu\mu}{}_{\lambda}-S_{\lambda}{}^{\nu\mu}+Q^{[\mu\nu]}{}_{\lambda}\Big) +\delta_{\lambda}^{\mu}\Big( 4 a_{3}(\phi) Q^{\nu}
    +2 a_{5}(\phi) q^{\nu}+2 c_{2}(\phi) S^{\nu}\Big)
    +\delta_{\lambda}^{\nu}\Big(  a_{5}(\phi) Q^{\mu}+2 a_{4}(\phi) q^{\mu}+ c_{3}(\phi) S^{\mu}\Big) \\
    &\quad +g^{\mu\nu}\Big(a_{5}(\phi) Q_{\lambda}+2 a_{4}(\phi) q_{\lambda}+c_{3}(\phi) S_{\lambda} \Big)
    +\Big( c_{2}(\phi) Q^{[\mu}+ c_{3}(\phi) q^{[\mu}+2 b_{3}(\phi) S^{[\mu}\Big) \delta^{\nu]}_{\lambda}  \\
    &\quad +\bigl(-2 a_{6}(\phi)+c_{6}(\phi)\bigr)\varepsilon^{\mu\nu\alpha\beta}Q_{\alpha\beta\lambda}
    +\bigl(2 b_5(\phi)-c_{6}(\phi)\bigr)\varepsilon^{\mu\nu\alpha\beta}S_{\alpha\beta\lambda} 
    -2 a_{6}(\phi)\varepsilon_{\lambda}{}^{\nu\alpha\beta}Q_{\alpha\beta}{}^{\mu}
    -c_{6}(\phi)\varepsilon_{\lambda}{}^{\nu\alpha\beta}S_{\alpha\beta}{}^{\mu}  \\
    &\quad - \varepsilon_{\lambda }{}^{\mu \nu \alpha}\bigl(b_4(\phi) S_{\alpha}+c_{4}(\phi) Q_{\alpha}+c_{5}(\phi) q_{\alpha}\bigr)
    +\Big( \frac{b_4(\phi)}{2}+c_{5}(\phi)\Big) t^{\mu}\delta_{\lambda}^{\nu} +\Big( -\frac{b_4(\phi)}{2}+2c_{4}(\phi)\Big) t^{\nu}\delta_{\lambda}^{\mu}
    +c_{5}(\phi) g^{\mu\nu}t_{\lambda} \Big)\,.
\end{aligned}
\end{equation}

\subsection{Metric equations}
\begin{equation}
\begin{aligned}
\mathcal{G}^{(\mathcal{C}_i)}{}_{\mu\nu}
\equiv &
\frac{\kappa}{2}\Bigg\{
\frac{1}{2}\mathcal{C}_2\Big(
Q_\nu\nabla_\mu\phi
+2S^\rho{}_{\rho\nu}\nabla_\mu\phi
+4S_\nu\nabla_\mu\phi
-2\nabla_\mu\nabla_\nu\phi
+Q_\mu\nabla_\nu\phi
+2S^\rho{}_{\rho\mu}\nabla_\nu\phi
\\
&\qquad
+4S_\mu\nabla_\nu\phi
-2\nabla_\nu\nabla_\mu\phi
-2g_{\mu\nu}q^\rho\nabla_\rho\phi
\Big)
+\mathcal{C}_4\Big(
\varepsilon_\nu{}^{\rho\delta\lambda}
S_{\delta\lambda\mu}\nabla_\rho\phi
+\varepsilon_\mu{}^{\rho\delta\lambda}
S_{\delta\lambda\nu}\nabla_\rho\phi
\Big)
\\
&\qquad
+\mathcal{C}_3\Big(
S_\nu\nabla_\mu\phi
+S_\mu\nabla_\nu\phi
-g_{\mu\nu}S^\rho\nabla_\rho\phi
\Big)
+\mathcal{C}_1\Big(
Q_\nu\nabla_\mu\phi
+Q_\mu\nabla_\nu\phi
+4g_{\mu\nu}S^\rho\nabla_\rho\phi
-2g_{\mu\nu}\nabla_\rho\nabla^\rho\phi
\Big)
\\
&\qquad
-2\mathcal{C}_1^{\prime}
g_{\mu\nu}\nabla_\rho\phi\nabla^\rho\phi
-2\mathcal{C}_2^{\prime}
\nabla_\mu\phi\nabla_\nu\phi
\Bigg\}.
\end{aligned}
\label{eq:metric-contribution-C}
\end{equation}
\begin{align*}
\mathcal{G}^{(a_i)}{}_{\mu\nu}
\equiv &
\frac{1}{2}\Bigg\{
a_1\Big(
-4q_\lambda Q_{\mu\nu}{}^\lambda
+2Q_{\mu\rho\delta}Q_\nu{}^{\rho\delta}
-g_{\mu\nu}Q_{\rho\delta\lambda}Q^{\rho\delta\lambda}
-4Q_{\rho\mu\delta}Q_\nu{}^{\rho\delta}
+2Q_{\mu\nu}{}^\rho Q_\rho
+8Q_{\mu\nu}{}^\rho S_\rho
-4g^{\rho\delta}\nabla_\delta Q_{\rho\mu\nu}
\Big)
\notag\\
&\qquad
+a_2\Big(
-2q_\lambda Q_{\mu\nu}{}^\lambda
-2q_\lambda Q_{\nu\mu}{}^\lambda
-2Q_\nu{}^{\delta\rho}Q_{\rho\mu\delta}
-g_{\mu\nu}Q_{\delta\rho\lambda}Q^{\rho\delta\lambda}
+Q_{\mu\nu}{}^\rho Q_\rho
+Q_{\nu\mu}{}^\rho Q_\rho
\notag\\
&\qquad\qquad
+4Q_{\mu\nu}{}^\rho S_\rho
+4Q_{\nu\mu}{}^\rho S_\rho
-2g^{\rho\delta}\nabla_\delta Q_{\mu\nu\rho}
-2g^{\rho\delta}\nabla_\delta Q_{\nu\mu\rho}
\Big)
\notag\\
&\qquad
+a_3\Big(
2Q_\mu Q_\nu
-4g_{\mu\nu}q_\rho Q^\rho
+g_{\mu\nu}Q_\rho Q^\rho
+8g_{\mu\nu}Q_\rho S^\rho
-4g_{\mu\nu}g^{\rho\delta}\nabla_\delta Q_\rho
\Big)
\notag\\
&\qquad
+a_4\Big(
-2q_\mu q_\nu
-g_{\mu\nu}q_\rho q^\rho
+q_\nu Q_\mu
+q_\mu Q_\nu
+4q_\nu S_\mu
+4q_\mu S_\nu
-2\nabla_\mu q_\nu
-2\nabla_\nu q_\mu
\Big)
\notag\\
&\qquad
+a_5\Big(
-2g_{\mu\nu}q_\rho q^\rho
+Q_\mu Q_\nu
+2Q_\nu S_\mu
+2Q_\mu S_\nu
+4g_{\mu\nu}q_\rho S^\rho
-2g_{\mu\nu}g^{\rho\delta}\nabla_\delta q_\rho
-\nabla_\mu Q_\nu
-\nabla_\nu Q_\mu
\Big)
\notag\\
&\qquad
+2a_6\Big(
2\varepsilon_\nu{}^{\rho\lambda f}
Q_{\lambda f}{}^\delta Q_{\rho\mu\delta}
+\varepsilon^{\rho\delta\lambda f}
Q_{\lambda\nu f}Q_{\rho\mu\delta}
-\varepsilon_\nu{}^{\delta\lambda f}
Q_{\lambda f}{}^\rho Q_{\rho\mu\delta}
-\varepsilon_\nu{}^{\rho\lambda f}
Q_{\lambda\delta f}Q_{\rho\mu}{}^\delta
\notag\\
&\qquad\qquad
+2\varepsilon_\mu{}^{\rho\lambda f}
Q_{\lambda f}{}^\delta Q_{\rho\nu\delta}
-\varepsilon_\mu{}^{\delta\lambda f}
Q_{\lambda f}{}^\rho Q_{\rho\nu\delta}
-\varepsilon_\mu{}^{\rho\lambda f}
Q_{\lambda\delta f}Q_{\rho\nu}{}^\delta
+\varepsilon_\nu{}^{\delta\lambda f}
Q_{\lambda\rho f}Q_{\mu\delta}{}^\rho
\notag\\
&\qquad\qquad
+\varepsilon_\mu{}^{\delta\lambda f}
Q_{\lambda\rho f}Q_{\nu\delta}{}^\rho
+\varepsilon_\nu{}^{\rho\lambda f}
Q_{\rho\mu\delta}S_{\lambda f}{}^\delta
+\varepsilon_\mu{}^{\rho\lambda f}
Q_{\rho\nu\delta}S_{\lambda f}{}^\delta
+\varepsilon^{\rho\delta\lambda f}
Q_{\rho\nu\delta}S_{\lambda f\mu}
\notag\\
&\qquad\qquad
+\varepsilon^{\rho\delta\lambda f}
Q_{\rho\mu\delta}S_{\lambda f\nu}
-\varepsilon_\nu{}^{\delta\lambda f}
Q_{\rho\mu\delta}S_{\lambda f}{}^\rho
-\varepsilon_\mu{}^{\delta\lambda f}
Q_{\rho\nu\delta}S_{\lambda f}{}^\rho
+\varepsilon_\nu{}^{\rho\delta\lambda}
\nabla_\lambda Q_{\rho\mu\delta}
+\varepsilon_\mu{}^{\rho\delta\lambda}
\nabla_\lambda Q_{\rho\nu\delta}
\Big)
\notag\\
&\qquad
-4a_1^{\prime}Q_{\mu\nu}{}^\rho\nabla_\rho\phi
-2a_2^{\prime}\Big(
Q_{\mu\nu}{}^\rho+Q_{\nu\mu}{}^\rho
\Big)\nabla_\rho\phi
-4a_3^{\prime}g_{\mu\nu}Q^\rho\nabla_\rho\phi
-2a_4^{\prime}\Big(
q_\nu\nabla_\mu\phi+q_\mu\nabla_\nu\phi
\Big)
\notag\\
&\qquad
-a_5^{\prime}\Big(
Q_\nu\nabla_\mu\phi
+Q_\mu\nabla_\nu\phi
+2g_{\mu\nu}q^\rho\nabla_\rho\phi
\Big)
+2a_6^{\prime}\Big(
\varepsilon_\nu{}^{\rho\delta\lambda}
Q_{\delta\mu\lambda}\nabla_\rho\phi
+\varepsilon_\mu{}^{\rho\delta\lambda}
Q_{\delta\nu\lambda}\nabla_\rho\phi
\Big)
\Bigg\}.
\label{eq:metric-contribution-a}
\end{align*}
\begin{equation*}
\begin{aligned}
\hspace{-2.5 cm} \mathcal{G}^{(b_i)}{}_{\mu\nu} &
\equiv
\frac{1}{2}\Bigg\{
b_1\Big(
-g_{\mu\nu}S_{\delta\lambda}{}^\rho S_\rho{}^{\delta\lambda}
+4S_{\mu\delta}{}^\rho S_{\nu\rho}{}^\delta
-2S_{\rho\delta\mu}S_\nu{}^{\rho\delta}
\Big)
+b_2\Big(
-2S_{\mu\delta}{}^\rho S_{\nu\rho}{}^\delta
-g_{\mu\nu}S_\lambda{}^{\delta\rho}S_{\rho\delta}{}^\lambda
\Big)\\
&\qquad+b_3\Big(
2S_\mu S_\nu-g_{\mu\nu}S_\rho S^\rho
\Big)
+b_4\Big(
\varepsilon_\nu{}^{\rho\delta\lambda}
S_{\delta\lambda\mu}S_\rho
+\varepsilon_\mu{}^{\rho\delta\lambda}
S_{\delta\lambda\nu}S_\rho
\Big)
-2b_5\varepsilon^{\rho\delta\lambda f}
S_{\lambda f\nu}S_{\rho\delta\mu}
\Bigg\}.
\end{aligned}
\label{eq:metric-contribution-b}
\end{equation*}
\begin{align*}
\mathcal{G}^{(c_i)}{}_{\mu\nu}
\equiv &
\frac{1}{4}\Bigg\{
c_1\Big(
-2g^{\delta i}Q_{i\nu\rho}S_{\mu\delta}{}^\rho
+2Q_{\nu\rho\delta}S_\mu{}^{\delta\rho}
+2g^{\rho\delta}Q_{\delta\rho\lambda}S_{\mu\nu}{}^\lambda
-Q_\rho S_{\mu\nu}{}^\rho
-2g^{\delta f}Q_{f\mu\rho}S_{\nu\delta}{}^\rho
+2Q_{\mu\rho\delta}S_\nu{}^{\delta\rho}
\notag\\
&\qquad
+2g^{\rho\delta}Q_{\delta\rho\lambda}S_{\nu\mu}{}^\lambda
-Q_\rho S_{\nu\mu}{}^\rho
-2Q_{\rho\nu\delta}S_\mu{}^{\rho\delta}
-2Q_{\rho\mu\delta}S_\nu{}^{\rho\delta}
-2g_{\mu\nu}Q_{\rho\delta\lambda}S^{\rho\lambda\delta}
-4S_{\mu\nu}{}^\rho S_\rho
-4S_{\nu\mu}{}^\rho S_\rho
\notag\\
&\qquad
+2g^{\delta\lambda}g_{\nu\rho}
\nabla_\lambda S_{\mu\delta}{}^\rho
+2g^{\delta\lambda}g_{\mu\rho}
\nabla_\lambda S_{\nu\delta}{}^\rho
\Big)
+2c_2\Big(
Q_\nu S_\mu
+Q_\mu S_\nu
-2g_{\mu\nu}q_\rho S^\rho
+4g_{\mu\nu}S_\rho S^\rho
-2g_{\mu\nu}g^{\rho\delta}\nabla_\delta S_\rho
\Big)
\notag\\
&\qquad
+c_3\Big(
Q_\nu S_\mu
+Q_\mu S_\nu
+8S_\mu S_\nu
-2g_{\mu\nu}q_\rho S^\rho
-2\nabla_\mu S_\nu
-2\nabla_\nu S_\mu
\Big)
\notag\\
&\qquad
+2c_4\Big(
2\varepsilon^{\rho f i\delta}
g_{\mu\nu}Q_{\rho\lambda f}S_{i\delta}{}^\lambda
+\varepsilon_\nu{}^{\rho\lambda f}
Q_\rho S_{\lambda f\mu}
+\varepsilon_\mu{}^{\rho\lambda f}
Q_\rho S_{\lambda f\nu}
-4\varepsilon_\rho{}^{f i\delta}
g_{\mu\nu}S_{i\delta}{}^\lambda S_{\lambda f}{}^\rho
+2\varepsilon^{\lambda f i\delta}
g_{\mu\nu}S_{i\delta\rho}S_{\lambda f}{}^\rho
\notag\\
&\qquad
+4\varepsilon_i{}^{\rho f\delta}
g_{\mu\nu}Q_{\rho f}{}^\lambda S_{\lambda\delta}{}^i
-4\varepsilon_i{}^{\rho f\delta}
g_{\mu\nu}Q_{\rho\lambda f}S_\delta{}^{\lambda i}
+2\varepsilon_\rho{}^{\lambda f i}
g_{\mu\nu}\nabla_i S_{\lambda f}{}^\rho
\Big)
+c_5\Big(
2q_\rho\varepsilon_\nu{}^{\rho\delta\lambda}S_{\delta\lambda\mu}
\notag\\
&\qquad+2q_\rho\varepsilon_\mu{}^{\rho\delta\lambda}S_{\delta\lambda\nu}
-2q_\nu\varepsilon_{\mu\rho}{}^{\delta\lambda}S_{\delta\lambda}{}^\rho
-2q_\mu\varepsilon_{\nu\rho}{}^{\delta\lambda}S_{\delta\lambda}{}^\rho
+\varepsilon_{\nu\rho}{}^{\delta\lambda}
Q_\mu S_{\delta\lambda}{}^\rho
+\varepsilon_{\mu\rho}{}^{\delta\lambda}
Q_\nu S_{\delta\lambda}{}^\rho
+2\varepsilon_{\nu\lambda}{}^{\rho f}
Q_{\rho\mu\delta}S_f{}^{\delta\lambda}
\notag\\
&\qquad
+2\varepsilon_{\mu\lambda}{}^{\rho f}
Q_{\rho\nu\delta}S_f{}^{\delta\lambda}
+2\varepsilon_\delta{}^{\rho\lambda f}
Q_{\mu\nu\rho}S_{\lambda f}{}^\delta
+2\varepsilon_\delta{}^{\rho\lambda f}
Q_{\nu\mu\rho}S_{\lambda f}{}^\delta
-\varepsilon_\nu{}^{\rho\lambda f}
Q_{\rho\mu\delta}S_{\lambda f}{}^\delta
-2\varepsilon_\delta{}^{\rho\lambda f}
Q_{\rho\mu\nu}S_{\lambda f}{}^\delta
\notag\\
&\qquad
-\varepsilon_\mu{}^{\rho\lambda f}
Q_{\rho\nu\delta}S_{\lambda f}{}^\delta
+\varepsilon_\nu{}^{\delta\lambda f}
Q_{\mu\rho\delta}S_{\lambda f}{}^\rho
+\varepsilon_\mu{}^{\delta\lambda f}
Q_{\nu\rho\delta}S_{\lambda f}{}^\rho
+\varepsilon_\nu{}^{\delta\lambda f}
Q_{\rho\mu\delta}S_{\lambda f}{}^\rho
+\varepsilon_\mu{}^{\delta\lambda f}
Q_{\rho\nu\delta}S_{\lambda f}{}^\rho
\notag\\
&\qquad
+4\varepsilon_{\nu\rho\lambda}{}^f
S_f{}^{\delta\lambda}S_{\mu\delta}{}^\rho
-2\varepsilon_{\nu\rho}{}^{\lambda f}
S_{\lambda f}{}^\delta S_{\mu\delta}{}^\rho
+2\varepsilon_\nu{}^{\delta\lambda f}
S_{\lambda f\rho}S_{\mu\delta}{}^\rho
+4\varepsilon_\delta{}^{\rho\lambda f}
S_{\lambda f}{}^\delta S_{\mu\rho\nu}
\notag\\
&\qquad
+4\varepsilon_{\mu\rho\lambda}{}^f
S_f{}^{\delta\lambda}S_{\nu\delta}{}^\rho
-2\varepsilon_{\mu\rho}{}^{\lambda f}
S_{\lambda f}{}^\delta S_{\nu\delta}{}^\rho
+2\varepsilon_\mu{}^{\delta\lambda f}
S_{\lambda f\rho}S_{\nu\delta}{}^\rho
+4\varepsilon_\delta{}^{\rho\lambda f}
S_{\lambda f}{}^\delta S_{\nu\rho\mu}
\notag\\
&\qquad
-4\varepsilon_{\nu\lambda}{}^{\delta f}
S_{\mu\delta}{}^\rho S_{\rho f}{}^\lambda
-4\varepsilon_{\mu\lambda}{}^{\delta f}
S_{\nu\delta}{}^\rho S_{\rho f}{}^\lambda
+2\varepsilon_\nu{}^{\delta\lambda f}
S_{\lambda f}{}^\rho S_{\rho\delta\mu}
\notag\\
&\qquad
+2\varepsilon_\mu{}^{\delta\lambda f}
S_{\lambda f}{}^\rho S_{\rho\delta\nu}
-2\varepsilon_{\nu\lambda}{}^{\delta f}
Q_{\mu\rho\delta}S_f{}^{\rho\lambda}
-2\varepsilon_{\mu\lambda}{}^{\delta f}
Q_{\nu\rho\delta}S_f{}^{\rho\lambda}
-2\varepsilon_{\nu\lambda}{}^{\delta f}
Q_{\rho\mu\delta}S_f{}^{\rho\lambda}
\notag\\
&\qquad
-2\varepsilon_{\mu\lambda}{}^{\delta f}
Q_{\rho\nu\delta}S_f{}^{\rho\lambda}
-4\varepsilon_{\nu\lambda}{}^{\delta f}
S_{\rho\delta\mu}S_f{}^{\rho\lambda}
-4\varepsilon_{\mu\lambda}{}^{\delta f}
S_{\rho\delta\nu}S_f{}^{\rho\lambda}
+4\varepsilon_{\nu\rho}{}^{\delta\lambda}
S_{\delta\lambda}{}^\rho S_\mu
\notag\\
&\qquad
+4\varepsilon_{\mu\rho}{}^{\delta\lambda}
S_{\delta\lambda}{}^\rho S_\nu
-4\varepsilon_{\nu\rho}{}^{\lambda f}
S_\lambda{}^{\delta\rho}\nabla_\mu g_{\delta f}
-2\varepsilon_{\nu\rho}{}^{\delta\lambda}
\nabla_\mu S_{\delta\lambda}{}^\rho
-4\varepsilon_{\mu\rho}{}^{\lambda f}
S_\lambda{}^{\delta\rho}\nabla_\nu g_{\delta f}
\notag\\
&\qquad
-2\varepsilon_{\mu\rho}{}^{\delta\lambda}
\nabla_\nu S_{\delta\lambda}{}^\rho
\Big)
+2c_6\Big(
2\varepsilon_\nu{}^{\rho\lambda f}
Q_{\rho\lambda}{}^\delta S_{\delta f\mu}
+2\varepsilon_\mu{}^{\rho\lambda f}
Q_{\rho\lambda}{}^\delta S_{\delta f\nu}
-2\varepsilon_\nu{}^{\rho\lambda f}
Q_{\rho\delta\lambda}S_{f\mu}{}^\delta
\notag\\
&\qquad
-2\varepsilon_\mu{}^{\rho\lambda f}
Q_{\rho\delta\lambda}S_{f\nu}{}^\delta
+2\varepsilon^{\rho\delta\lambda f}
S_{\lambda f\nu}S_{\rho\delta\mu}
-2\varepsilon_\nu{}^{\delta\lambda f}
S_{\lambda f}{}^\rho S_{\rho\delta\mu}
-2\varepsilon_\mu{}^{\delta\lambda f}
S_{\lambda f}{}^\rho S_{\rho\delta\nu}
\notag\\
&\qquad
+\varepsilon_\nu{}^{\delta\lambda f}
g_{\mu\rho}\nabla_f S_{\delta\lambda}{}^\rho
+\varepsilon_\mu{}^{\delta\lambda f}
g_{\nu\rho}\nabla_f S_{\delta\lambda}{}^\rho
\Big)
+2c_1^{\prime}\Big(
S_{\mu\nu}{}^\rho\nabla_\rho\phi
+S_{\nu\mu}{}^\rho\nabla_\rho\phi
\Big)
\notag\\
&\qquad
-4c_2^{\prime}g_{\mu\nu}S_\rho\nabla^\rho\phi
-2c_3^{\prime}\Big(
S_\nu\nabla_\mu\phi
+S_\mu\nabla_\nu\phi
\Big)
-4c_4^{\prime}
\varepsilon_\delta{}^{\rho\lambda f}
g_{\mu\nu}S_{\lambda f}{}^\delta\nabla_\rho\phi
\notag\\
&\qquad
-2c_5^{\prime}\Big(
\varepsilon_{\nu\rho}{}^{\delta\lambda}
S_{\delta\lambda}{}^\rho\nabla_\mu\phi
+\varepsilon_{\mu\rho}{}^{\delta\lambda}
S_{\delta\lambda}{}^\rho\nabla_\nu\phi
\Big)
+2c_6^{\prime}\Big(
\varepsilon_\nu{}^{\rho\delta\lambda}
S_{\delta\lambda\mu}\nabla_\rho\phi
+\varepsilon_\mu{}^{\rho\delta\lambda}
S_{\delta\lambda\nu}\nabla_\rho\phi
\Big)
\Bigg\}.
\label{eq:metric-contribution-c}
\end{align*}

\section{Connection solutions and auxiliary functions}\label{Aizi}
\subsection{At least one of $a_i$, $b_i$, or $c_i$ is nonzero}\label{AiziPart1}
The general coefficients $A_1$,$A_2$,$A_3$, $A_4$ of torsion and nonmetricity that appear in \eqref{torphi}, \eqref{nonmetphi} and elsewhere are related to the functions present in the action \eqref{action} in the following way (in the next paragraphs, the other possibilities are reported)
\begin{align*}
    A_1 =& \frac{1}{z_{17}} \Big( -4 (z_{12} z_{3} z_{6} -  z_{11} z_{4} z_{6} -  z_{12} z_{2} z_{7} + z_{10} z_{4} z_{7} + z_{11} z_{2} z_{8} -  z_{10} z_{3} z_{8}) \kappa \mathcal{C}_1 \\
    &+ 2 (- z_{12} z_{15} z_{2} + z_{11} z_{16} z_{2} + z_{12} z_{14} z_{3} -  z_{10} z_{16} z_{3} -  z_{11} z_{14} z_{4} + z_{10} z_{15} z_{4}\\&
+ z_{16} z_{3} z_{6} -  z_{15} z_{4} z_{6} -  z_{16} z_{2} z_{7} + z_{14} z_{4} z_{7}+ z_{15} z_{2} z_{8} -  z_{14} z_{3} z_{8}) \kappa \mathcal{C}_2\\
&+ (z_{12} z_{3} z_{6} + z_{16} z_{3} z_{6} -  z_{11} z_{4} z_{6} -  z_{15} z_{4} z_{6} -  z_{12} z_{2} z_{7} -  z_{16} z_{2} z_{7} \\
    &+ z_{10} z_{4} z_{7} + z_{14} z_{4} z_{7} + z_{11} z_{2} z_{8} + z_{15} z_{2} z_{8} -  z_{10} z_{3} z_{8} -  z_{14} z_{3} z_{8}) \kappa \mathcal{C}_3 \\
    &- 2 (z_{12} z_{15} z_{6} -  z_{11} z_{16} z_{6} -  z_{12} z_{14} z_{7} + z_{10} z_{16} z_{7} + z_{11} z_{14} z_{8} -  z_{10} z_{15} z_{8}) \kappa \mathcal{C}_4\\
    &+ 2 (z_{12} z_{15} z_{2} -  z_{11} z_{16} z_{2} -  z_{12} z_{14} z_{3} + z_{10} z_{16} z_{3} + z_{11} z_{14} z_{4} -  z_{10} z_{15} z_{4}\\
    &+ z_{16} z_{3} z_{6} -  z_{15} z_{4} z_{6} -  z_{16} z_{2} z_{7} + z_{14} z_{4} z_{7} + z_{15} z_{2} z_{8} -  z_{14} z_{3} z_{8}) \mathcal{A}^{\prime} \Big)\\
    & \hspace{-2 cm} A_2 =\frac{1}{z_{17}} \Big(4 (z_{12} z_{3} z_{5} -  z_{11} z_{4} z_{5} -  z_{1} z_{12} z_{7} + z_{1} z_{11} z_{8} + z_{4} z_{7} z_{9} -  z_{3} z_{8} z_{9}) \kappa \mathcal{C}_1\\
   &+ 2 \bigl(- z_{12} z_{13} z_{3} + z_{11} z_{13} z_{4} -  z_{16} z_{3} z_{5} + z_{15} z_{4} z_{5} -  z_{13} z_{4} z_{7} + z_{13} z_{3} z_{8}\\
   &+ z_{1} (z_{12} z_{15} -  z_{11} z_{16} + z_{16} z_{7} -  z_{15} z_{8}) + z_{16} z_{3} z_{9} -  z_{15} z_{4} z_{9}\bigr) \kappa \mathcal{C}_2 \\
   &+ (- z_{12} z_{3} z_{5} -  z_{16} z_{3} z_{5} + z_{11} z_{4} z_{5}+ z_{15} z_{4} z_{5} + z_{1} z_{12} z_{7} + z_{1} z_{16} z_{7}\\
   &-  z_{13} z_{4} z_{7} -  z_{1} z_{11} z_{8} -  z_{1} z_{15} z_{8} + z_{13} z_{3} z_{8} -  z_{4} z_{7} z_{9} + z_{3} z_{8} z_{9}) \kappa \mathcal{C}_3\\
   &+ (2 z_{12} z_{15} z_{5} - 2 z_{11} z_{16} z_{5} - 2 z_{12} z_{13} z_{7} + 2 z_{11} z_{13} z_{8} + 2 z_{16} z_{7} z_{9}- 2 z_{15} z_{8} z_{9}) \kappa \mathcal{C}_4 \\
   &+ (-2 z_{1} z_{12} z_{15} + 2 z_{1} z_{11} z_{16} + 2 z_{12} z_{13} z_{3}- 2 z_{11} z_{13} z_{4} - 2 z_{16} z_{3} z_{5} + 2 z_{15} z_{4} z_{5} \\
   &+ 2 z_{1} z_{16} z_{7}- 2 z_{13} z_{4} z_{7} - 2 z_{1} z_{15} z_{8} + 2 z_{13} z_{3} z_{8}- 2 z_{16} z_{3} z_{9} + 2 z_{15} z_{4} z_{9}) \mathcal{A}^{\prime} \Big)\\
    & \hspace{-1 cm}
    A_3 =  \frac{1}{z_{17}} \Big(-4 (z_{12} z_{2} z_{5} -  z_{10} z_{4} z_{5} -  z_{1} z_{12} z_{6}+ z_{1} z_{10} z_{8} + z_{4} z_{6} z_{9} -  z_{2} z_{8} z_{9}) \kappa \mathcal{C}_1\\
    &+ 2 \bigl(z_{12} z_{13} z_{2} -  z_{10} z_{13} z_{4} + z_{16} z_{2} z_{5} -  z_{14} z_{4} z_{5} + z_{13} z_{4} z_{6} -  z_{13} z_{2} z_{8} \\
    &+ z_{1} (- z_{12} z_{14} + z_{10} z_{16} -  z_{16} z_{6} + z_{14} z_{8}) -  z_{16} z_{2} z_{9} + z_{14} z_{4} z_{9}\bigr) \kappa \mathcal{C}_2\\
    &+ (z_{12} z_{2} z_{5} + z_{16} z_{2} z_{5} -  z_{10} z_{4} z_{5} -  z_{14} z_{4} z_{5} -  z_{1} z_{12} z_{6} -  z_{1} z_{16} z_{6}\\
    &+ z_{13} z_{4} z_{6} + z_{1} z_{10} z_{8} + z_{1} z_{14} z_{8}-  z_{13} z_{2} z_{8} + z_{4} z_{6} z_{9} -  z_{2} z_{8} z_{9}) \kappa \mathcal{C}_3\\
    &+ (-2 z_{12} z_{14} z_{5} + 2 z_{10} z_{16} z_{5} + 2 z_{12} z_{13} z_{6} - 2 z_{10} z_{13} z_{8} - 2 z_{16} z_{6} z_{9} \\
    &+ 2 z_{14} z_{8} z_{9}) \kappa \mathcal{C}_4 + (2 z_{1} z_{12} z_{14} - 2 z_{1} z_{10} z_{16} - 2 z_{12} z_{13} z_{2} + 2 z_{10} z_{13} z_{4} \\
    &+ 2 z_{16} z_{2} z_{5} - 2 z_{14} z_{4} z_{5} - 2 z_{1} z_{16} z_{6} + 2 z_{13} z_{4} z_{6} + 2 z_{1} z_{14} z_{8}\\
    &- 2 z_{13} z_{2} z_{8} + 2 z_{16} z_{2} z_{9} - 2 z_{14} z_{4} z_{9}) \mathcal{A}^{\prime} \Big)\\
    & \hspace{-2 cm}
  A_4 = \frac{1}{z_{17}} \Big(4 (z_{11} z_{2} z_{5} -  z_{10} z_{3} z_{5} -  z_{1} z_{11} z_{6} + z_{1} z_{10} z_{7} + z_{3} z_{6} z_{9} -  z_{2} z_{7} z_{9}) \kappa \mathcal{C}_1\\
  &+ 2 \bigl(- z_{11} z_{13} z_{2} + z_{10} z_{13} z_{3} -  z_{15} z_{2} z_{5} + z_{14} z_{3} z_{5} -  z_{13} z_{3} z_{6} + z_{13} z_{2} z_{7} \\
  &+ z_{1} (z_{11} z_{14} -  z_{10} z_{15} + z_{15} z_{6} -  z_{14} z_{7}) + z_{15} z_{2} z_{9} -  z_{14} z_{3} z_{9}\bigr) \kappa \mathcal{C}_2 \\
  &+ (- z_{11} z_{2} z_{5} -  z_{15} z_{2} z_{5} + z_{10} z_{3} z_{5} + z_{14} z_{3} z_{5} + z_{1} z_{11} z_{6} + z_{1} z_{15} z_{6} \\
  &-  z_{13} z_{3} z_{6} -  z_{1} z_{10} z_{7} -  z_{1} z_{14} z_{7} + z_{13} z_{2} z_{7} -  z_{3} z_{6} z_{9} + z_{2} z_{7} z_{9}) \kappa \mathcal{C}_3\\
  &+ (2 z_{11} z_{14} z_{5} - 2 z_{10} z_{15} z_{5} - 2 z_{11} z_{13} z_{6} + 2 z_{10} z_{13} z_{7} + 2 z_{15} z_{6} z_{9}\\
  &- 2 z_{14} z_{7} z_{9}) \kappa \mathcal{C}_4 + (-2 z_{1} z_{11} z_{14} + 2 z_{1} z_{10} z_{15} + 2 z_{11} z_{13} z_{2} - 2 z_{10} z_{13} z_{3}\\
  &- 2 z_{15} z_{2} z_{5} + 2 z_{14} z_{3} z_{5} + 2 z_{1} z_{15} z_{6} - 2 z_{13} z_{3} z_{6} - 2 z_{1} z_{14} z_{7} \\
  &+ 2 z_{13} z_{2} z_{7} - 2 z_{15} z_{2} z_{9} + 2 z_{14} z_{3} z_{9}) \mathcal{A}^{\prime} \Big)\,,
  \end{align*}
with
\begin{equation*}
\begin{aligned}[c]
 z_1 =& -3\,b_4 - 4\,b_5\,,\\
z_2 =& -2 \mathcal{A} + 2\,b_1 + 2\,b_2\,,\\
z_3 =& -4 c_4 - c_5 - c_6\,,\\
  z_4 =& - c_4 -  \frac{5}{2} c_5 + \frac{1}{2} c_6\,,\\
   z_5 =& 4 \mathcal{A} -  c_1 + 3 c_3\,,\\
    z_6 =& 6 c_5 - 2 c_6\,,\\
    z_7 =& \mathcal{A} + 2 \,a_2 + 2 \,a_4 + 4 \,a_5\,,\\
     z_8 =& \frac{1}{2} \mathcal{A} + 2 a_1 + \,a_2 + 5 \,a_4 + \,a_5\,,
\end{aligned}
\hspace{3 cm}
\begin{aligned}[c]
 z_9 =& -4 \mathcal{A} + 2\,b_1 - \,b_2 + 3\,b_3 - c_1 + 3 c_3\,,\\
 z_{10} =& 3\,b_4 + 4\,b_5 + 6 c_5 - 2 c_6\,,\\
  z_{11} =& - \mathcal{A} + 2 a_2 + 2 a_4 + 4 a_5 + \frac{1}{2} c_1 + 2 c_2 + \frac{1}{2} c_3\,,\\
   z_{12}=&\frac{3}{2} \mathcal{A} + 2 a_1 + a_2 + 5 a_4 + a_5 -  \frac{1}{4} c_1 + \frac{1}{2} c_2 + \frac{5}{4} c_3\,,\\
    z_{13} =& -2\,b_1 +\,b_2 - 3\,b_3 + 2 c_1 + 6 c_2\,,\\
     z_{14} =& -3\,b_4 - 4\,b_5 + 12 c_4 + 4 c_6\,,\\
      z_{15} =& 4 a_1 + 16 a_3 + 2 a_5 -  \frac{1}{2} c_1 - 2 c_2 -  \frac{1}{2} c_3\,,\\
      z_{16} =& - \frac{1}{2} \mathcal{A} + 2 a_2 + 4 a_3 + 5 a_5 + \frac{1}{4} c_1 -  \frac{1}{2} c_2 -  \frac{5}{4} c_3\,,
\end{aligned}
\end{equation*}
\begin{equation*}
\begin{aligned}
  z_{17} =& 2 \bigl(z_{10} z_{16} z_3 z_5 -  z_{10} z_{15} z_4 z_5 -  z_1 z_{10} z_{16} z_7 + z_{10} z_{13} z_4 z_7 + z_{12} (z_{15} z_2 z_5 -  z_{14} z_3 z_5 -  z_1 z_{15} z_6 \\
  &+ z_{13} z_3 z_6 + z_1 z_{14} z_7 -  z_{13} z_2 z_7) + z_1 z_{10} z_{15} z_8 -  z_{10} z_{13} z_3 z_8 + z_{11} (- z_{16} z_2 z_5 + z_{14} z_4 z_5 \\
  &+ z_1 z_{16} z_6 -  z_{13} z_4 z_6 -  z_1 z_{14} z_8 + z_{13} z_2 z_8) -  z_{16} z_3 z_6 z_9 + z_{15} z_4 z_6 z_9 + z_{16} z_2 z_7 z_9\\
  &-  z_{14} z_4 z_7 z_9 -  z_{15} z_2 z_8 z_9 + z_{14} z_3 z_8 z_9\bigr)\,.
   \end{aligned}
\end{equation*}

\subsection{If $a_i= b_i= c_i =0$}\label{AiziPart2}
Here we show the coefficients $A_1$,$A_2$,$A_3$, $A_4$ of torsion and nonmetricity that appears in \eqref{torphi}, \eqref{nonmetphi} when $a_i= b_i= c_i =0$.
If $a_i= b_i= c_i =0$ and $\mathcal{C}_i = 0$ then we have
\begin{equation}\label{Aiex1g}
    \begin{aligned}
    &A_3 + 4 A_1 = \frac{\mathcal{A}^{\prime}}{\mathcal{A}}, \hspace{2 cm}  A_2 =0, \hspace{2cm}  A_4=0\,.
    \end{aligned}
\end{equation}
If $a_i= b_i= c_i =0$ and $\mathcal{C}_i \neq 0$ then we have
\begin{equation}
    \begin{aligned}
    &A_1= - \frac{ \bigl(\kappa(6 \mathcal{C}_2 + \mathcal{C}_3) - 2 \mathcal{A}^{\prime}\bigr)}{8 \mathcal{A}}, \hspace{2 cm}  A_2 =- \frac{\kappa \,\mathcal{C}_4 }{2 \mathcal{A}} \\
    &A_3 = \frac{5 \kappa\,\bigl(4 \mathcal{C}_2 + \mathcal{C}_3\bigr) }{8 \mathcal{A}}, \hspace{2cm}  A_4= -  \frac{\kappa\,\bigl(4 \mathcal{C}_2 + \mathcal{C}_3\bigr)}{4 \mathcal{A}}\,,
    \end{aligned}
\end{equation}
where the constraint $\mathcal{C}_1 = \frac{1}{16}(-4\, \mathcal{C}_2 +3\, \mathcal{C}_3)$ has already been imposed.

\subsection{Other functions}\label{AiziPart3}
The functions used in the field equations throughout the main text are defined as follows:
\begin{align*}
\mathcal{F}_{1}(\mathcal{A},\mathcal{A}^{\prime},\mathcal{C}_i,\mathcal{C}_i^{\prime})
&\coloneqq
- \tfrac{3 \kappa}{128 \mathcal{A}} \Bigl(\kappa(48 \mathcal{C}_2^2 + 24 \mathcal{C}_2 \mathcal{C}_3 - \mathcal{C}_3^2 + 64 \mathcal{C}_4^2 ) - 32 \mathcal{C}_3 \mathcal{A}^{\prime}\Bigr),\\
\mathcal{F}_{2}
&\coloneqq
-\tfrac{1}{2} \mathcal{F}_{1}
,\\
\mathcal{F}_{3}
&\coloneqq
-\tfrac{1}{\kappa} \mathcal{F}_{1},\\
\mathcal{F}_{4}
&\coloneqq
-\tfrac{1}{2 \kappa} \Big(\mathcal{F}_{1}^{\prime} + \tfrac{\mathcal{A}^{\prime}}{\mathcal{A}}\mathcal{F}_{1}\Big),\\
\mathcal{F}_{5}(\mathcal{Z}_{1},\mathcal{Z}_{4},\mathcal{Z}_{6},\mathcal{Z}_{9})
&\coloneqq
\mathcal{Z}_{1}+\mathcal{Z}_{4}+\mathcal{Z}_{6}+\mathcal{Z}_{9},\\
\mathcal{F}_{6}(\mathcal{Z}_{0},\mathcal{Z}_{5},\mathcal{Z}_{7},\mathcal{Z}_{8})
&\coloneqq
\mathcal{Z}_{0}+\mathcal{Z}_{5}+\mathcal{Z}_{7}+\mathcal{Z}_{8},\\
\mathcal{F}_{7}(\mathcal{Z}_{2},\mathcal{Z}_{10},\mathcal{C}_i)
&\coloneqq
\mathcal{Z}_{10}+\mathcal{Z}_{2}-\tfrac{1}{2}\kappa\mathcal{C}_2,\\
\mathcal{F}_{8}(\mathcal{Z}_{3},\mathcal{Z}_{11})
&\coloneqq
\mathcal{Z}_{3}+\mathcal{Z}_{11},\\
\mathcal{F}_{9}(\mathcal{F}_{6},\mathcal{F}_{7},\mathcal{F}_{8},\mathcal{A},\mathcal{C}_i,A_i,A_i^{\prime})
&\coloneqq
\mathcal{F}_{6}
+\left(4A_1+A_3-A_4\right)\mathcal{F}_{7}
+\left(6A_1+A_3-2A_4\right)\mathcal{F}_{8}
-\tfrac{\kappa\mathcal{C}_1}{2}\left(12A_1+4A_3+A_4\right) \notag\\
&\quad
+\mathcal{A}\Bigg[
4A_1^2-A_2^2+2A_1A_3-2A_1A_4
+\tfrac{1}{4}A_3^2-\tfrac{1}{2}A_3A_4-\tfrac{1}{8}A_4^2
+4A_1^{\prime}+A_3^{\prime}-\tfrac{1}{4}A_4^{\prime}
\Bigg],\\
\mathcal{F}_{10}(\mathcal{F}_{5},\mathcal{F}_{7},\mathcal{A},A_i,A_i^{\prime})
&\coloneqq
\mathcal{F}_{5}
-2\left(2A_1+A_3\right)\mathcal{F}_{7}
+\mathcal{A}\Bigg[
8A_1^2-2A_2^2+4A_1A_3+2A_1A_4 \\
&\qquad
+\tfrac{1}{2}A_3^2+\tfrac{1}{2}A_3A_4-\tfrac{1}{4}A_4^2
-4A_1^{\prime}-A_3^{\prime}-\tfrac{1}{2}A_4^{\prime}
\Bigg],\\
\mathcal{F}_{11}(\mathcal{A}^{\prime},\mathcal{C}_i,A_i)
&\coloneqq
-\left(\tfrac{3}{2}\mathcal{C}_3+\tfrac{6\mathcal{A}^{\prime}}{\kappa}\right)A_1
-3\mathcal{C}_4A_2
-\left(2\mathcal{C}_1+\tfrac{1}{2}\mathcal{C}_2+\tfrac{3\mathcal{A}^{\prime}}{2\kappa}\right)A_3
-\left(\tfrac{1}{2}\mathcal{C}_1+\tfrac{5}{4}\mathcal{C}_2-\tfrac{3\mathcal{A}^{\prime}}{4\kappa}\right)A_4,\\
\mathcal{F}_{12}(\mathcal{A}^{\prime},\mathcal{C}_i,a_i^{\prime},b_i^{\prime},c_i^{\prime},A_i,A_i^{\prime})
&\coloneqq
\tfrac{1}{\kappa}
\Bigg[
a_1^{\prime}\left(2A_3^2+A_3A_4+\tfrac{5}{4}A_4^2\right)
+a_2^{\prime}\left(\tfrac{1}{2}A_3^2+\tfrac{5}{2}A_3A_4+\tfrac{7}{8}A_4^2\right) \notag\\
&\hspace{1.0cm}
+a_3^{\prime}\left(8A_3^2+4A_3A_4+\tfrac{1}{2}A_4^2\right)
+a_4^{\prime}\left(\tfrac{1}{2}A_3^2+\tfrac{5}{2}A_3A_4+\tfrac{25}{8}A_4^2\right) \notag\\
&\hspace{1.0cm}
+a_5^{\prime}\left(2A_3^2+\tfrac{11}{2}A_3A_4+\tfrac{5}{4}A_4^2\right)
\Bigg] \notag\\
&\quad
+\tfrac{3}{\kappa}
\Bigg[
\left(A_1^2-A_2^2\right)b_1^{\prime}
-\left(\tfrac{1}{2}A_1^2+A_2^2\right)b_2^{\prime}
+\tfrac{3}{2}A_1^2b_3^{\prime}
+3A_1A_2b_4^{\prime}
+4A_1A_2b_5^{\prime}
\Bigg] \notag\\
&\quad
+\tfrac{3}{4\kappa}
\Bigg[
A_1\left(2A_3-A_4\right)c_1^{\prime}
+2A_1\left(4A_3+A_4\right)c_2^{\prime}
+A_1\left(2A_3+5A_4\right)c_3^{\prime} \notag\\
&\hspace{1.5cm}
+4A_2\left(4A_3+A_4\right)c_4^{\prime}
+2A_2\left(2A_3+5A_4\right)c_5^{\prime}
+2A_2\left(2A_3-A_4\right)c_6^{\prime}
\Bigg] \notag\\
&\quad
+\tfrac{3\mathcal{A}^{\prime}}{\kappa}
\left(
-4A_1^2+A_2^2-\tfrac{1}{4}A_3^2+\tfrac{1}{4}A_3A_4+\tfrac{1}{8}A_4^2-2A_1A_3+A_1A_4
\right) \notag\\
&\quad
-\left(\tfrac{3}{2}\mathcal{C}_3+\tfrac{6\mathcal{A}^{\prime}}{\kappa}\right)A_1^{\prime}
-3\mathcal{C}_4A_2^{\prime}
-\left(2\mathcal{C}_1+\tfrac{1}{2}\mathcal{C}_2+\tfrac{3\mathcal{A}^{\prime}}{2\kappa}\right)A_3^{\prime}
-\left(\tfrac{1}{2}\mathcal{C}_1+\tfrac{5}{4}\mathcal{C}_2-\tfrac{3\mathcal{A}^{\prime}}{4\kappa}\right)A_4^{\prime},\\
\mathcal{F}_{13}(\mathcal{F}_{9},\mathcal{F}_{10},\mathcal{F}_{12},\mathcal{A},\mathcal{A}^{\prime})
&\coloneqq
\mathcal{F}_{12}
+\tfrac{\mathcal{A}^{\prime}}{2\kappa\mathcal{A}}
\left(4\mathcal{F}_{9}+\mathcal{F}_{10}\right),\\
\mathcal{F}_{14}(\mathcal{C}_i,c_i,A_i)
&\coloneqq
-\kappa\mathcal{C}_4
-\left(4c_4+c_5+c_6\right)A_3
-\left(c_4+\tfrac{5}{2}c_5-\tfrac{1}{2}c_6\right)A_4,\\
\mathcal{F}_{15}(\mathcal{A},\mathcal{A}^{\prime},\mathcal{C}_i,a_i,A_i)
&\coloneqq
\kappa\mathcal{C}_2-\mathcal{A}^{\prime}
+\left(\mathcal{A}+2a_2+2a_4+4a_5\right)A_3
+\left(\tfrac{1}{2}\mathcal{A}+2a_1+a_2+5a_4+a_5\right)A_4,\\
\mathcal{F}_{16}(\mathcal{A},\mathcal{A}^{\prime},\mathcal{C}_i,a_i,c_i,A_i)
&\coloneqq
\kappa\left(\mathcal{C}_2+\tfrac{1}{2}\mathcal{C}_3\right)
+\mathcal{A}^{\prime}
+\left(-\mathcal{A}+2a_2+2a_4+4a_5+\tfrac{1}{2}c_1+2c_2+\tfrac{1}{2}c_3\right)A_3 \notag\\
&\quad
+\left(\tfrac{3}{2}\mathcal{A}+2a_1+a_2+5a_4+a_5-\tfrac{1}{4}c_1+\tfrac{1}{2}c_2+\tfrac{5}{4}c_3\right)A_4,\\
\mathcal{F}_{17}(\mathcal{A},\mathcal{C}_i,a_i,c_i,A_i)
&\coloneqq
\kappa\left(2\mathcal{C}_1-\tfrac{1}{2}\mathcal{C}_3\right)
+\left(4a_1+16a_3+2a_5-\tfrac{1}{2}c_1-2c_2-\tfrac{1}{2}c_3\right)A_3 \notag\\
&\quad
+\left(-\tfrac{1}{2}\mathcal{A}+2a_2+4a_3+5a_5+\tfrac{1}{4}c_1-\tfrac{1}{2}c_2-\tfrac{5}{4}c_3\right)A_4,\\
\mathcal{F}_{18}(\mathcal{A},a_i^{\prime},A_i,A_i^{\prime})
&\coloneqq
\tfrac{\mathcal{A}}{8}\left(2A_3^2-4A_3A_4-A_4^2\right)
-2A_3a_1^{\prime}-A_4a_2^{\prime} \notag\\
&\quad
-2\left(4A_3+A_4\right)a_3^{\prime}
-\left(A_3+\tfrac{5}{2}A_4\right)a_5^{\prime}
+\mathcal{A}\left(A_3^{\prime}-\tfrac{1}{4}A_4^{\prime}\right),\\
\mathcal{F}_{19}(\mathcal{A},a_i^{\prime},A_i,A_i^{\prime})
&\coloneqq
\tfrac{\mathcal{A}}{4}\left(2A_3^2+2A_3A_4-A_4^2\right)
-2A_4a_1^{\prime}
-\left(2A_3+A_4\right)a_2^{\prime} \notag\\
&\quad
-\left(2A_3+5A_4\right)a_4^{\prime}
-\left(4A_3+A_4\right)a_5^{\prime}
-\mathcal{A}\left(A_3^{\prime}+\tfrac{1}{2}A_4^{\prime}\right),\\
\mathcal{F}_{20}(\mathcal{A},\mathcal{A}^{\prime},A_i,\mathcal{C}_i)
&\coloneqq
-\left(2\mathcal{C}_1+\tfrac{1}{2}\mathcal{C}_2\right)A_3 -\left(\tfrac{1}{2}\mathcal{C}_1+\tfrac{5}{4}\mathcal{C}_2\right)A_4 +\tfrac{3\mathcal{A}'}{4\kappa}\left(A_4-2A_3\right),\\
\mathcal{F}_{21}(\mathcal{A},\mathcal{A}', \mathcal{C}_i, \mathcal{C}_i^{\prime}, a_i, a_i', A_i, A_i^{\prime})
&\coloneqq
-\tfrac{\mathcal{A}'}{\mathcal{A}} \left[ \left(2\mathcal{C}_1+\tfrac{1}{2}\mathcal{C}_2\right)A_3 +\left(\tfrac{1}{2}\mathcal{C}_1+\tfrac{5}{4}\mathcal{C}_2\right)A_4 +2\mathcal{C}_1' +\tfrac{1}{2}\mathcal{C}_2' \right] \nonumber\\ &-\left(2\mathcal{C}_1+\tfrac{1}{2}\mathcal{C}_2\right)A_3' -\left(\tfrac{1}{2}\mathcal{C}_1+\tfrac{5}{4}\mathcal{C}_2\right)A_4'+\tfrac{1}{\kappa} \Bigg\{ \Big[ 2A_3^2+A_3A_4+\tfrac{5}{4}A_4^2 \nonumber\\ &
-\tfrac{\mathcal{A}'}{\mathcal{A}}\left(4A_3+A_4\right) \Big]a_1' -\tfrac{\mathcal{A}'}{\mathcal{A}}a_1 \left( 2A_3^2+A_3A_4+\tfrac{5}{4}A_4^2 +4A_3'+A_4' \right) \Bigg\} \nonumber\\ &+\tfrac{1}{\kappa} \Bigg\{ \left[ \tfrac{1}{2}A_3^2+\tfrac{5}{2}A_3A_4+\tfrac{7}{8}A_4^2 -\tfrac{\mathcal{A}'}{\mathcal{A}} \left(A_3+\tfrac{5}{2}A_4\right) \right]a_2'  -\tfrac{\mathcal{A}'}{\mathcal{A}}a_2 \Big( \tfrac{1}{2}A_3^2+\tfrac{5}{2}A_3A_4 \nonumber\\ &
+\tfrac{7}{8}A_4^2+A_3'+\tfrac{5}{2}A_4' \Big) \Bigg\}+\tfrac{1}{\kappa} \Bigg\{ \left[ 8A_3^2+4A_3A_4+\tfrac{1}{2}A_4^2 -\tfrac{\mathcal{A}'}{\mathcal{A}} \left(16A_3+4A_4\right) \right]a_3' \nonumber\\ & -\tfrac{\mathcal{A}'}{\mathcal{A}}a_3 \left( 8A_3^2+4A_3A_4+\tfrac{1}{2}A_4^2 +16A_3'+4A_4' \right) \Bigg\} +\tfrac{1}{\kappa} \Bigg\{ \Big[\tfrac{1}{2}A_3^2+\tfrac{5}{2}A_3A_4 \nonumber\\ &
+\tfrac{25}{8}A_4^2 -\tfrac{\mathcal{A}'}{\mathcal{A}} \left(A_3+\tfrac{5}{2}A_4\right) \Big]a_4' -\tfrac{\mathcal{A}'}{\mathcal{A}}a_4 \left( \tfrac{1}{2}A_3^2+\tfrac{5}{2}A_3A_4+\tfrac{25}{8}A_4^2 +A_3'+\tfrac{5}{2}A_4' \right) \Bigg\} \nonumber\\ &+\tfrac{1}{\kappa} \Bigg\{ \left[ 2A_3^2+\tfrac{11}{2}A_3A_4+\tfrac{5}{4}A_4^2 -\tfrac{\mathcal{A}'}{\mathcal{A}} \left(4A_3+\tfrac{11}{2}A_4\right) \right]a_5' \nonumber\\ &\hspace{2.1cm} -\tfrac{\mathcal{A}'}{\mathcal{A}}a_5 \left( 2A_3^2+\tfrac{11}{2}A_3A_4+\tfrac{5}{4}A_4^2 +4A_3'+\tfrac{11}{2}A_4' \right) \Bigg\},\\
\mathcal{F}_{22}(\mathcal{A},\mathcal{C}_i,b_i,A_i)
&\coloneqq
-\kappa\mathcal{C}_4
-\left(3b_4+4b_5\right)A_1
+2\left(-\mathcal{A}+b_1+b_2\right)A_2,\\
\mathcal{F}_{23}(\mathcal{A},\mathcal{A}^{\prime},\mathcal{C}_i,c_i,A_i)
&\coloneqq
\kappa\mathcal{C}_2-\mathcal{A}^{\prime}
+\left(4\mathcal{A}-c_1+3c_3\right)A_1
+2\left(3c_5-c_6\right)A_2,\\
\mathcal{F}_{24}(\mathcal{A},\mathcal{A}^{\prime},\mathcal{C}_i,b_i,c_i,A_i)
&\coloneqq
\kappa\left(\mathcal{C}_2+\tfrac{1}{2}\mathcal{C}_3\right)
+\mathcal{A}^{\prime}
+\left(-4\mathcal{A}+2b_1-b_2+3b_3-c_1+3c_3\right)A_1 \notag\\
&\quad
+\left(3b_4+4b_5+6c_5-2c_6\right)A_2,\\
\mathcal{F}_{25}(\mathcal{C}_i,b_i,c_i,A_i)
&\coloneqq
\kappa\left(2\mathcal{C}_1-\tfrac{1}{2}\mathcal{C}_3\right)
+\left(-2b_1+b_2-3b_3+2c_1+6c_2\right)A_1 \notag\\
&\quad
+\left(-3b_4-4b_5+12c_4+4c_6\right)A_2,\\
\mathcal{F}_{26}(\mathcal{A},\mathcal{C}_i,\mathcal{C}_i^{\prime},b_i,c_i,c_i^{\prime},A_i,A_i^{\prime})
&\coloneqq
-\kappa\left[
\left(2\mathcal{C}_2+\tfrac{3}{2}\mathcal{C}_3\right)A_1
+2\mathcal{C}_4A_2
+\mathcal{C}_1^{\prime}
\right] 
+A_1^2\Big(4\mathcal{A}-3b_1+\tfrac{3}{2}b_2 \notag\\
&\quad
-\tfrac{9}{2}b_3+2c_1-6c_3\Big)-2A_1A_2\left(3b_4+4b_5+6c_5-2c_6\right)
+A_2^2\left(-\mathcal{A}+b_1+b_2\right) \notag\\
&\quad
-A_1\left(c_1^{\prime}+3c_2^{\prime}\right)
-2A_2\left(3c_4^{\prime}+c_6^{\prime}\right)
+\left(4\mathcal{A}-c_1-3c_2\right)A_1^{\prime}
-2\left(3c_4+c_6\right)A_2^{\prime},\\
\mathcal{F}_{27}(\mathcal{A},\mathcal{C}_i,\mathcal{C}_i^{\prime},b_i,c_i,c_i^{\prime},A_i,A_i^{\prime})
&\coloneqq
\kappa\left[
\left(8\mathcal{C}_2+3\mathcal{C}_3\right)A_1
+2\mathcal{C}_4A_2
-\mathcal{C}_2^{\prime}
\right]
+A_1^2\left(8\mathcal{A}+6b_1-3b_2+9b_3-8c_1+24c_3\right) \notag\\
&\quad
+2A_1A_2\left(3b_4+4b_5+24c_5-8c_6\right)
+2A_2^2\left(-\mathcal{A}+b_1+b_2\right) \notag\\
&\quad
+A_1\left(c_1^{\prime}-3c_3^{\prime}\right)
-2A_2\left(3c_5^{\prime}-c_6^{\prime}\right)
+\left(-4\mathcal{A}+c_1-3c_3\right)A_1^{\prime}
-2\left(3c_5-c_6\right)A_2^{\prime},\\
\mathcal{F}_{28}(\mathcal{A}^{\prime},\mathcal{C}_i,A_i)
&\coloneqq
-\tfrac{3}{2}\left(\mathcal{C}_3+\tfrac{4\mathcal{A}^{\prime}}{\kappa}\right)A_1
-3\mathcal{C}_4A_2,\\
\mathcal{F}_{29}(\mathcal{A},\mathcal{A}^{\prime},\mathcal{C}_i,\mathcal{C}_i^{\prime},b_i,b_i^{\prime},c_i,c_i^{\prime},A_i,A_i^{\prime})
&\coloneqq
-\tfrac{3}{2}\mathcal{C}_3
\left(A_1^{\prime}+\tfrac{\mathcal{A}^{\prime}}{\mathcal{A}}A_1\right)
-3\mathcal{C}_4
\left(A_2^{\prime}+\tfrac{\mathcal{A}^{\prime}}{\mathcal{A}}A_2\right)
-\tfrac{\mathcal{A}^{\prime}}{2\mathcal{A}}
\left(4\mathcal{C}_1^{\prime}+\mathcal{C}_2^{\prime}\right) \notag\\
&\quad
+\tfrac{3\mathcal{A}^{\prime}}{\kappa\mathcal{A}}
\Bigg[
A_1^2\left(-b_1+\tfrac{1}{2}b_2-\tfrac{3}{2}b_3\right)
+A_2^2\left(b_1+b_2\right)
-A_1A_2\left(3b_4+4b_5\right)
\Bigg] \notag\\
&\quad
+\tfrac{3}{\kappa}
\Bigg[
\left(A_1^2-A_2^2\right)b_1^{\prime}
-\left(\tfrac{1}{2}A_1^2+A_2^2\right)b_2^{\prime}
+\tfrac{3}{2}A_1^2b_3^{\prime}
+3A_1A_2b_4^{\prime}
+4A_1A_2b_5^{\prime}
\Bigg] \notag\\
&\quad
-\tfrac{3\mathcal{A}^{\prime}}{2\kappa\mathcal{A}}
\Bigg[
A_1\left(c_1^{\prime}+4c_2^{\prime}+c_3^{\prime}\right)
+A_1^{\prime}\left(c_1+4c_2+c_3\right) \notag\\
&\hspace{3.1cm}
+2A_2\left(4c_4^{\prime}+c_5^{\prime}+c_6^{\prime}\right)
+2A_2^{\prime}\left(4c_4+c_5+c_6\right)
\Bigg],\\
\mathcal{Z}_{0}(\mathcal{C}_i,\mathcal{C}_i^{\prime},A_i)
&\coloneqq
\tfrac{\kappa}{4}
\left[
6\left(4\mathcal{C}_1-\mathcal{C}_3\right)A_1
-8\mathcal{C}_4A_2
-\mathcal{C}_2\left(2A_3+5A_4\right)
-4\mathcal{C}_1^{\prime}
\right],\\
\mathcal{Z}_{1}(\mathcal{C}_i,\mathcal{C}_i^{\prime},A_i)
&\coloneqq
\tfrac{\kappa}{2}
\left[
6\left(2\mathcal{C}_2+\mathcal{C}_3\right)A_1
+4\mathcal{C}_4A_2
+\left(2\mathcal{C}_1+\mathcal{C}_2\right)\left(4A_3+A_4\right)
-2\mathcal{C}_2^{\prime}
\right],\\
\mathcal{Z}_{2}(a_i,A_i)
&\coloneqq
-\left(a_2+a_4+2a_5\right)A_3
-\tfrac{1}{2}\left(2a_1+a_2+5a_4+a_5\right)A_4,\\
\mathcal{Z}_{3}(a_i,A_i)
&\coloneqq
-\left(2a_1+8a_3+a_5\right)A_3
-\tfrac{1}{2}\left(2a_2+4a_3+5a_5\right)A_4,\\
\mathcal{Z}_{4}(a_i,a_i^{\prime},A_i,A_i^{\prime})
&\coloneqq
12A_1A_3\left(a_2+a_4+2a_5\right)
+A_3^2\left(4a_1+3a_2+16a_3+3a_4+8a_5\right) \notag\\
&\quad
+6A_1A_4\left(2a_1+a_2+5a_4+a_5\right)
+2A_3A_4\left(2a_1+a_2+4a_3+3a_4+2a_5\right) \notag\\
&\quad
+\tfrac{A_4^2}{4}\left(-6a_1-a_2+4a_3-15a_4+2a_5\right) \notag\\
&\quad
-2A_4a_1^{\prime}
-\left(2A_3+A_4\right)a_2^{\prime}
-\left(2A_3+5A_4\right)a_4^{\prime}
-\left(4A_3+A_4\right)a_5^{\prime} \notag\\
&\quad
-2\left(a_2+a_4+2a_5\right)A_3^{\prime}
-\left(2a_1+a_2+5a_4+a_5\right)A_4^{\prime},\\
\mathcal{Z}_{5}(a_i,a_i^{\prime},A_i,A_i^{\prime})
&\coloneqq
6A_1A_3\left(2a_1+8a_3+a_5\right)
-\tfrac{A_3^2}{2}\left(a_2+a_4+2a_5\right) \notag\\
&\quad
+3A_1A_4\left(2a_2+4a_3+5a_5\right)
-\tfrac{A_3A_4}{2}\left(10a_1+3a_2+36a_3+5a_4+10a_5\right) \notag\\
&\quad
-\tfrac{A_4^2}{8}\left(10a_1+23a_2+36a_3+25a_4+50a_5\right) \notag\\
&\quad
-2A_3a_1^{\prime}
-A_4a_2^{\prime}
-2\left(4A_3+A_4\right)a_3^{\prime}
-\left(A_3+\tfrac{5}{2}A_4\right)a_5^{\prime} \notag\\
&\quad
-\left(2a_1+8a_3+a_5\right)A_3^{\prime}
-\tfrac{1}{2}\left(2a_2+4a_3+5a_5\right)A_4^{\prime},\\
\mathcal{Z}_{6}(b_i,A_i)
&\coloneqq
3A_1^2\left(2b_1-b_2+3b_3\right)
+2A_1A_2\left(3b_4+4b_5\right)
+2A_2^2\left(b_1+b_2\right),\\
\mathcal{Z}_{7}(b_i,A_i)
&\coloneqq
-\tfrac{3}{2}A_1^2\left(2b_1-b_2+3b_3\right)
-2A_1A_2\left(3b_4+4b_5\right)
+A_2^2\left(b_1+b_2\right),\\
\mathcal{Z}_{8}(c_i,c_i^{\prime},A_i,A_i^{\prime})
&\coloneqq
6A_1^2\left(c_1+3c_2\right)
+12A_1A_2\left(3c_4+c_6\right) \notag\\
&\quad
-\tfrac{A_1}{4}\left(2A_3+5A_4\right)\left(c_1+6c_2+3c_3\right)
-2A_2A_3\left(c_4+c_5\right) \notag\\
&\quad
-A_2A_4\left(14c_4+5c_5+3c_6\right)
-A_1\left(c_1^{\prime}+3c_2^{\prime}\right)
-2A_2\left(3c_4^{\prime}+c_6^{\prime}\right) \notag\\
&\quad
-\left(c_1+3c_2\right)A_1^{\prime}
-2\left(3c_4+c_6\right)A_2^{\prime},\\
\mathcal{Z}_{9}(c_i,c_i^{\prime},A_i,A_i^{\prime})
&\coloneqq
6A_1^2\left(-c_1+3c_3\right)
+12A_1A_2\left(3c_5-c_6\right) \notag\\
&\quad
+\tfrac{A_1}{2}\left(4A_3+A_4\right)\left(c_1+6c_2+3c_3\right)
+8A_2A_3\left(c_4+c_5\right) \notag\\
&\quad
+A_2A_4\left(2c_4-7c_5+3c_6\right)
+A_1\left(c_1^{\prime}-3c_3^{\prime}\right)
-2A_2\left(3c_5^{\prime}-c_6^{\prime}\right) \notag\\
&\quad
+\left(c_1-3c_3\right)A_1^{\prime}
-2\left(3c_5-c_6\right)A_2^{\prime},\\
\mathcal{Z}_{10}(c_i,A_i)
&\coloneqq
\tfrac{1}{2}\left(c_1-3c_3\right)A_1
-\left(3c_5-c_6\right)A_2,\\
\mathcal{Z}_{11}(c_i,A_i)
&\coloneqq
-\left(c_1+3c_2\right)A_1
-2\left(3c_4+c_6\right)A_2.
\end{align*}


\section{Comparison with Ref.~\cite{Rigouzzo:2022yan}}\label{AppendixC}
The action \eqref{action} and the general scalar-field metric-affine action of \cite{Rigouzzo:2022yan} provide equivalent parametrizations of the same class of theories. In deriving the parity-odd part of this map, the four-dimensional Schouten identities for the Levi-Civita-tensor contractions have been used. In contrast, the relations involving the derivative couplings in our notation can be integrated to recover the corresponding functions $A_i$ of Ref.~\cite{Rigouzzo:2022yan}, up to additive constants multiplying total divergences. After accounting for the different conventions and integrating by parts the divergence couplings of \cite{Rigouzzo:2022yan}, the relations between the functions in the two papers are the following (l.h.s. current paper notation = r.h.s. \cite{Rigouzzo:2022yan} notation):
\begin{align*}
\mathcal{A} &= \kappa \Omega^2, & \mathcal{B} &= \tilde{K}, & \mathcal{V} &= V,\\
        a_1 &=\kappa  \left(2 B_4-\tfrac{\Omega^2}{4}\right) , 
        & a_2 &=\kappa  \left(2 B_5+\tfrac{\Omega^2}{2}\right), 
        & a_3 &=\kappa  \left(2 B_1-\tfrac{5 B_4}{9}+\tfrac{B_5}{9}+\tfrac{\Omega^2}{4}\right),\\[2pt]
        a_4 &=2 \kappa  \left(B_2-\tfrac{4 B_4}{9}-\tfrac{B_5}{9}\right) , 
        & a_5 &=\kappa  \left(2 B_3+\tfrac{4 B_4}{9}-\tfrac{8 B_5}{9}-\tfrac{\Omega^2}{2}\right), &a_6 &=2 \kappa  D_2,\\[2pt]
        b_1 &= \kappa  \left(-16 C_2+\tfrac{16 C_4}{3}-\Omega^2\right), 
        & b_2 &=\kappa  \left(-32 C_2-\tfrac{16 C_4}{3}+2 \
        \Omega^2\right) ,\\[4pt]
        b_3 &=\kappa  \left(8 C_1-\tfrac{16 C_4}{3}+4 \Omega^2\right), 
        & b_4 &=8 \kappa  \left(C_3-\tfrac{D_1}{3}\right) ,
        & b_5 &= 2 \kappa  D_1,\\[2pt]
         c_1 &=\kappa  \left(4 E_5-2 \Omega^2\right) ,
        & c_2 &= \kappa  \left(4 E_1-\tfrac{4 E_5}{3}+2 \Omega^2\right),
        & c_3 &=\kappa  \left(4 E_3+\tfrac{4 E_5}{3}-2 \Omega^2\right),\\[2pt]
        c_4 &= \kappa  \left(\tfrac{2 D_3}{3}+4 E_2\right),
        & c_5 &= \kappa  \left(-\tfrac{2 D_3}{3}+4 E_4\right),
        & c_6 &=-2 \kappa  D_3 ,\\[2pt]
        \mathcal{C}_1 &=2 A_4^{\prime}-2 \Omega \Omega^{\prime}, &\mathcal{C}_2 =& 2 A_3^{\prime}+2 \Omega \Omega^{\prime}, & \mathcal{C}_3 & =4 A_2^{\prime}-8 \Omega \Omega^{\prime} ,\\[2pt]
        \mathcal{C}_4 & = 4 A_1^{\prime}\,.
\end{align*}

\bibliographystyle{utphys}
\bibliography{ref}
\end{document}